\documentclass[useAMS,usenatbib]{mn2e}
\usepackage{times}

\usepackage{amssymb}
\usepackage{psfig}

\title[Chemo-photometric evolution of spiral galaxies]{Energetic 
constraints to chemo-photometric evolution of spiral galaxies}

\author[A. Buzzoni]{Alberto Buzzoni\\
INAF - Osservatorio Astronomico di Bologna, Via Ranzani 1, 40127 Bologna, Italy
}

\begin{document}

\date{Accepted ... Received ... in original form}

\pagerange{\pageref{firstpage}--\pageref{lastpage}} \pubyear{2011}

\maketitle

\label{firstpage}

\begin{abstract}
The problem of chemo-photometric evolution of late-type 
galaxies is dealt with relying on prime physical arguments of energetic 
self-consistency between chemical enhancement of galaxy mass, through 
nuclear processing inside stars, and luminosity evolution of the system.
Our analysis makes use of the Buzzoni template galaxy models along the Hubble morphological sequence. 
The contribution of Type {\sc ii} and {\sc i}a SNe is also accounted for in our 
scenario. 
Chemical enhancement is assessed in terms of the so-called 
``yield metallicity'' ($\cal Z$), that is the metal abundance of processed 
mass inside stars, as constrained by the galaxy photometric history.
For a Salpeter IMF, ${\cal Z} \propto t^{0.23}$
being nearly insensitive to the galaxy star formation history.
The ISM metallicity can be set in terms of ${\cal Z}$, and just modulated
by the gas fraction and the net fraction of returned stellar mass ($f$). 
For the latter, a safe upper limit can be placed, such as $f \lesssim 0.3$ at 
any age.

The comparison with the observed age-metallicity relation allows us to to set a firm 
upper limit to the Galaxy birthrate, such as $b \lesssim 0.5$, and to
the chemical enrichment ratio $\Delta Y/\Delta Z \lesssim 5$.
About four out of five stars in the solar vicinity are found 
to match the expected $\cal Z$ figure within a factor of two, a feature 
that leads us to conclude that star formation in the Galaxy must have
proceeded, all the time, in a highly contaminated environment where returned  
stellar mass is in fact the prevailing component to gas density. 

The possible implication of the Milky Way scenario for the more general picture of 
late-type galaxy evolution is dicussed moving from three relevant 
relationships, as suggested by the observations. Namely, {\it i)} the down-sizing 
mechanism appears to govern star formation in the local Universe; {\it ii)} 
the ``delayed'' star formation among low-mass galaxies, as implied by the inverse 
$b$-$M_{\rm gal}$ dependence, naturally leads to a more copious gas fraction 
when moving from giant to dwarf galaxies; {\it iii)} although lower-mass galaxies 
tend more likely to take the look of later-type spirals, it is mass, not  
morphology, that drives galaxy chemical properties.
Facing the relatively flat trend of $\cal Z$ vs.\ galaxy type, the increasingly 
poorer gas metallicity, as traced by the $[O/H]$ abundance of H{\sc ii} regions 
along the Sa~$\to$~Im Hubble sequence, seems to be mainly the result of the 
softening process, that dilute enriched stellar mass within a larger fraction of 
residual gas.

The problem of the residual lifetime for spiral galaxies as active 
star-forming systems has been investigated. If returned mass is 
left as the main (or unique) gas supplier to the ISM, as implied by the Roberts 
timescale, then star formation might continue only
at a maximum birthrate $b_{\rm max} \ll f/(1-f) \lesssim 0.45$, for a Salpeter IMF.
As a result, only massive ($M_{\rm gal} \gtrsim 10^{11}~M_\odot$) Sa/Sb 
spirals may have some chance to survive $\sim 30$\% or more beyond a Hubble time.
Things may be worse, on the contrary, for dwarf systems, that seem currently on 
the verge of ceasing their star formation activity unless to drastically 
reduce their apparent birthrate below the $b_{\rm max}$ threshold.
\end{abstract}



\section{Introduction \label{s_i}}

A fair estimate of ``galaxy metallicity'' has been widely recognized as a
basic issue when assessing the distinctive evolutionary properties of late-type 
galaxies (LTGs), including all disk-dominated systems along the Hubble morphological
sequence. However, since both spiral and irregular galaxies consist of a composite 
stellar population, the question cannot easily be settled, as it depends on the specific 
criterion we want to adopt to identify a value of $Z$ (or equivalently of [Fe/H])
ideally representative of the stellar system as a whole.

For example, the youngest stars closely track the chemical composition of gas clouds 
within a galaxy, and we could therefore refer to their value of $Z$ as the
galaxy {\it actual} metallicity ($Z_o$). On the other hand, this value may 
not account for the composition of earlier generations of stars, that 
however still provide a sizable contribution to galaxy total luminosity.
In this regard, one would better like to consider an {\it effective} 
metallicity estimate ($Z_{eff}$), by weighting the composing stellar populations 
with their actual photometric contribution or with the number of star members in 
each group. 

These arguments yet emerge {\it ``in nuce''} from the seminal analysis of 
\citet{schmidt63}, but only \citet{arimoto87} extensively dealt with the problem of a fair estimate of the  
effective metallicity in their LTG models using bolometric luminosity as a weighting 
factor to account for the relative contribution of the composing stellar populations. 
With this criterion, by considering a collection of theoretical simple stellar 
populations (SSPs) with a spread in the values of Z, \citet{buzzoni95} showed that 
the integrated spectral energy distribution (SED) for the aggregate still resembled 
the spectrum of just one SSP with $\log Z_{\rm eff} = \langle \log Z \rangle$.
From her side, \citet{greggio97} further refined these arguments, at least in the 
early-type galaxy framework, emphasizing the difference between average methods such 
as to lead to $[<Fe/H>]$ vs.\ $<[Fe/H]>$.
A direct estimate of both $Z_o$ and $Z_{\rm eff}$ is provided by \citet[][hereafter AJ91]{arimoto91}
in their LTG models.  As expected, the effective metallicity is found to be systematically
lower by a 20-30\% with respect to the actual one.

Apart from any operational definition, however, one has to remark that the classical 
approach to galaxy chemical evolution, usually pursued in the literature, mainly proceeds 
from a dynamical point of view, where chemical composition is tightly related to the kinematical 
properties of the different stellar populations sharing the galaxy morphological 
design. \citet{sandage87} and \citet{koeppen90} are striking examples, in this regard, 
of the empirical and theoretical focus to the problem, respectively.
After all, just on these basis \citet{baade44} led to his first definition of Pop I 
and II stars.

However, an even more direct physical argument ties metal abundance to galaxy luminosity 
evolution. For any gram of metals eventually produced in a galaxy, in fact, one gram of 
Hydrogen has been burnt at some time releasing a fixed amount of energy.
As a consequence,{\it a straightforward relationship must be in place between the actual amount of 
enriched mass in a galaxy (M$_{\rm YZ}$), in the form of metals plus Helium in excess to 
the primordial abundance, and the past luminosity evolution.} The importance of this
constraint, often overlooked even in the recent literature, has been first raised   
by \citet{wirth81} in the context of early-type galaxy evolution, and more explicitely 
assessed by \citet{pagel97}. This argument also found a more definite theoretical settlement 
through the so-called ``Fuel consumption theorem'', as proposed by \citet{rb83,rb86}.

Given the range of interesting implications for the study of galaxy evolution,
we want here to further investigate the problem of the energetic constraint
to chemo-photometric evolution of galaxies, with special attention to the more
composite scenario of late-type systems. Our analysis will rely on the template galaxy
models developed in \citet{buzzoni05,buzzoni02}. To better catch the real intention of this paper, we
would like to warn the reader that our discussion will proceed as far as possible on
analytical bases. This will forcedly require some educated simplifications to 
the problem (after careful physical assessing however of the neglected complications),
with the aim of more neatly let to emerge the leading relationships that govern galaxy 
evolution, and in particular the way these systems (and the Universe as a whole) constrained 
their chemical composition at the different cosmic epochs.

\section{Theoretical fundamentals} 

In order to better single out the relevant physical boundaries involved in our problem,
we need to consider basically the two main photometric contributions, that constrain 
overall galaxy luminosity evolution. From one hand, in fact, we can recognize
the ``smooth'' photometric contribution by stars along their quiescent 
(namely ``hydrostatic'') evolutionary stages, according to a given IMF.
On the other hand, one has also to consider the Supernova bursting events, 
that interleave galaxy evolution with short but extremely relevant episodes
of mass processing.

Following \citet{clayton83}, H-burning cycles release 26.731 MeV per $^4$He particle, 
while a supplementary amount of 7.274 MeV per $^{12}$C nucleus is produced by the 3$\alpha$ 
He-burning cycle. Overall, the energy budget of the nuclear processes that lead to 
Carbon synthesis in stars reaches a total of
\begin{equation}
W_* = 6.98\ 10^{18}~{\rm erg~g}^{-1}.
\end{equation}
This specific 
energy output accounts for the ``quiescent'' luminosity emission of a galaxy provided by 
the non-explosive evolutionary stages of stars along the entire IMF mass range. 

A supplementary important source of energy and ISM metal enrichment should however comprise
also the SN explosive events. A quite different elemental pattern characterizes 
the ISM chemical enrichment in case of SNe{\sc i}a, meant to be the main synthesizers of 
the Fe-Ni elemental group \citep{chevalier76,nomoto80}, and SNe{\sc ii}, which stem on the 
contrary from the core collapse of high-mass stars and mainly supply Oxygen and other heavier 
$\alpha$ elements \citep[e.g.][]{matteucci86,trimble91}.
Nevertheless, both these events may actually be treated coherently in terms of specific 
energy release as they take in charge the post-Carbon nuclear burning up to Iron 
synthesis (and beyond). Again, following \citet{clayton83}, whatever the chemical path of 
nuclear reactions, the atomic binding energy difference implies to release 60.605~MeV 
per $^{56}$Fe nucleus or, equivalently,
\begin{equation}
W_{\rm SN} = 1.03\ 10^{18}~{\rm erg~g}^{-1}.
\label{eq:wsn}
\end{equation}

Compared to ``quiescent'' nuclear reaction chains, for fixed amount
of released energy, explosive nucleosynthesis is therefore roughly a factor of 
seven more efficient in the whole $Z$ enrichment of the Universe,
selectively involving, however, only the heaviest metal supply 
\citep[see, e.g.][for a brief but timely focus on this subject]{matteucci04}.

\subsection{Processed mass by ``quiescent'' luminosity evolution} 
 
Let us explore first the general case of a system consisting of a composite stellar population 
(CSP), as a result of a roughly continual feeding of fresh stellar generations along the 
whole photometric history.
The nature itself of the problem suggests to consider the total bolometric luminosity of 
the CSP as a convolution integral of several simple stellar populations (SSPs) continually 
distributed in time according to a given star formation rate (SFR), namely
\begin{equation}
{\cal L}_{\rm CSP} = \int {\cal \ell}_{\rm SSP}(\tau) \otimes {\rm SFR}\, d\tau.
\label{convol}
\end{equation}

In \citet{buzzoni05} we have shown that SSP bolometric luminosity (${\cal \ell}_{\rm SSP}$) 
smoothly evolves with time according to a simple power law such as 
${\cal \ell}_{\rm SSP}(t) \propto t^{-\alpha}$; for a Salpeter IMF, the power index 
$\alpha = 0.77$, is virtually independent from the SSP chemical 
composition.\footnote{Our SSP evolution assumes a \citet{salpeter55} IMF
such as ${\cal N}(M) \propto M^{-s}$, with the power index $s=2.35$ along the accounted 
stellar mass range from 0.1 to 120~M$_\odot$. See \citet{buzzoni05,buzzoni02} for 
further details.\label{foot1}} 
Our template galaxy models considered a disk SFR with a characteristic 
birthrate, $b = SFR_o / \langle SFR \rangle$ \citep[see, for instance][]{miller79},
depending on the morphological type \citep{kennicutt94}.
The distinctive gas~$\to$~star conversion efficiency implies for the star formartion rate
a power-law time dependence such as ${\rm SFR}(t) \propto t^{-\beta}$, so that
$b = (1-\beta)$ \citep[see][for a full discussion]{buzzoni05}.

Within this framework, the total luminosity of a CSP at time $t$ becomes therefore
\begin{equation}
{\cal L}_{\rm CSP}(t) = C\ \int_0^t \tau^{-\alpha}\ (t-\tau)^{-\beta}\, d\tau,
\label{lcsp}
\end{equation}
with a normalization constant $C$ depending on the total mass of the aggregate.
Providing that both $\alpha < 1$ and $\beta < 1$, the integral equation has 
the Euler Beta function as a straightforward analytical solution. More explicitely,
it can also be written as
\begin{equation}
{\cal L}_{\rm CSP}(t) = {{\Gamma(1-\alpha)\Gamma(1-\beta)}\over 
{\Gamma(2-\alpha-\beta)}}\ C\,t^{1-\gamma},
\end{equation}
being $\Gamma$ the Gamma function and $\gamma = \alpha + \beta$. Compared to a reference epoch $t_o$, 
one can simply write
\begin{equation}
{\cal L}_{\rm CSP}(t) = {\cal L}_{\rm CSP}(t_o) \left({t\over t_o}\right)^{1-\gamma}.
\label{eq:lcsp}
\end{equation}

If we integrate the CSP emitting power ${\cal L}_{\rm CSP}$ over the entire galaxy life,
until time $T$, this eventually gets the global energy $\cal E$ released by the nuclear 
processes to sustain galaxy luminosity until that age. 
\begin{equation}
{\cal E}(T) = \int_0^T {\cal L}_{\rm CSP}(t)\, dt = {\cal E}_o\ \left({T\over t_o}\right)^{2-\gamma},
\end{equation}
where, ${\cal E}_o = [{\cal L}_{\rm CSP}(t_o)\times t_o]/(2-\gamma)$.

Clearly, at any time the latter quantity {\it must} balance the amount of processed mass 
M$_{\rm YZ}$. With respect to the total CSP stellar mass (that is the total mass ever converted
in stars, ${\rm M}^*(T) = \int_0^T {\rm SFR}(t)\ dt$), we have that 
\begin{equation}
{{\rm M}_{\rm YZ}^* \over {\rm M}^*}\Big|_T =  
{ W_*^{-1}{\cal E}(T) \over {\int_0^T {\rm SFR}(t)\ dt}} = {{\cal E}_o \over {W_*\, {\rm M}_o}}\ 
\left({T\over t_o}\right)^{1-\alpha},
\label{eq:stars}
\end{equation}
where $W_*$ is the energy-mass conversion factor set by the nuclear burning, according to our previous 
arguments.\footnote{Note that $W_*$ directly relates to the total efficiency $\epsilon$ of nuclear
reactions. According to the Einstein equation, i.e.\  ${\cal E} = \epsilon\,\Delta Mc^2$,
we have that $W_* = {\cal E}/\Delta M = \epsilon\,c^2$. This eventually leads to a mass $\to$ energy 
conversion efficiency of $\epsilon = 0.0078$.\label{foot2}} Again, we set 
${\rm M}_o = {\rm M}^*(t_o) = \int_0^{to} {\rm SFR}(\tau)\ d\tau$ for notation convenience.
Interestingly enough, eq.~(\ref{eq:stars}) shows that the way metals enhance
with time in a CSP does {\it not} depend on the SFR details, being fully modulated on the contrary 
by the IMF slope, which eventually constrains the CSP luminosity evolution through the power 
index $\alpha$. Furthermore, in the same equation, note that a parameter 
${\cal F} = ({\cal E}_o/M_o)/W_*$ can also be regarded as a ``burning efficiency factor'', 
which is a measure of how deeply processed mass has been exploited to produce luminosity.

\subsubsection{Trading mass vs.\ luminosity: the galaxy M/L ratio}

Reference stellar mass (${\rm M}_o$) and the corresponding CSP luminosity 
(${\cal L}_{\rm CSP}(t_o)$) are, of course, strictly related depending on the SFR details. 
Relying on the \citet{buzzoni05,buzzoni02} theoretical framework, we computed in Table~\ref{t1} the 
expected value of the $M/L$ ratio at 15~Gyr for different CSPs (col.~3 in the table) along 
a range of values of the SFR power index, $\beta$ (col.~1) and stellar birthrate $b$ (col.~2). 
These values are representative of the 
full range of scenarios that characterize disk evolution for the different LTG models. 
From the values of $M/L$ one can easily derive the expected CSP bolometric luminosity, 
${\cal L}_{\rm CSP}(t_o)$, as well as the value of ${\cal E}_o$ at the reference age 
(col.~4 and 5, respectively, of Table~\ref{t1}). Finally, the burning efficiency factor, 
${\cal F}$ is also computed in column 6.

As both the total stellar mass and luminosity respond in the same way to the SFR, one
has that 
\begin{equation}
{M\over {\cal L}}\Big|_t = {M\over {\cal L}}\Big|_{t_o} \left({t\over t_o}\right)^\alpha,
\end{equation}
that is {\it time evolution of galaxy M/L ratio is a nearly universal law, that 
only depends on the observing wavelength} and, for bolometric and a Salpeter IMF, 
scales aproximately as $M/L \propto t^{3/4}$.

The SFR enters, on the contrary, by setting the absolute value of $M/L$,
at a given age; for example, when compared to a simple star burst, a smooth star 
formation will result, on average, in a larger number of young bright stars of high mass
at a given age. This means, in general, that $M/L$ decreases with decreasing $\beta$.
In this regard, a convenient fit to the data of Table~\ref{t1} provides:
\begin{equation}
(M/L)_{15} = {1\over {(0.23+b)}} +0.19,
\label{eq:mlb}
\end{equation}
with an rms of $\pm 0.04$ on the predicted ratio.

\begin{table}
\scriptsize
\caption{Reference values at $t = 15$~Gyr for a $10^{11}M_\odot$ theoretical CSP 
with changing star-formation properties$^{(a)}$}
\label{t1}
\begin{tabular}{rccccc}
\hline
$\beta$ & b & ${\rm M}_*/{\cal L}_{\rm CSP}$ &  ${{\cal L}_{\rm CSP}}^{(b)}$  & ${\cal E}_o/M_o$ & ${\cal F}^{(c)}$ \\
        &   & [M$_\odot$/L$_\odot$]     &     [L$_\odot$] & $\times 10^{17}$ [erg g$^{-1}$] &  \\
        &   &                           &                                   &         &  \\
\hline
 0.8  & 0.2 & 2.42 & 0.41 & 8.77 & 0.126 \\
 0.6  & 0.4 & 1.82 & 0.55 & 8.03 & 0.115 \\
 0.4  & 0.6 & 1.44 & 0.69 & 7.65 & 0.110 \\
 0.2  & 0.8 & 1.19 & 0.84 & 7.50 & 0.108 \\
 0.0  & 1.0 & 1.02 & 0.98 & 7.33 & 0.105 \\
--0.2 & 1.2 & 0.89 & 1.12 & 7.21 & 0.103 \\
--0.4 & 1.4 & 0.80 & 1.25 & 7.06 & 0.101 \\
--0.6 & 1.6 & 0.72 & 1.38 & 6.94 & 0.099 \\
--0.8 & 1.8 & 0.66 & 1.51 & 6.84 & 0.098 \\
--1.0 & 2.0 & 0.61 & 1.64 & 6.77 & 0.097 \\
\hline
\noalign{$^{(a)}$ Assuming SFR~$\propto t^{-\beta}$, and a birthrate $b = 1-\beta$}
\noalign{$^{(b)}$ For a 1~M$_\odot$ CSP}
\noalign{$^{(c)}$ Burning efficiency factor ${\cal F} = ({\cal E}_o/M_o)\,K^{-1}$}
\end{tabular}
\end{table}

\subsection{Supernova processing}

The ``quiescent'' evolution of galaxy luminosity, as described by previous equations, does not
take into account the extra-energy release (and the corresponding extra-metal production) by 
the SN events. As well known, supernovae display a composite genesis. From one hand, in fact,
Type {\sc ii} SNe  basically deal with the C/O core collapse of individual stars with  
conveniently high mass such as to exceed the Chandrasekhar limit near
the end of their photometric evolution. On the other hand, a much more composite family of
objects (shortly grouped into Type {\sc i}a SNe) has to be reconducted to 
binary-star evolution. In this case, the key mechanism deals with the mass transfer between
the system members, where a primary (originally more massive) star, in its way to the 
final white-dwarf (WD) death, acts as a fresh-matter supplier to boost evolution 
of the secondary member. The latter star will therefore be prompted, at some stage, 
to feed back the WD with fresh fuel until triggering (under appropriate physical 
conditions) the fatal detonation on the WD surface.

As well known, the complexity of this scenario, characterized by a number of unknown 
or poorly constrained physical parameters (i.e.\ orbit parameters, total mass of the system,
and mass ratio of the members), is the main source of uncertainty even in more 
explicit and sophisticated models of galaxy chemical evolution 
\citep[see, in particular,][for a brilliant assessment of the problem]{matteucci86}. 
All the more reason, let us try here a simplified approach, just to single out the few 
leading quantities that constrain our relevant physical output.

\subsubsection{Type {\sc ii} SNe}

Considering first the core-collapse events, originating by the explosion
of stars with $M_{\rm up} \gtrsim 8 \pm 1$~M$_\odot$ \citep[see, e.g.][]{smartt09} 
one has that the expected total number of events per unit SSP total mass can be written as
\begin{equation}
{n_{\rm SNII}}(\rm SSP) = \left({2-s}\over {1-s}\right)\,{{[M^{1-s}]_{Mup}^{120}}\over
{[M^{2-s}]_{0.1}^{120}}},
\end{equation}
where we assume a power-law IMF according to footnote~\ref{foot1} remarks.
For a Salpeter case (i.e.\ $s = 2.35$), and in force of the fact that galaxy
SFR does not change much on a time-scale comparable with the lifetime of stars with 
$M \ge M_{\rm up}$ \citep[namely a few $10^7$~yr at most, see e.g. eq.~3 in][]{buzzoni02}, 
we have
\begin{equation}
{n_{\rm SNII}}(t) = 0.126\, M_{\rm up}^{-1.35} \,{\rm SFR}(t) \qquad [{\rm yr}^{-1} (M_\odot {\rm yr}^{-1})^{-1}]
\label{eq:sn200}
\end{equation}
For $M_{\rm up} = 8 M_\odot$, this leads to a rate of $7.6\,10^{-3}$ SN{\sc ii} events 
per year for a $SFR = 1$~M$_\odot$~yr$^{-1}$. This figure perfectly matches the empirical rate of 
$(7.5\pm 2.5)\,10^{-3}$ independently obtained by \citet{scannapieco05} on the basis of the 
observed SN rate density within $z \le 1$ by \citet{dahlen04} compared against the 
cosmic SFR density as measured by \citet{giavalisco04}.

\begin{figure}
\centerline{
\psfig{file=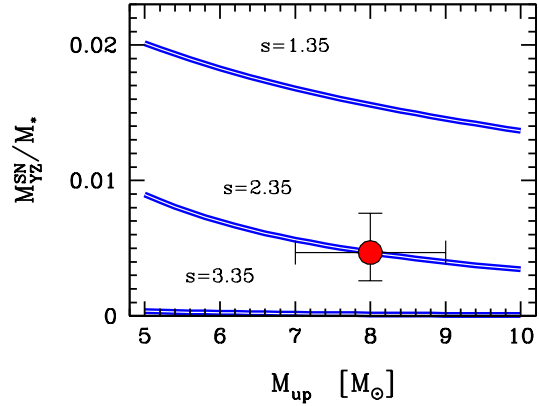,width=0.9\hsize}
}
\caption{The expected fraction of enriched stellar mass provided by the
SN~{\sc ii} events according to three different IMF power-law indices
(Salpeter case is for $s = 2.35$), and for a different value of
the SN triggering mass, $M_{\rm up}$. The consensus case is marked by
the big dot on the Salpeter curve. Vertical error bars derive from
eq.~(\ref{eq:sn2}).
}
\label{f1}
\end{figure}

If $E_{51}^{SNII}$ is the mean released energy associated to each SN event (in unit of 10$^{51}$ erg), 
and recalling the SN specific energy release as previously discussed, then the fraction of 
enriched stellar mass (M$_{\rm YZ}^{\rm SN}$) along time results
\begin{equation}
{{\rm M}_{\rm YZ}^{\rm SNII} \over {\rm M}^*} =  n_{\rm SNII} E_{51}^{II} {{10^{51}}\over 
{W_{\rm SN}\, M_\odot}} = 0.061\,M_{\rm up}^{-1.35}\, E_{51}^{SNII},
\label{eq:sn20}
\end{equation}
being $W_{\rm SN}$ the energy-mass conversion factor set by the explosive nuclear burning as in 
eq.~\ref{eq:wsn}, and M$_\odot$ the mass of the Sun. The output of eq.~(\ref{eq:sn20})
is displayed in graphical form in Fig.~\ref{f1}, exploring also different IMF slopes.
For $M_{\rm up} = 8 \pm 1$~M$_\odot$, and $E_{51}^{SNII} = 1.3 \pm 0.3$
\citep{thielemann96,hashimoto95,kasen07}
one derives
\begin{equation}
{{\rm M}_{\rm YZ}^{\rm SNII} \over {\rm M}^*} = 0.0047^{+29}_{-21}.
\label{eq:sn2}
\end{equation}
As for eq.~(\ref{eq:stars}), note that both M$_{\rm YZ}^{\rm SNII}$ and M$^*$ depend on 
$\int {\rm SFR}(t) dt$, so that the M$_{\rm YZ}^{\rm SNII}/$M$^*$ ratio does {\it neither} depend 
on time {\it nor} on the SFR details, being fully constrained by the IMF slope alone (and by the 
preferred value of $M_{\rm up}$, as well), as displayed in Fig.~\ref{f1}.  
As a consequence, the appearence of SN{\sc ii} on the scene of galaxy evolution just enters
with a straight offset to metallicity.\footnote{Given the special nature of the 
explosive SN nuclesynthesis, the reported $E_{51}^{SNII}$ value should account for the 
emitted luminosity, as obtained for instance from the
time integration of the SN lightcurve \citep[e.g.]{milone88}, plus the kinetic energy 
associated to the SN burst. The latter contribution is actually by far the prevailing one,
and it exceeds by about two orders of magnitude the emitted luminosity itself \citep{woosley05}.}

\subsubsection{Type {\sc i}a SNe}

From the theoretical side, \citet{greggio83} have first argued that the SN{\sc i}a 
population originates from the mix of two stellar components. In fact, the
SN detonation may set on both on He-rich WDs (mainly coming from the evolution of
relatively low-mass stars, about 2-3~M$_\odot$), and on the surface of C/O-rich WDs
(the final output of more massive stars, about 5-8~M$_\odot$)\footnote{In any case, 
the upper mass of SN{\sc i} progenitors must be edged by the on-set 
mass (M$_{\rm up}$) of core-collapsed SNe{\sc ii}.}
This difference, together with the binary nature of the physical mechanism that
drives the explosion, makes the total SN{\sc i}a rate to depend (at least) on two reference 
parameters. From one hand, simple evolutionary arguments on the stellar lifetime
sets a minimum characteristic ``delay time'' ($\Delta_{\rm SNI}$) for the SN burst, which may follow 
the star formation event (and the first SN{\sc ii} burst) by a few 10$^7$~yrs for 
CO-rich WDs \citep{tornambe86} up to several 10$^8$~yrs for He-rich WDs \citep{greggio83}. 
On the other hand, the broad range of 
geometric and physical parameters of the binary system may even further displace in time the 
SN event, blurring the contribution of the two feeding channels into a composite
delay-time distribution, namely $\phi_{(\Delta)}(t)$, that we eventually have to deal with.

From the observing side, \citet{mannucci05} have firmly demonstrated that SN{\sc i}a
rate among spiral galaxies tightly relates to the corresponding rate of core-collapsed SNe, 
the latter meant to closely trace the galaxy SFR. This feature evidently sets important
contrains to the $\Delta_{\rm SNI}$ distribution, leading \citet{greggio05} to conclude that, in a 
SSP ``the distribution function of the delay times decreases with time. This means that the 
majority of SN{\sc i}a precursors are relatively short lived''.  On this line, \citet{maoz10} argued that 
over 50-85\% of the whole SN{\sc i}a population in a SSP are expected to explode within the first Gyr 
from the starburst. On the other hand, the residual presence of Type {\sc I}a SNe even
among quiescent early-type galaxies, that is in lack of any evident star formation activity,
calls for a few very delayed events and therefore for a skewed long-term tail of the 
delay-time distribution \citep{dallaporta73,mannucci05,panagia07,brandt10}. Theoretical figures 
for the SSP case indicate that 
$\phi_{(\Delta)} \propto t^{-\eta}$, with $\eta \sim 1.2$ \citep{maoz10}, or $\sim 2.3$, as in the 
\citet{greggio05} or \citet{matteucci01} models.\footnote{Note that, in all cases, $\eta > \beta$.
This assures $\phi_{(\Delta)}$ to fade with time at a quicker rate compared to the reference
SFR for our model galaxies. In force of this argument, we can therefore expect
the bulk of SN{\sc i}a to safely trace the SFR all the time.}

Similarly to what already attempted by \citet{scannapieco05}, we adopt for our discussion the 
\citet{mannucci05} empirical parameterization for the observed SN{\sc i}a rate (see eq.~2 therein). 
One can therefore envisage a prevailing term directly related to galaxy SFR and an ``extended'' 
SN fraction (to recall the \citealt{scannapieco05} original notation), that collects the most 
delayed events and can therefore be meant to scale at any time with $\int {\rm SFR}(t) dt$, that 
is with the total stellar mass of a galaxy.

For the first component, the direct SFR dependence is obtained through the core-collapsed SN rate as from 
eq.~(\ref{eq:sn200}) assuming, with \citet{mannucci05}, that
\begin{equation}
n_{\rm SNI} = 
\begin{array}{l}
~~~\\
0.35~n_{\rm SNII}.\\
\pm \phantom{3}8
\end{array}
\end{equation}
By relying then on eq.~(\ref{eq:sn20}), this leads to
\begin{equation}
{{\rm M}_{\rm YZ}^{\infty} \over {\rm M}^*} = 0.35\,\left({{E_{51}^{SNI}}\over {E_{51}^{SNII}}}\right)\, 
{{\rm M}_{\rm YZ}^{\rm II} \over {\rm M}^*},
\label{eq:prompt}
\end{equation}
where the associated energy for a typical SN{\sc i} event (again, in unit of $10^{51}$~erg) is 
of the same order than for SNe{\sc ii}, namely ${E_{51}^{SNI}} = 1.5 \pm 0.5$ 
\citep{nomoto84,khokhlov93,reinecke02,hillebrandt03}. 

However, the l.h. term of eq.~(\ref{eq:prompt}) has to be regarded as a somewhat asymptotic value 
for the enriched mass fraction as it considers virtually no delay in the SN{\sc i}a appearence,
being both enriched and total stellar mass perfectly tuned with SFR.
If we set SNe{\sc i}a to come on the galaxy stage just after the first 
SN{\sc ii} metal ``glitch'', say for $\Delta_{\rm SNI} \gtrsim 50$~Myr or so, i.e.\ the reference 
lifetime for a 5~M$_\odot$ star, then one has to scale previous equation by a factor 
$(1-\tau^{\beta-1})$, with $\tau = t/\Delta_{\rm SNI}$.
After this correction, we eventually obtain
\begin{equation}
{{\rm M}_{\rm YZ}^{\rm SNI} \over {\rm M}^*} = 0.0019^{+15}_{-13}\,(1-\tau^{\beta-1}),
\end{equation}
where the error bars have been conservatively computed by logarithmic error propagation.

To this contribution one has further to add the long-term ``extended'' SN component whose number of events, 
according to \citet{mannucci05}, amounts to 
\begin{equation}
n_{\rm SNI}^{{\rm ``ext''}} = 
\begin{array}{l}
~~~\\
~\,4.7~10^{-14}   \qquad [{\rm yr}^{-1} {\rm M}_\odot^{-1}].\\
\pm \phantom{.}9
\end{array}
\label{eq:tardy}
\end{equation}
This rate is nicely confirmed, within the errors, also by \citet{sullivan06}.

In case of a star-forming CSP with power-law SFR, the mass-specific event number that
occurred up to age $T$ can be written as
\begin{eqnarray}
{{N_{\rm SNI}}\over {{\rm M}^*}}\Big|_T^{{\rm ``ext''}} & = & n_{\rm SNI}^{{\rm ``ext''}}\,\left[
{{\int_{\Delta}^T \int_0^{t} {\rm SFR}(\tau)\,d\tau\,dt}\over {\int_0^T {\rm SFR}(\tau)\,d\tau}}\right] \hfill \nonumber \\
 \quad    & \quad  & \quad \nonumber \\
     & = & \Delta_{\rm SNI}\, n_{\rm SNI}^{{\rm ``ext''}}
\,{{\tau \,(1 - \tau^{\beta-2})} \over {2-\beta}}. 
\label{eq:nsn1}
\end{eqnarray}

\begin{figure}
\centerline{
\psfig{file=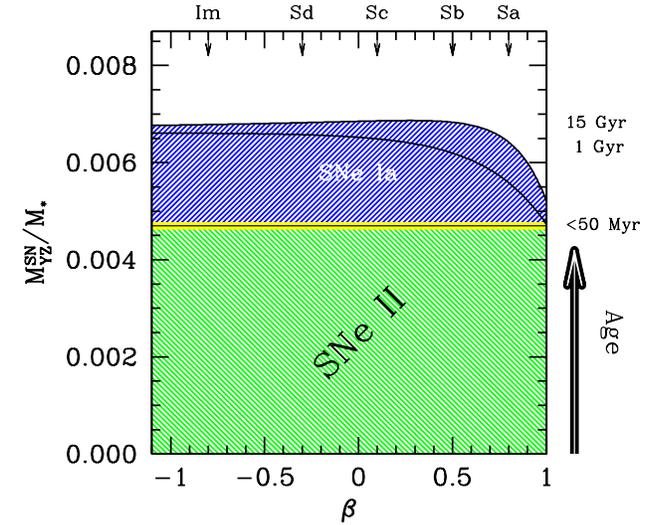,width=\hsize}
}
\caption{The chemical enrichment from SN contribution according
to the different components. Note the prevailing role of core-collapsed objects
(i.e.\ SN~{\sc ii}) along the early stages of galaxy evolution ($t \lesssim 50~$Myr,
see labels to the right axis). The total SN~{\sc i}a component adds then a further 0.2~dex 
to galaxy metallicity, mostly within the first Gyr of life.
}
\label{snrall}
\end{figure}

The corresponding amount of processed mass directly derives as
\begin{equation}
{{{\rm M}_{\rm YZ}^{\rm SNI}} \over {\rm M}^*}\Big|^{{\rm ``ext''}} = 
{{N_{\rm SNI}}\over {{\rm M}^*}}\Big|_T^{{\rm ``ext''}}  E_{51}^{SNI} {{10^{51}}\over 
{W\, M_\odot}}.
\label{eq:ext}
\end{equation}
By replacing the relevant quantities, we eventually obtain
\begin{equation}
{{{\rm M}_{\rm YZ}^{\rm SNI}} \over {\rm M}^*}\Big|^{{\rm ``ext''}} =  
{\Delta_{\rm SNI} \over 2.9\,10^{13}} \,{{\tau\,[1 - \tau^{-(1+b)}]}\over {1+b}},
\end{equation}
providing to express the delay time $\Delta_{\rm SNI}$ in years.

In conclusion, the relative amount of processed mass related to the whole SN activity in a 
CSP of SFR power-law index $\beta$ can be summarized as
\begin{equation}
{{\rm M}_{\rm YZ}^{\rm SN} \over {\rm M}^*} = 0.0047 +\left\{
\begin{array}{l}
0 \hfill {\rm for~} \tau < 1\\
~~\\
0.0019\,A(\tau,\Delta_{\rm SNI},b) \qquad \hfill {\rm for~} \tau \ge 1,
\end{array}
\right.
\label{eq:snrall}
\end{equation}
with

\begin{equation}
A(\tau,\Delta_{\rm SNI},b) = (1-\tau^{-b}) + {\Delta_{\rm SNI} \over 5.5\,10^{10}}\,{{\tau\,[1 - \tau^{-(1+b)}]}\over {1+b}}
\label{eq:a}
\end{equation}

In Fig.~\ref{snrall} we displayed the relevant output in graphical form, for different ages 
and values of the SFR power index $\beta$. In the plot we also marked the
representative values of $\beta$ for spiral galaxies of different Hubble morphological
type, as derived from the \citet{buzzoni05} templates.

\subsection{Metals versus Helium: the enrichment ratio}

With the help of data in Table~\ref{t1}, through eq.~(\ref{eq:lcsp}) and (\ref{eq:stars}),
it is immediate to derive both disk luminosity (i.e.\ ${\cal L}_{\rm CSP}$) and the amount of
chemically enriched mass (${\rm M}_{\rm YZ}$) along galaxy life.

One may even rely on the expected enrichment ratio 
$R = \Delta Y/\Delta Z$ to single out the fraction of Helium and heavy 
elements eventually produced. The $R$ parameter can be tuned up either empirically,
for example via the observation of Galactic and extragalactic H{\sc ii} clouds 
\citep{lequeux79,serrano81,pagel92,maciel01,peimbert02,casagrande07}, or theoretically, from stellar evolution 
theory \citep{maeder92,marigo03}. Some unsettled discrepancies 
still exist, in this regard, with theory easing in general a more
copius production of metals within the stellar population, thus implying a systematically
lower value of $R$ compared to the the observational evidence.
As first pointed out by \citet{mallik85}, this discrepancy might be reconciled from
the theoretical side by acting on the IMF upper edge, such as to decrease the expected 
contribution of the high-mass stellar component. Allover, an admitted range between 
$R = 3 \pm 1$ is accepted in the literature \citep[e.g.][]{tosi96,maciel01,fukugita06}, and will be 
considered here.

We could therefore introduce the concept of galaxy {\it yield} 
metallicity and Helium content defined, respectively, as
\begin{equation}
\left\{
\begin{array}{ll}
{\cal Z} = {{(M_{YZ}^* + M_{YZ}^{\rm SN})} \over M_*}\,/(1 + R) & \\
         &      \\
{\cal Y} = {\rm Y}_p + R \,{\cal Z}, & \\
\end{array}
\right.
\label{eq:yield}
\end{equation}
where ${\cal Y} - {\rm Y}_p$ is the Helium abundance, in excess to the primordial 
content (Y$_p$).

\section{Relevant model output}

The ``yield'' abundance of eq.~(\ref{eq:yield}) can be regarded as the actual mean
composition of the bright stars in a CSP at a given epoch. Realistically,
$({\cal Z}, {\cal Y})$ give a measure of the {\it maximum} amount of ISM contamination
in the evidently extreme case of ``diluting'' the entire stellar mass 
into a vanishing residual fraction of fresh primordial gas. Accordingly,
$\cal Z$ might also be regarded as the maximum value allowed to $Z_o$, for newly born 
stars.

Clearly, the real process that leads to ISM chemical enrichment follows from 
a much more composite scenario, being the result of a continual and subtly entangled 
interplay between gas and stars inside a galaxy \citep{tinsley80}. Actually, this is 
the central subject of the many important theoretical contributions to the study of 
chemical evolution of galaxies that have been succeeding along the last few decades
\citep[see, e.g.][for an exaustive introduction to the subject]{matteucci03}.

Nonetheless, even facing such a sophisticated reference framework, our simplified
approach to chemical enrichment may still be worth of some attention, as definition of
yield metallicity provides us with a straightforward and physically inherent tool 
to assess the maximum metal abundance that can 
be reached inside a galaxy at any time. With the study of $({\cal Z}, {\cal Y})$ we 
therefore place a firm constrain to galaxy chemical history in terms of the experienced 
luminosity evolution. 
A number of interesting considerations stem from the analysis of the theoretical
output of our exercise, mainly dealing with the expected age-metallicity relation
(AMR) and with the aged problem of the G-dwarf metallicity distribution.

\subsection{The age-metallicity relation}

By combining eq.~(\ref{eq:stars}) and (\ref{eq:snrall}) with  eq.~(\ref{eq:yield}) we 
easily obtain in explicit form the expected AMR for different evolutionary scenarios. 
With little arithmetic, recalling the ${\cal E}_o$ and $M_o$ definitions, the AMR can 
eventually be set in terms of the integrated properties of the galaxy stellar population:
\begin{equation}
{\cal Z}(t) = {0.071 \over {(1+R)\,(0.23+b)\,(M/L)_{15}}}\,t_9^{0.23} + {\cal Z}^{\rm SN},
\label{eq:calz}
\end{equation}
providing to express the reference bolometric M/L ratio\footnote{Again, the ratio 
refers to the {\it stellar} mass, that is the total mass ever converted to stars.} 
at 15 Gyr in solar unit (as from Table~\ref{t1}), and time $t_9$ in Gyr.\footnote{A less 
direct, but still analytically manageable, form for [Fe/H] vs.\ $t$ could also be 
obtained, in case, by recalling the basic definition:
$[Fe/H] = \log {\cal Z} - \log[0.77-{\cal Z}(1+R)] +1.54$, where we assumed 
(Z$_\odot$, Y$_\odot$) = (0.02, 0.28) for the Sun, and a primordial Helium abundance 
$Y_p = 0.23$.}

Quite interestingly, note from eq.~(\ref{eq:calz}) {\it that time evolution of ${\cal Z}$ 
is almost insensitive to the SFR details}, once considering the
$M/L$ vs.\ $b$ inverse dependence, as suggested in eq.~(\ref{eq:mlb}).
This is a not so obvious property of the relation, and derives from the nearly constant 
burning efficiency factor
${\cal F}$ of Table~\ref{t1}, that contrasts the amount of released energy from 
mass processing (namely, the ${\cal E}_o/M_o$ parameter) with the maximum energy
allowed per unit mass (i.e.\ $W_*$ in eq.~\ref{eq:stars}).
One sees, from Table~\ref{t1}, that $\cal F$ varies by less than 20\% when moving
from a ``bursting'' star formation event to a roughly constant or even increasing SFR 
with time.
This makes both ${\rm M}_{\rm YZ}$ and ${\rm M}^*$ to scale with $b$ in a similar way.
For this reason, the same aproximate ${\cal Z} \propto t^{1/4}$ dependence has to be
expected for a wide range of evolutionary scenarios, including the case of SSP star 
bursts.

\begin{table}
\scriptsize
\caption{Photometric properties, reference $M/L$ ratio and yield metallicity
at $t = 15$~Gyr for galaxies of different morphological type$^{(a)}$}
\label{t2}
\begin{tabular}{lccccc}
\hline
Hubble &  \multicolumn{4}{c}{\hrulefill ~~~ Disk component~~~\hrulefill} &  Whole galaxy$^{(b)}$ \\
type   & $\beta$ & ${\rm M}_*/L $ & (Bol-V)~~(U-V)~~(B-V)~~(V-K) & ${\cal Z}$ & $M/L$  \\
\hline
Sa &  ~~0.8 & 2.42 & --0.78~~~~~0.60~~~~~0.63~~~~~2.70 & 0.033 & 4.48 \\
Sb &  ~~0.5 & 1.62 & --0.78~~~~~0.44~~~~~0.56~~~~~2.59 & 0.030 & 3.06 \\
Sc &  ~~0.1 & 1.10 & --0.79~~~~~0.34~~~~~0.51~~~~~2.51 & 0.028 & 1.86 \\
Sd & --0.3  & 0.84 & --0.80~~~~~0.28~~~~~0.48~~~~~2.46 & 0.027 & 1.05 \\
Im & --0.8  & 0.66 & --0.81~~~~~0.24~~~~~0.46~~~~~2.42 & 0.026 & 0.66 \\
\hline
\noalign{$^{(a)}$ From \citet{buzzoni05} template galaxy models}
\noalign{$^{(b)}$ Including the spheroid component (i.e.\ bulge + halo)}
\end{tabular}
\end{table}

The SN contribution is accounted for, in eq.~(\ref{eq:calz}), by the term  
${\cal Z}^{\rm SN} =  {{M_{YZ}^{\rm SN}} \over M_*}/(1 + R)$, as from
eq.~(\ref{eq:snrall}), where its explicit age dependence, according to the 
different evolutionary regimes, is constrained by the $A(\tau,\Delta,b)$ function,
as in eq.~(\ref{eq:a}).
As shown in Fig.~\ref{snrall}, the effect of SN mass processing quickly
reaches a steady contribution to ${\cal Z}$ yet within the first Gyr of the galaxy
life. For this reason, we can therefore safely set the SN enrichment to 
\begin{equation}
{\cal Z}^{\rm SN} \simeq 0.0066/(1+R)
\end{equation}
for any practical application in the framework of our discussion.\footnote{To a closer
analysis of Fig.~\ref{snrall}, our aproximation tends to slightly overestimate the right
${\cal Z}^{\rm SN}$ figure for more ``bursting'' SFRs, as in the case of Sa galaxies.
This has no larger impact, however, on our conclusions, as even a fully crude lower estimate
such as ${\cal Z}^{\rm SN} \simeq 0.0047/(1+R)$ would lead current [Fe/H] predictions for
these galaxies to change by much less than 0.03~dex. See Sec.~4 for an appropriate
context to assess this issue.} 
After accounting for the eq.~(\ref{eq:mlb}) fit, the AMR of eq.~(\ref{eq:calz})
reduces to
\begin{equation}
{\cal Z}(t) = \left[{0.068 \over {1+0.18\,b}}\,t_9^{0.23} + 0.0066\right]\,(1+R)^{-1}.
\label{eq:calz3}
\end{equation}
The equation confirms that chemical enrichment of a galaxy disk is a potentially 
fast process. With $R=3$, for instance, the solar metallicity is quickly reached yet 
within the first few Gyrs of the galaxy life, and a current figure about
${\cal Z} \simeq 1.5\,Z_\odot$ after one Hubble-time evolution.  

\begin{figure}
\centerline{
\psfig{file=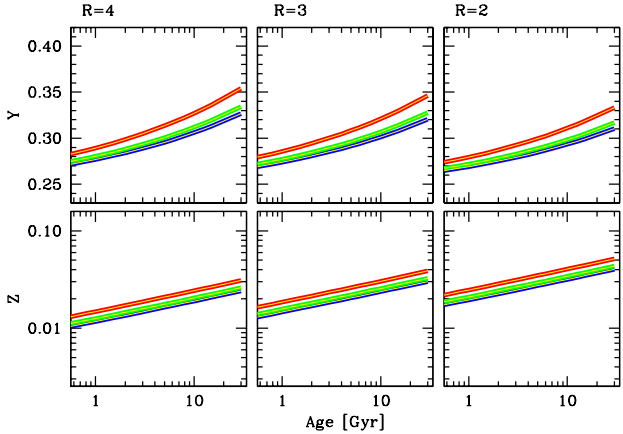,width=\hsize}
}
\caption{Yield chemical abundance for the CSP set of Table~3. Theoretical
output is from eq.~(\ref{eq:yield}), assuming a range
for the enrichment ratio $R = \Delta Y/\Delta Z = 3 \pm 1$, according to the
label in each panel. Note the nearly insensitive response of
both $\cal Y$ and $\cal Z$ to the SFR details.}
\label{zevol}
\end{figure}

The detailed evolution of yield metallicity along the entire Hubble 
morphological sequence can easily be obtained from the corresponding $b$ values, 
as summarized in Table~\ref{t2}. A more general display of the (${\cal Z},{\cal Y}$)
evolution, for CSPs of different SFR power index $\beta$, is displayed in Fig.~\ref{zevol}
along the assumed range for the enrichment ratio $R$. The relevant case for $R=3$
is also reported in detail in Table~\ref{t3}.

\subsubsection {The AMR in the Galaxy}

\begin{figure*}
\centerline{
\psfig{file=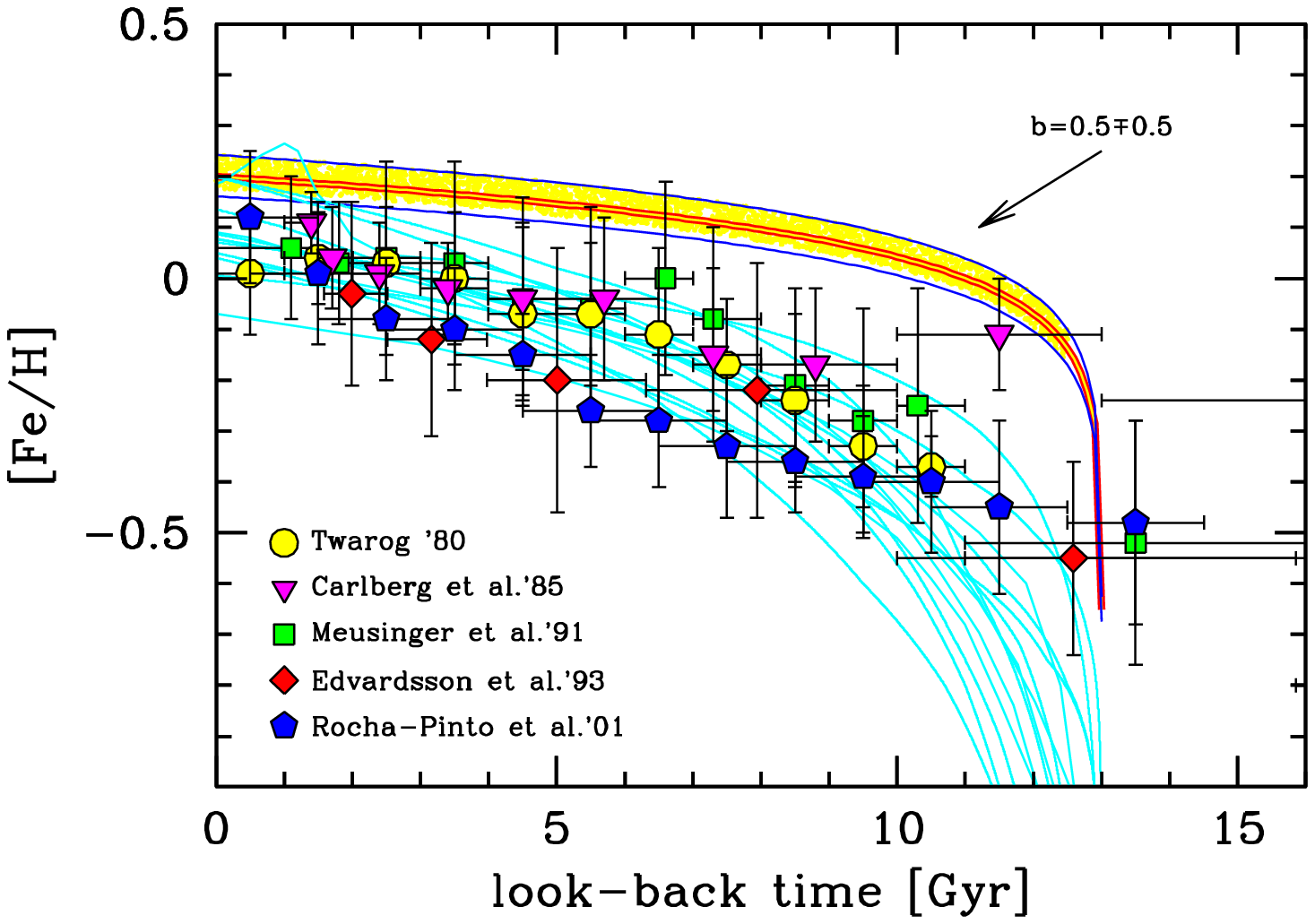,width=0.72\hsize}
}
\caption{The solar-neighbourhood AMR, according to different 
observing surveys, as labelled in the plot. In particular, the works of 
\citet{twarog80} (dots), \citet{carlberg85} (triangles), \citet{meusinger91}
(squares), \citet{edvardsson93} (diamonds), and \citet{rochapinto00} (pentagons)
have been considered. Observations are compared with model
predictions from several theoretical codes (the shelf of thin curves; 
see Fig.~\ref{slopes} for a detailed source list).
The expected evolution of yield metallicity, according to eq.~(\ref{eq:calz3}),
is also displayed, assuming $[Fe/H] = \log {\cal Z}/Z_\odot$, and
a birthrate $b = 0.5\pm 0.5$, as labelled on the curve. A current age of 13~Gyr
is adopted for the Milky Way. See text for a full discussion.}
\label{tutti}
\end{figure*}

A summary of the predicted chemical evolution of the Galaxy from some of the most 
recognized models in the recent literature is proposed in Fig.~\ref{tutti}.
In particular, the models by \citet{matteucci89}, \citet{wyse89}, \citet{carigi94}, 
\citet{pardi94}, \citet{prantzos95}, \citet{timmes95}, \citet{giovagnoli95}, 
\citet{pilyugin96}, \citet{mihara96},  as well as by \citet{chiappini97}, 
\citet{portinari98}, \citet{boissier99}, and \citet{alibes01} are considered, 
along with our $\cal Z$ law for $b = 0.5\pm 0.5$, from eq.~(\ref{eq:calz3}).  

\begin{table}
\scriptsize
\caption{Yield chemical abundance (${\cal Z}$~~,~~${\cal Y}$) for CSPs with different 
SFR and enrichment ratio $R = (\Delta Y/\Delta Z) = 3^{(\dagger)}$}
\label{t3}
\begin{tabular}{lccccc}
\hline
{\large R = 3} & \multicolumn{5}{c}{{\large Flat or decreasing SFR with time}} \\
\\
    & \multicolumn{5}{c}{\hrulefill {\large $\quad \beta \quad $} \hrulefill } \\
{\large t$_9$}  & 0.8 & 0.6 & 0.4 & 0.2 & 0.0 \\ 
 $[$Gyr$]$  & (${\cal Z}$~~,~~${\cal Y}$) & (${\cal Z}$~~,~~${\cal Y}$) & (${\cal Z}$~~,~~${\cal Y}$) & (${\cal Z}$~~,~~${\cal Y}$) & (${\cal Z}$~~,~~${\cal Y}$) \\
\hline
1   & 0.019 0.286 & 0.017 0.281 & 0.016 0.279 & 0.016 0.278 & 0.016 0.277  \\
1.5 & 0.020 0.290 & 0.019 0.286 & 0.018 0.283 & 0.018 0.282 & 0.017 0.281  \\
2   & 0.021 0.294 & 0.020 0.289 & 0.019 0.287 & 0.019 0.286 & 0.018 0.285  \\
3   & 0.023 0.300 & 0.022 0.295 & 0.021 0.292 & 0.020 0.291 & 0.020 0.289  \\
4   & 0.025 0.305 & 0.023 0.299 & 0.022 0.296 & 0.022 0.295 & 0.021 0.293  \\
5   & 0.026 0.308 & 0.024 0.302 & 0.023 0.299 & 0.023 0.298 & 0.022 0.296  \\
6   & 0.027 0.311 & 0.025 0.305 & 0.024 0.302 & 0.023 0.300 & 0.023 0.299  \\
8   & 0.029 0.317 & 0.027 0.310 & 0.025 0.306 & 0.025 0.305 & 0.024 0.303  \\
10  & 0.030 0.321 & 0.028 0.314 & 0.027 0.310 & 0.026 0.309 & 0.026 0.307  \\
12.5& 0.032 0.326 & 0.029 0.318 & 0.028 0.314 & 0.028 0.312 & 0.027 0.311  \\
15  & 0.033 0.330 & 0.031 0.322 & 0.029 0.317 & 0.029 0.316 & 0.028 0.314  \\
\hline
{\large R = 3}  & \multicolumn{5}{c}{{\large Increasing SFR with time}} \\
\\
    & \multicolumn{5}{c}{\hrulefill {\large $\quad \beta \quad $} \hrulefill } \\
{\large t$_9$}  & --0.2 & --0.4 & --0.6 & --0.8 & --1.0 \\
 $[$Gyr$]$  & (${\cal Z}$~~,~~${\cal Y}$) & (${\cal Z}$~~,~~${\cal Y}$) & (${\cal Z}$~~,~~${\cal Y}$) & (${\cal Z}$~~,~~${\cal Y}$) & (${\cal Z}$~~,~~${\cal Y}$) \\
\hline
1   & 0.016 0.276 & 0.015 0.276 & 0.015 0.275 & 0.015~ 0.274 & 0.015~ 0.274 \\
1.5 & 0.017 0.281 & 0.017 0.280 & 0.016 0.279 & 0.016~ 0.278 & 0.016~ 0.278 \\
2   & 0.018 0.284 & 0.018 0.283 & 0.017 0.282 & 0.017~ 0.281 & 0.017~ 0.281 \\
3   & 0.020 0.288 & 0.019 0.287 & 0.019 0.286 & 0.019~ 0.286 & 0.018~ 0.285 \\
4   & 0.021 0.292 & 0.020 0.291 & 0.020 0.290 & 0.020~ 0.289 & 0.020~ 0.289 \\
5   & 0.022 0.295 & 0.021 0.294 & 0.021 0.293 & 0.021~ 0.292 & 0.021~ 0.291 \\
6   & 0.023 0.298 & 0.022 0.296 & 0.022 0.295 & 0.022~ 0.295 & 0.021~ 0.294 \\
8   & 0.024 0.302 & 0.024 0.301 & 0.023 0.300 & 0.023~ 0.299 & 0.023~ 0.298 \\
10  & 0.025 0.306 & 0.025 0.304 & 0.024 0.303 & 0.024~ 0.302 & 0.024~ 0.301 \\
12.5& 0.026 0.309 & 0.026 0.308 & 0.026 0.307 & 0.025~ 0.306 & 0.025~ 0.305 \\
15  & 0.028 0.313 & 0.027 0.311 & 0.027 0.310 & 0.026~ 0.309 & 0.026~ 0.308 \\
\hline
\noalign{$^{(\dagger)}$ The values of (${\cal Z}',{\cal Y}'$) for any other enrichment ratio $R'$ can be derived from the $R = 3$ case,
as ${\cal Z}' = [4/(1+R')]\,{\cal Z}$ and ${\cal Y}' = {\cal Y} + [(R'-3)/(R'+1)]{\cal Z}$.}
\end{tabular}
\end{table}

The whole set of theoretical AMRs is compared in the figure with the empirical 
AMR in the solar neighborhood, as obtained by different stellar samples, probing 
the young O-B stars and the unevolved F-G subgiant population. In particular, six 
important contributions still provide the reference framework for this analysis, namely
the work by \citet{twarog80}, \citet{carlberg85}, \citet{meusinger91}, 
\citet{edvardsson93}, \citet{rochapinto00}, and the exhaustive Geneva-Kopenhagen catalog
of stars in the solar neighbourhood \citep{nordstrom04}, recently revised by
\citet{holmberg07}.

Just a glance to Fig.~\ref{tutti} makes clear the wide spread of theoretical
predictions, that span nearly a factor of 3 (namely 0.3-0.5~dex in [Fe/H] range) in 
the expected Galaxy metallicity at a given epoch. For example, newly born stars 
at $t=10$~Gyr yet exceed the solar value in the \citet{matteucci89} model,
being as rich as $[Fe/H] = +0.11$; this is roughly twice the value of the 
\citet{portinari98} model, which predicts on the contrary $[Fe/H] = -0.13$. 
Instead of any further speculative attempt to elect the ``right'' solution among
the many envisaged scenarios 
\citep[see, however,][for an interesting discussion on this issue]{romano05}, 
we could rather look at the different theoretical outputs of Fig.~\ref{tutti}
as a plain evidence of the inherent uncertainty of any ``deterministic'' approach 
to Galaxy chemical evolution.

\begin{figure}
\centerline{
\psfig{file=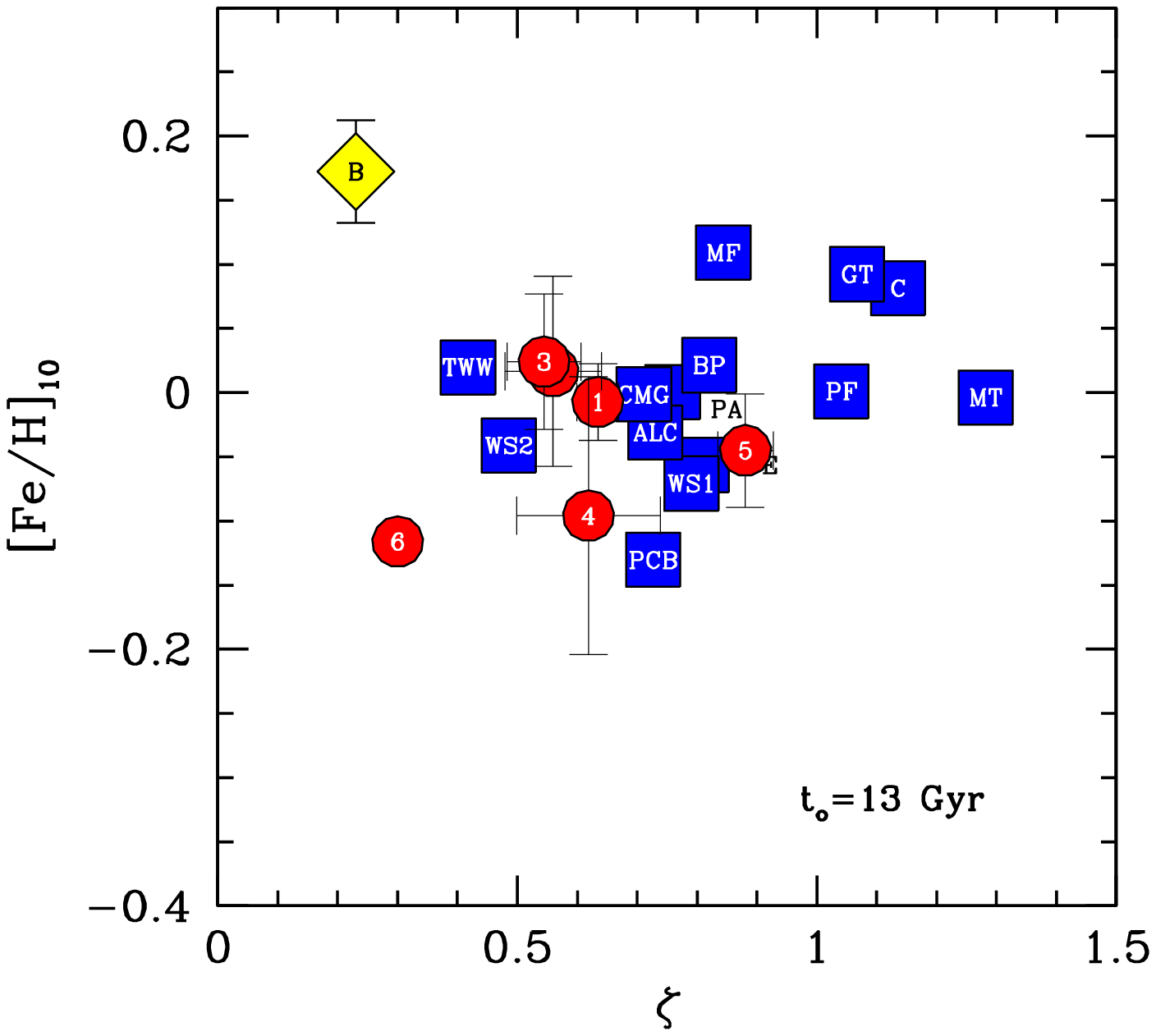,width=0.9\hsize}
}
\caption{Representative parameters for the observed (dots) and predicted (squares) 
AMR for the solar neighborhood. The theoretical works by
\citet{matteucci89} (labelled as ``MF'' on the plot), \citet{wyse89} (WS1 and WS2, 
respectively for a Schmidt law with $n = 1$ and 2), \citet{carigi94} (C), 
\citet{pardi94} (PF), \citet{prantzos95} (PA), \citet{timmes95} (TWW), 
\citet{giovagnoli95} (GT), \citet{pilyugin96} (PE), \citet{mihara96}
(MT), \citet{chiappini97} (CMG), \citet{portinari98} (PCB), 
\citet{boissier99} (BP), and \citet{alibes01} (ALC) have been acounted for,
including the expected output from the $\cal Z$ law of eq.~(\ref{eq:calz3}) 
(square marker labelled as ``B''). The Milky Way is assumed to
be 13~Gyr old for every model. The theoretical data set is also compared with 
the observations, according to the surveys of \citet{twarog80} (dot number ``1''), 
\citet{carlberg85} (``2''), \citet{meusinger91} (``3''), \citet{edvardsson93} (``4''), 
\citet{rochapinto00} (``5''), and \citet{holmberg07} (``6'').
An AMR in the form $[Fe/H] = \zeta \log t_9 + \omega$, is assumed, throughout. 
The displayed quantities are therefore the slope coefficient $\zeta$, vs.\ the 
expected metallicity at $t = 10$~Gyr, namely $[Fe/H]_{10} = \zeta +\omega$. 
Note that models tend, on average, to predict a sharper chemical evolution
with respect to the observations. This leads, in most cases, to a steeper slope 
$\zeta$ and a slightly higher value for $[Fe/H]_{10}$. See the text for a discussion
of the implied evolutionary scenario.}
\label{slopes}
\end{figure}

It may be useful, however, to try a comparison between models and observations in 
terms of the same fitting parameters, by aproximating the different AMRs with a 
general power-law evolution such as $Z \propto t^\zeta$ or, equivalently, 
$[Fe/H] = \zeta \log t_9 + \omega$.\footnote{Of course, this aproximation holds 
as far as the SN term could be neglected, that is for ${\cal Z}(t) \gg {\cal Z}^{\rm SN}$.}
In this notation, $\zeta$ is the slope of the [Fe/H] vs.\ $\log t$ relationship, 
as derived from the fit, while $[Fe/H]_{10} = \zeta +\omega$ provides the nominal 
metal abundance of newly born stars at $t = 10$~Gyr. Table~\ref{t4} and 
Fig.~\ref{slopes} summarize the results of our fit for the whole set of models analyzed,
comparing with the corresponding parameters from eq.~(\ref{eq:calz3}) for $R=3$
and $b = 0.5\pm 0.5$.

As a striking feature, the plot clearly shows the stronger chemical
evolution generally implied by the theoretical codes. The latter predict, in fact, 
a steeper slope for the $[Fe/H]$ vs.\ $\log t$ relation, exceeding the observed $[Fe/H]_{10}$ 
metallicity and predicting a larger value of $\zeta$. 
To some extent, this untuned behaviour adds further arguments to the delicate role 
of the gas-infall mechanisms, so extensively invoked by modern theory. Actually, 
while sustaining star formation, any external gas supply may even ease metal 
enhancement, rather than moderate it by diluting processed mass into a larger 
amount of fresh gas \citep{larson76,firmani92}.

\begin{table}
\scriptsize
\caption{Fitting parameters for Galaxy theoretical and empirical AMR.$^{(a)}$}
\label{t4}
\begin{tabular}{lcc}
\hline
\multicolumn{3}{c}{{\normalsize Yield metallicity}}\\
Source & $\zeta$ & [Fe/H]$_{10}$ \\
\hline
This work$^{(b)}$    &  {\normalsize {\bf 0.23}} &  {\normalsize $ {\bf +0.17\pm 0.04}$} \\
\hline
\hline
\multicolumn{3}{c}{{\normalsize Models}}\\
Source & $\zeta$ & [Fe/H]$_{10}$ \\
\hline
\citet{matteucci89}  &  0.84 &   $+0.11$ \\
\citet{wyse89} (n=1) &  0.79 &   $-0.07$ \\
\phantom{Wyse \&} '' \phantom{Silk (198} (n=2)     &  0.49 &   $-0.04$ \\
\citet{carigi94}     &  1.14 &   $+0.08$ \\
\citet{pardi94}      &  1.04 &   $+0.00$ \\
\citet{prantzos95}   &  0.76 &   $+0.00$ \\
\citet{timmes95}     &  0.42 &   $+0.02$ \\
\citet{giovagnoli95} &  1.06 &   $+0.09$ \\
\citet{pilyugin96}   &  0.81 &   $-0.06$ \\
\citet{mihara96}     &  1.28 &   $-0.00$ \\
\citet{chiappini97}  &  0.71 &   $-0.00$ \\
\citet{portinari98}  &  0.73 &   $-0.13$ \\
\citet{boissier99}   &  0.82 &   $+0.02$ \\
\citet{alibes01}     &  0.73 &   $-0.03$ \\
\hline
\hline
\multicolumn{3}{c}{{\normalsize Observations}}\\
\multicolumn{3}{l}{\underline{{\small Milky Way 13 Gyr old}}}\\
 & $\zeta$ & [Fe/H]$_{10}$ \\
\hline
\citet{twarog80}     &  $0.63\pm 0.04$ &  $-0.01\pm 0.03$\\
\citet{carlberg85}   &  $0.56\pm 0.08$ &  $+0.02\pm 0.07$\\
\citet{meusinger91}  &  $0.54\pm 0.06$ &  $+0.02\pm 0.05$\\
\citet{edvardsson93} &  $0.62\pm 0.12$ &  $-0.10\pm 0.11$\\
\citet{rochapinto00} &  $0.88\pm 0.04$ &  $-0.05\pm 0.04$\\
\citet{holmberg07}   &  $0.30\pm 0.01$ &  $-0.12\pm 0.01$\\
\hline
{\bf Mean of all surveys} & {\normalsize ${\bf 0.59\pm 0.08}$} & {\normalsize $ {\bf -0.04\pm 0.03}$}\\
\hline
\hline
\noalign{$^{(a)}$ Assuming $[Fe/H] = \zeta \log t_9 + \omega$, with
$t_9$ in Gyr. By definition, $[Fe/H]_{10} = \zeta+\omega$}
\noalign{$\qquad$  is the expected metallicity at $t = 10$~Gyr.}
\noalign{$^{(b)}$ From eq.~(\ref{eq:calz}), for $R=3$. $[Fe/H]_{10}$ range limits from 
varying galaxy birthrate $b = 0.5\pm 0.5$.}  
\end{tabular}
\end{table}

\subsubsection{Constraints to the Galaxy SFR}

For its physical definition, yield metallicity provides an upper envelope to 
the observed AMR, throughout (see, again, Fig.~\ref{slopes}). In this regard, it may 
be useful to further explore this concept, in order to better clarify the link between
implied ${\cal Z}$ evolution and the observed galaxy AMR. 

If we assume that a net fraction $f(t)$ of the stellar
mass is returned to the ISM within time $t$\footnote{Note that fraction $f(t)$ has to 
be intended as the {\it net} balance of the global amount of processed mass returned 
to the ISM minus the fraction of enriched mass engaged again in later star formation 
processes.\label{foot8}}, by polluting the residual component of fresh
primordial gas ($M_{\rm gas}$), then the resulting ISM (or, shortly, the ``gas'') 
metallicity, $Z_{\rm gas}$, must be

\begin{equation}
Z_{\rm gas}(t) = {{f\,M_{\rm YZ}\,(1+R)^{-1}}\over {M_{\rm gas} + f\,M_*}}
=  {{f{\cal Z}} \over {\left({M_{\rm tot}/ M_*}\right)} + (f-1)}.
\label{eq:zgas}
\end{equation}
This quantity can be regarded as the representative metallicity of newly born stars.
If ${\cal S} = M_*/M_{\rm tot}$ is the mass fraction of ``processed'' mass and, 
accordingly, ${\cal G} = 1-{\cal S}$ the fresh
(i.e.\ ``unprocessed'') gas fraction then, with little arithmetic, eq.~(\ref{eq:zgas}) reduces to 
\begin{equation}
\left({Z_{\rm gas} \over {\cal Z}}\right) = \left[{{\cal G}\over{f(1-{\cal G})}} +1\right]^{-1}
= \left[{{1-{\cal S}}\over {f{\cal S}}} +1\right]^{-1}.
\label{eq:zgas2}
\end{equation} 
As expected, eq.~(\ref{eq:zgas2}) shows that, when fresh gas vanishes (i.e.\ ${\cal G} \to 0$), then 
$Z_{\rm gas}$ tends to ${\cal Z}$. 

According to eq.~(\ref{eq:zgas}), the total (i.e.\ fresh + processed) gas fraction 
can eventually be written as
\begin{equation}
{\cal G}_{\rm tot} =  {\cal G} +f\,{\cal S} = {\cal G}\,(1-f) + f.
\label{eq:gtot}
\end{equation}
Whatever the specific model assumptions about the mass recycling mechanisms,
it is clear from the equation that $f$ ends up by driving ${\cal G}_{\rm tot}$ 
when the fresh-gas reservoir is going to be exhausted.
This happens, of course, at the very late stages of galaxy evolution or.
Relying on this definition, eq.~(\ref{eq:zgas2}) can also be conveniently arranged in
an alternative form such as
\begin{equation}
\left({Z_{\rm gas} \over {\cal Z}}\right) = \left({f\over{1-f}}\right)\,\left({{1-{\cal G}_{\rm tot}}\over
{\cal G}_{\rm tot}}\right)
\label{eq:zgas3}
\end{equation} 

As, by definition, $f \le 1$ a further interesting property of eq.~(\ref{eq:zgas2}) is that
$({Z_{\rm gas}/{\cal Z}}) \le {\cal S}$.
We can take advantage of this simple constraint to set two important boundary conditions 
to galaxy SFR. 
First of all, as ${\cal S} \propto t^{1-\beta}$, then it must be 
\begin{equation}
\log Z_{\rm gas} - \log {\cal Z} \ge (1-\beta)\,\log t + {\rm const}.
\end{equation}

The l.h.\ side of the relation (or its equivalent form in terms of [Fe/H]) simply 
contrasts the observed AMR in the solar neighbourhood with the expected ``yield'' 
metallicity evolution of eq.~(\ref{eq:calz}). 
The corresponding analytical functions, for the Milky Way case, can be found in 
Table~\ref{t4}. In particular, by comparying the relevant slopes, $\zeta$, of 
the [Fe/H] vs.\ $\log t$ relations one eventually concludes that 
$\langle \zeta^{\rm AMR} - \zeta^{\rm yield}\rangle \ge (1-\beta)$, which implies, for
a current age of 13~Gyr, that
\begin{equation}
b = (1-\beta) \lesssim 0.4 \pm 0.1.
\label{eq:birthb}
\end{equation}

The latter constraint simply derives from the fact that, evidently,
$({Z_{\rm gas}/{\cal Z}}) \le (1-{\cal G})$, as well.
Therefore, at any epoch, the mass fraction of fresh gas in the galaxy
could {\it at most} be
\begin{equation}
{\cal G} \le 1 - \left({Z_{\rm gas} \over {\cal Z}}\right).
\label{eq:gmax}
\end{equation}
Again, by relying on the fitting functions of Table~\ref{t4}, for a 13 Gyr old
Milky Way one sees that $\log (Z_{\rm gas}/{\cal Z}) \simeq [Fe/H]_{\rm AMR} - [Fe/H]_{\rm yield}
= -0.17\pm 0.05 +log[(1+R)/4]$, so that the current mass fraction of primordial gas
in the Galaxy (${\cal G}_o$) must locate within the range    
\begin{equation}
0 \le {\cal G}_o \le 
\begin{array}{l}
~~~\\
0.17\,(4.9-R).\\
\pm \phantom{0}8
\end{array}
\label{eq:go}
\end{equation}
By the way, this also implies that $R \lesssim 5$, 
thus setting an useful constraint to the Galaxy enrichment ratio, too.

Overall, once compared with the representative SFR properties of the other spiral galaxies 
along the Hubble morphological sequence \citep{roberts94}, these figures are consistent
with an intermediate Sb type for the Milky Way \citep[e.g.][]{vanderkruit86}.
In particular, according to \citet{buzzoni05}, the allowed range for the SFR 
leads to an integrated disk color about $(B-V) \sim 0.60 \pm 0.05$,
in agreement with the emipirical evidence \citep{flynn06,melchior07}.

\begin{figure*}
\centerline{
\psfig{file=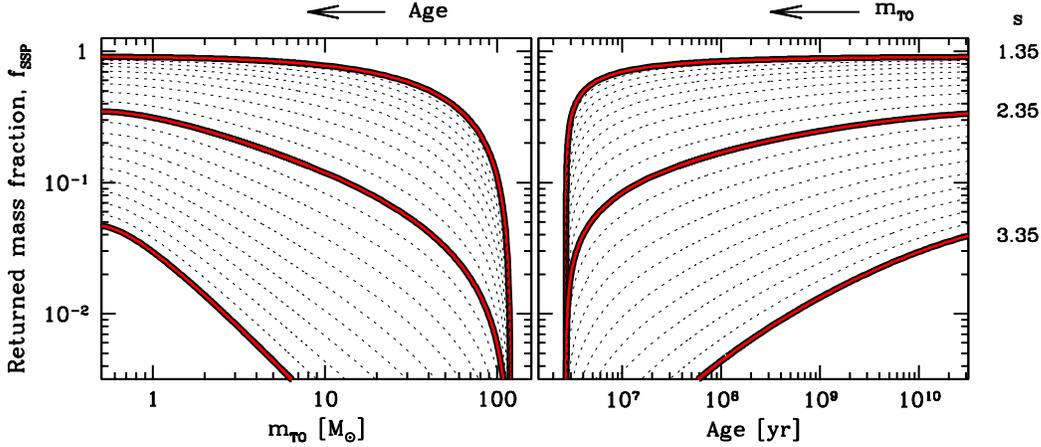,width=0.8\hsize}
}
\caption{The expected fraction $F_{\rm SSP} = M_{\rm ret}/M_{\rm tot}$ of stellar
mass returned to the ISM for a set of SSPs, with changing the IMF power-law
index ($s$), as labelled to the right. The evolution is traced both vs.\ SSP 
age and the main-sequence Turn Off stellar mass ($m_{\rm TO}$), the latter via eq.~(3)
of \citet{buzzoni02}. 
The relevant case for a Salpeter IMF ($s=2.35$) and two indicative variants
for giant- ($s=1.35$) and dwarf-dominated ($s=3.35$) SSPs are singled out
in both panels, while the thin dotted curves account for the intermediate cases,
by steps of $\Delta s = 0.05$ in the IMF index.}
\label{retmass}
\end{figure*}

\subsubsection{The returned mass fraction}

The exact knowledge of the way processed stellar mass is returned to the ISM, and 
the disaggregated contribution of the different elemental species 
is clearly the real core of any theoretical effort aiming at tracing the physical 
details of galaxy chemical evolution. In our straight approach, this concept is 
fully condensed into returned mass function $f(t)$, which is the ultimate key to 
quantitatively assess eq.~(\ref{eq:zgas2}).
The function includes, in principle, a formidable wealth of input physics, dealing with 
chemical yields for individual stars of different mass, the quiescent and explosive 
mass-loss properties of stars, the galaxy vs.\ external environment interplay etc. 
Such a wide range of ingredients, and their related uncertainties, gives actually 
reason of the remarkable differences among the Milky Way models reviewed in previous 
section.

A very plain, but still instructive approach to the problem of ISM pollution
by returned stellar mass has been investigated by \citet{kennicutt94},
following the classical scheme of the instantaneous gas recycling extensively 
practiced in the past literature \citep[see, e.g.][]{larson80}. 
Although ``old-fashioned'' and evidently inadequate for any detailed assessment 
of specific elemental abundances \citep[e.g.][]{arimoto86}, this method may however 
still accomodate, although with some important refinements, within our analysis 
for setting a safe upper limit to $f$.
Following the classical treatment \citep[see, e.g.][]{tinsley80}, we
can consider a SSP of total mass $M_{\rm tot}$, and an IMF stellar mass 
range between 0.1 and 120~$M_\odot$. With these constraints, the fraction of mass 
returned to the ISM within the lifetime of Turn Off stars, of mass $m_{TO}$, 
quantifies in
\begin{equation}
F_{\rm SSP} = {M_{\rm ret} \over M_{\rm tot}} = {{(2-s)}\over {[m^{2-s}]_{0.1}^{120}}}\int_{m(TO)}^{120} \Delta_m(m) m^{-s} dm.
\label{eq:fssp}
\end{equation}
This estimate requires to know the $\Delta_m(m)$ function, that is the mass amount 
lost by stars of initial mass $m$, during their Post-MS evolution, so that 
$\Delta_m = m - m_{\rm fin}$. Following the \citet{kennicutt94} arguments,
we could rely on fully empirical facts, and take the observed $\Delta_m$
relation, as from the Galactic open clusters. For example, according to 
\citet{weidemann00}, a nice fit can be obtained for stars within the range
$0.5~M_\odot \lesssim m \lesssim 8~M_\odot$, such as
\begin{equation}
\Delta_m = 
\begin{array}{l}
~~~\\
0.916\,m  - 0.44\\
\phantom{0}\pm 4\phantom{...000} \pm 2
\end{array}
\label{eq:w00}
\end{equation}
For higher masses, if we assume that a Chandrasekhar core of $1.4~M_\odot$ is
left all the way by the SN events, then
\begin{equation}
\Delta_m = (m-1.4).
\end{equation}
for $m \gtrsim 8~M_\odot$.
With these figures, the results of our calculations are shown in Fig.~\ref{retmass},
exploring the trend of $F_{\rm SSP}$ with varying SSP age or, equivalently the stellar mass 
marking the MS Turn Off point of the population.
Note that $F_{\rm SSP}$ smoothly increases with SSP age, and for a Salpeter IMF it 
reaches a ``ceiling'' value about $F_{\rm SSP} = 0.32$ after one Hubble 
time.\footnote{One has also to be aware, however, of the extreme sensitivity of 
this upper limit to the IMF power index, as evident from the curve grid 
in Fig.~\ref{retmass}. A relevant case in this regard is that of a 
``diet'' Salpeter IMF, as parameterized for instance by \citet{kennicutt83} or 
\citet{kroupa93} to better account for the observed flattening in the dwarf-star 
number counts at sub-solar mass. By means of eq.~(\ref{eq:fssp}) one could verify that
this IMF closely traces the standard Salpeter case being however capped at a slightly higher
value of $F_{\rm SSP} \sim 0.46$.}

On the other hand, as far as a CSP with a more entangled star-formation history is concerned, 
we have to expect the total fraction of returned mass, $F_{\rm CSP}(t)$, to be the result
of a convolution with the SFR, such as
\begin{equation}
F_{\rm CSP}(t) = \int_0^t F_{\rm SSP}(\tau) \otimes {\rm SFR}(\tau) d\tau.
\end{equation}
This leads, in general, to a lower value for $F_{\rm CSP}(t)$, compared to the
corresponding SSP case. In addition, for a proper assessment of the gas budget
in the galaxy one has also to consider that part of the returned mass is in fact 
to be recycled again into fresh stars, its exact amount being indeed a crucial
distinctive property of any theoretical scheme aimed at tracing galaxy chemical evolution.
Our conclusion is therefore that, in any case, it must be 
$f(t) \le F_{\rm CSP}(t) \le F_{\rm SSP}(t) \lesssim 0.3$.

\subsection{Mixing processes and metallicity spread}

One common feature of all these models, however, is that 
$f(t)$, and the physical conditions that led to establish the galaxy AMR are
univocally defined at any epoch. As a consequence, by definition, 
{\it no spread in [Fe/H] is envisaged by theory among coeval stars, throughout.}
On the other hand, observations of unevolved nearby stars clearly point 
to some degree of metallicity dispersion, perhaps slightly amplified by
observational uncertainties and the adopted age/metallicity classification 
procedure \citep{rochapinto00}. For example, by relying on the 
\citet{edvardsson93} survey, we derive $\sigma[Fe/H] = 0.23$~dex from the distribution 
of the individual stars around the mean AMR locus. This is about twice the typical 
value originally found by \citet{twarog80}, and can be contrasted with 
$\sigma[Fe/H] = 0.16$~dex, as obtained for the \citet{rochapinto00} 
stars, or with $\sigma[Fe/H] = 0.18$~dex, displayed by the \citet{holmberg07} 
extended star sample. Whether this spread is really the general case for spiral 
galaxies is a still open and debated question \citep[e.g.][]{koeppen90,wyse06,prantzos08}.

\begin{figure}
\centerline{
\psfig{file=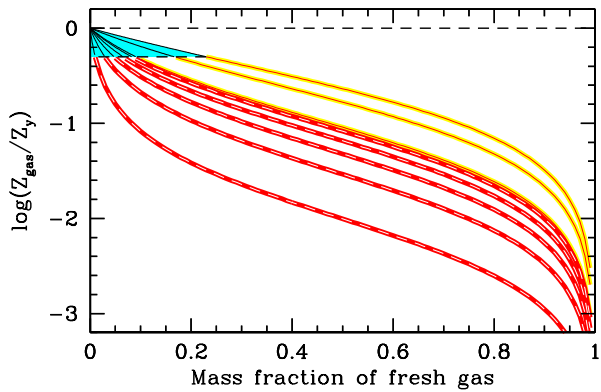,width=\hsize}
}
\caption{The output of eq.~(\ref{eq:zgas2}) is displayed vs.\ the mass
fraction of fresh primordial gas ($\cal G$). The two shelves of curves
assume the fraction $f$ of returned stellar mass as a free parameter.
The lower shelf (dashed curves) is for $f = 0.01 \to 0.09$ by steps of 
$\Delta f = 0.02$, while the upper shelf of curves (solid lines) cover the 
$f$ range from 0.1 to 0.3, by steps of $\Delta f = 0.1$. The shaded area within 
$-0.3 \le \log (Z_{\rm gas}/{\cal Z}) \le 0$ single out the allowed $f$-$\cal G$
combinations that may account for the observed metallicity spread of
stars in the solar neighborhood.
}
\label{gf}
\end{figure}

Again, the envisaged ${\cal Z}$ evolution can help, we believe, to set some 
interesting constraints to this problem, as well.
Following eq.~(\ref{eq:zgas2}), Fig.~\ref{gf} summarizes the expected response of 
the $Z_{\rm gas}/{\cal Z}$ ratio to just the two reference parameters that
modulate galaxy chemical evolution in our framework, namely ${\cal G}$, and $f$.
When regarded into a broader context, our plot displays expected 
relationaship of the global mass fraction of fresh gas in the galaxy. 
Alternatively, this ``macro'' view can be complemented by a ``micro'' view, 
being the ${\cal G}$ parameter intended as the local gas density of star-forming 
regions, such as ${\cal G} \equiv \rho_{\rm gas}/\rho_{\rm tot}$, and featuring 
therefore the specific mixing conditions that embed star formation.
One sees from the figure that, in principle, a wide range (a few dex!) may 
be allowed to $Z_{\rm gas}$, at every epochs, as coeval stars could naturally 
spread in the plot along an $f = {\rm const}$ locus, in consequence of the 
micro-environment conditions proper to the different star-forming regions inside 
a galaxy. 

\begin{figure}
\centerline{
\psfig{file=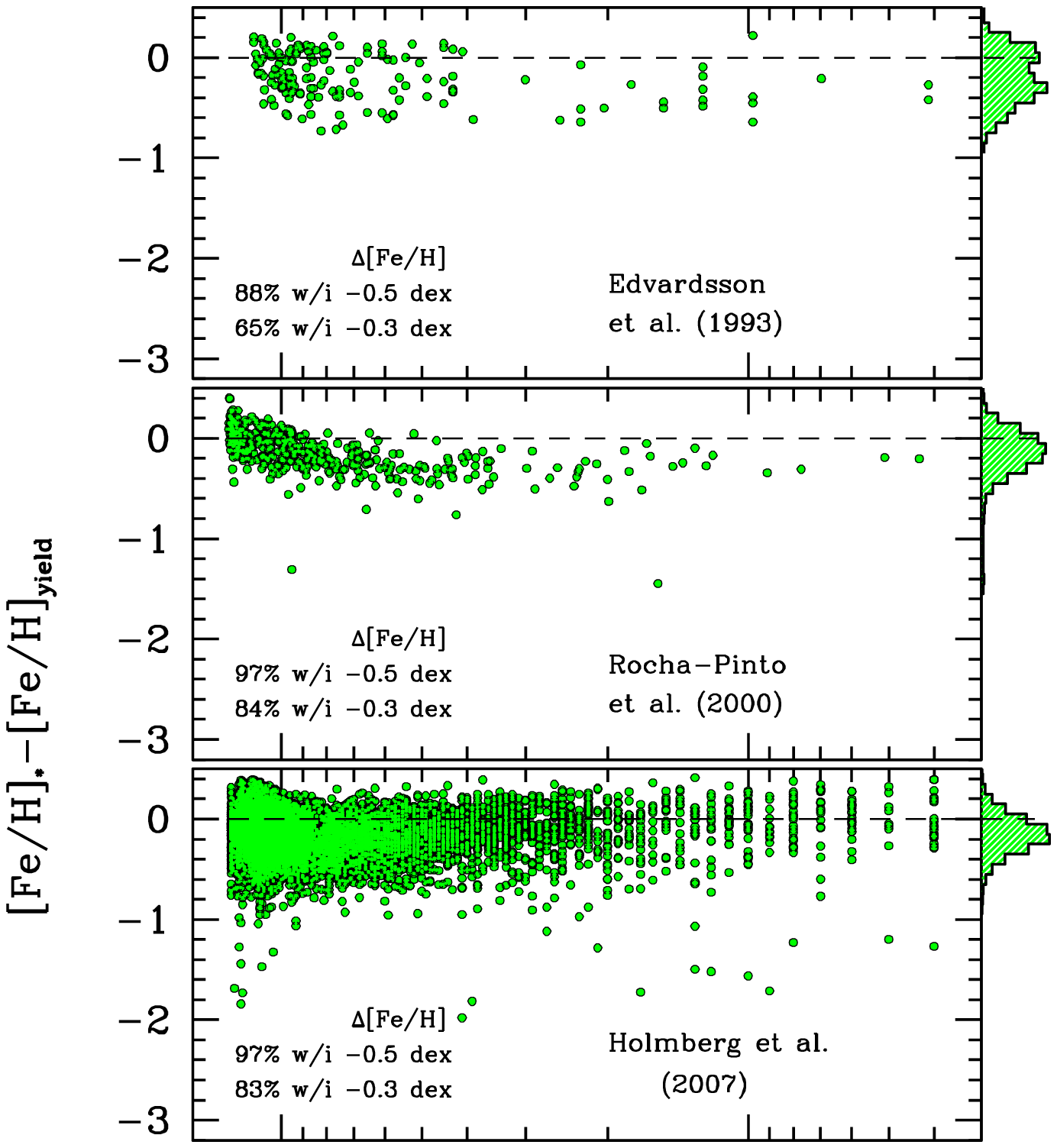,width=\hsize}
}
\centerline{
\psfig{file=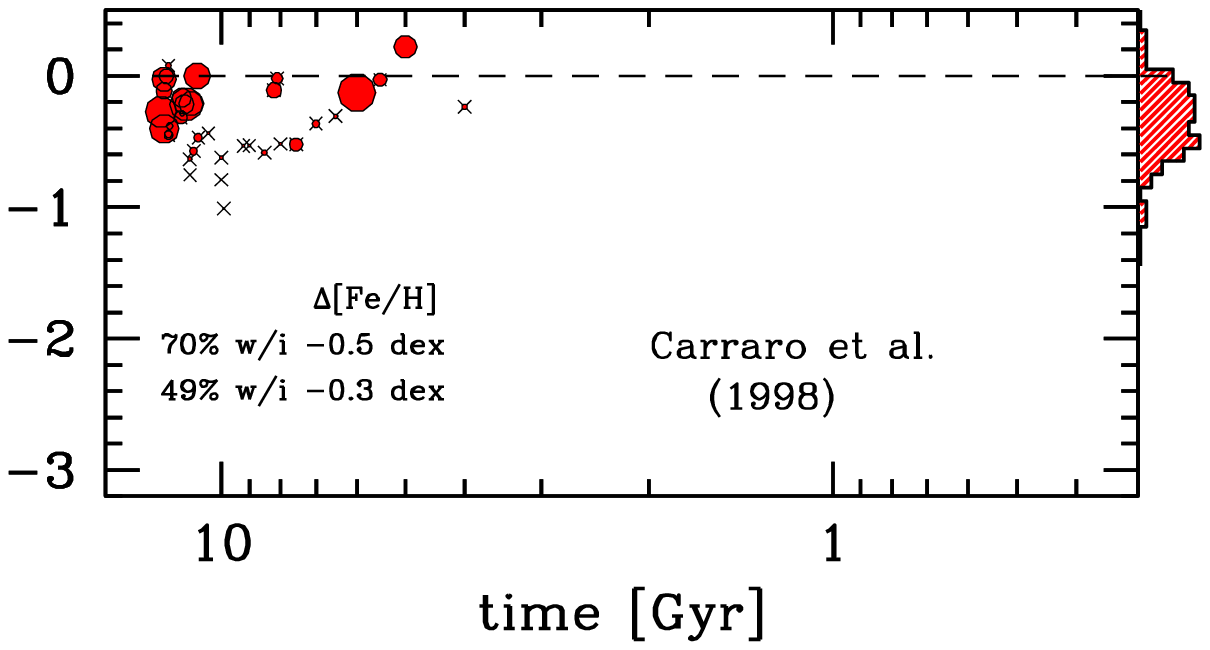,width=\hsize}
}
\caption{The observed metallicity distribution of stars
in the solar neighbourhood. The displayed quantity is 
$\Delta[Fe/H] = [Fe/H]_* - [Fe/H]_{{\rm yield}} \equiv \log (Z_{\rm gas}/{\cal Z})$.
Different stellar samples have been
considered from the work of \citet{edvardsson93}, 
\citet{rochapinto00}, and \citet{holmberg07}, from top to bottom, 
as labelled in each panel. The lower panel displays the same trend
for a sample of open stellar clusters, after \citet{carraro98}.
The different Galacticentric distance for these systems is indicatively
marked by the dot size (i.e.\ the smallest the dot, the largest the
distance from the Galaxy center). For each distribution, 
the relative fraction of stars within a -0.3 and -0.5~dex $\Delta[Fe/H]$
residual is reported in each panel. A cumulative histogram of the data
is also displayed, along the right axis, in arbitrary linear units.
}
\label{allstars2}
\end{figure}

The \citet{edvardsson93}, \citet{rochapinto00}, and \citet{holmberg07} star samples offer
once more an important feedback in this regard, as shown in Fig.~\ref{allstars2}.
For the first sample, 161 stars are nominally younger than 13 Gyr, the assumed age 
of the Milky Way; the overall statistics of their individual
$\Delta [Fe/H] = [Fe/H]_{{\rm yield}} - [Fe/H]_*$ distribution indicates that
88\% of the stars are comprised within $\Delta [Fe/H] \le 0.50$~dex, 
while a fraction of 65\% lie within $\Delta [Fe/H] \le 0.30$~dex. With the same 
procedure, for the 412 relevant stars in the 
\citet{rochapinto00} sample we obtain a fraction of 97\% and 84\%, respectively, within
the same $\Delta [Fe/H]$ limits. Similar figures (namely 97\% and 83\%, respectively) 
apply also for the 13141 eligible stars in the Geneva-Kopenhagen calatog.
Allover, this means that over 4/5 of all newly-born stars in 
the solar vicinity never departed by more than 0.3~dex from ${\cal Z}$, {\it at any epoch.}

Facing a so tight $\Delta [Fe/H] \equiv \log ({\cal Z}/Z_{\rm gas})$ star
distribution,\footnote{Strictly speaking, for its delayed supply to the ISM, Iron may not 
be a fair proxy of the whole metallicity among very metal-poor (i.e. ``pristine'') stars
in the Galaxy (see the following Sec.~3.2.2 for a most explicit assessment of this issue). This
assumption carries therefore some uncertanties for stars within $t\la 1$~Gyr in the
plots of Fig.~\ref{allstars2}.} one may actually wonder {\it why stars in the Galaxy appear so closely 
``tuned'' with the yield metallicity, everytime.}
 
In spite of any more or less sophisticated approach to this crucial issue, 
just a glance to Fig.~\ref{gf} makes evident that the observed range for the 
$Z_{\rm gas}/{\cal Z}$ ratio univocally implies (at least in the solar vicinity) that
star formation must have proceeded within ISM conditions characterized, all the way, 
by {\it a scanty gas abundance (${\cal G} \lesssim 0.5$, see again Fig.~\ref{gf})}.

Overall, this feature seems to add further strength to well recognized non-standard 
scenarios, where the plain dependence of SFR on gas density, like
in a classical \citet{schmidt59} law, might more effectively be replaced by some
positive feed-back to other triggering mechanisms, either related to disk dynamical 
properties, inefficient gas mixing or molecular cloud bunching 
\citep{larson78,wyse89,dopita94,oey00,boissier03,scalo04,nittler05,vanzee06}. 

A further interesting piece of information can be added to the emerging scenario
when probing the $\Delta [Fe/H]$ distribution at different
Galactocentric distance. This can be done, for instance, by relying on \citet{carraro98},
who collected homogenous data for a set of 37 open clusters. The evident advantage,
in this case, is that age and metallcity can be derived in a more direct way, based
on the observed c-m diagram of the whole stellar members of each cluster.\footnote{One 
has to remark, however, the evident bias of the \citet{carraro98} sample against
very old open clusters likely due, according to authors' discussion, to the disruptive 
mechanisms along the past dynamical history of the Galaxy \citep{friel95}.}
The \citet{carraro98} results are reported in the lower panel of Fig.~\ref{allstars2},
marking the cluster distance from Galaxy center with a different dot size
(i.e.\ the closer the cluster, the bigger its marker).

\begin{figure}
\centerline{
\psfig{file=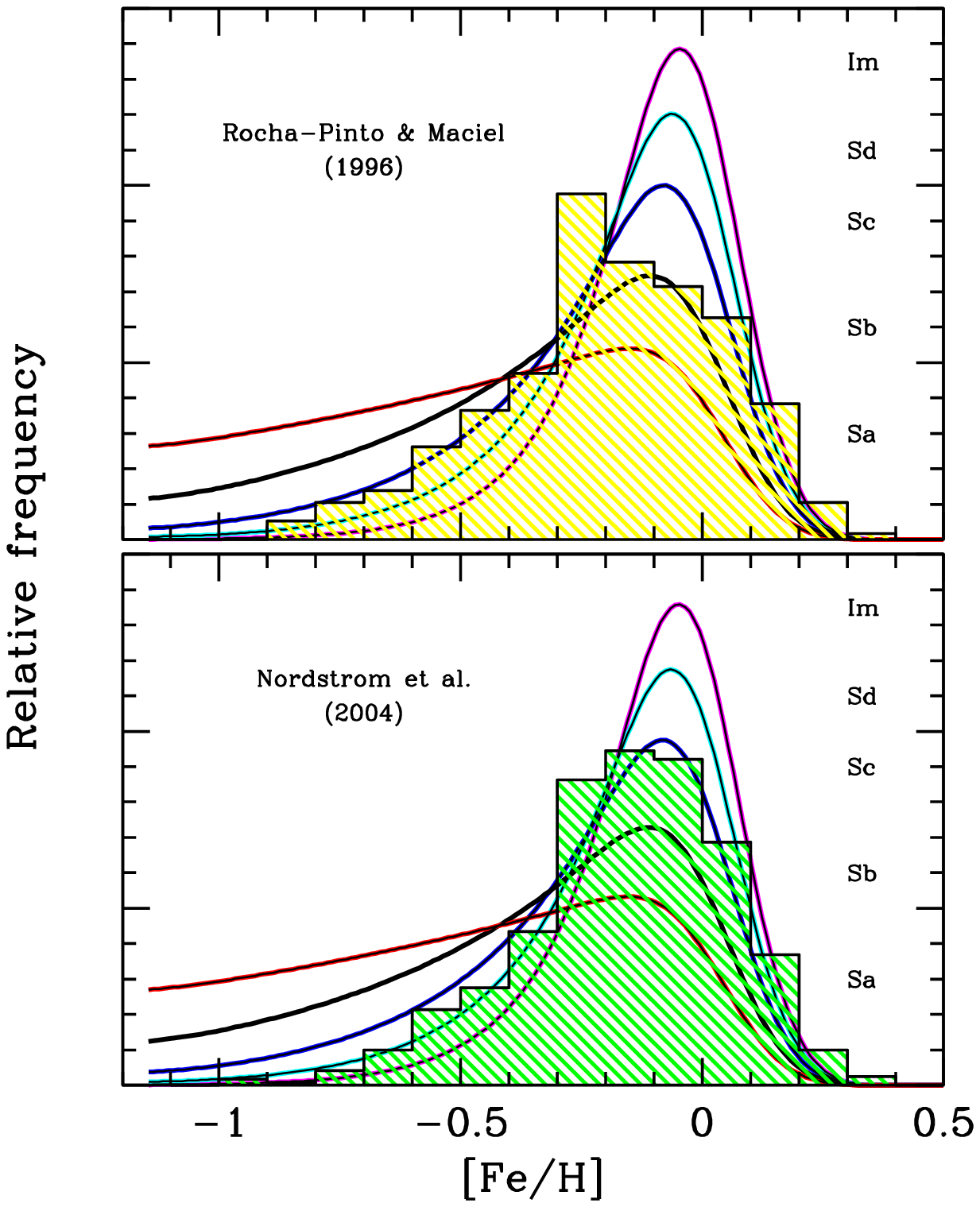,width=0.9\hsize}
}
\caption{The $[Fe/H]$ distribution of G-dwarf stellar samples in the solar
neighborhood (shaded histograms), according to \citet{rochapinto00} (upper panel), 
and \citet{nordstrom04} (lower panel), is compared with the expected 
yield-metallicity distribution accounting for the different star-formation 
histories as for the disk stellar population of \citet{buzzoni05} template 
galaxies along the Hubble sequence (solid curves, as labelled on the plot).
A roughly constant stellar distribution in the $Z$ domain leads to
an implied birthrate of the order of $b \simeq 0.6$, as pertinent to
Sbc galaxies (see Table~2 for a reference).
}
\label{gdwarfs}
\end{figure}

The local properties of disk metallicity are consistently traced also by the open cluster
population, whith a clear additional evidence, however, for less pronounced chemical 
enhancement processes in the most peripheral stellar systems, for which a systematically 
larger departure from ${\cal Z}$ is displayed in Fig.~\ref{allstars2}. 

\subsubsection{The ``G-dwarf'' problem}

In its essence, the tuned distribution of the $\Delta [Fe/H]$ parameter, as
in previous discussion, has much to do with the classical ``G-dwarf'' problem 
\citep{vandenbergh62,pagel75,twarog80}, that is an evident lack, nowadays, in the
[Fe/H] (or $\log Z$) domain of the ``desired'' population of extremely metal-deficient 
unevolved stars, reminiscent of the earliest evolutionary phases of the disk 
formation in the Galaxy. 

According to our SFR parameterization, for $b \lesssim 0.4$, an expected fraction of 
$\sim (50\,10^6/15\,10^9)^b \gtrsim 10\%$ of G-dwarf stars should have produced in the 
disk of our galaxy within the lapse ($t \lesssim 50$~Myr) of the first type-II SN burst.
According to Fig.~\ref{snrall}), for these stars a metallicity as poor as
$[Fe/H] \sim -1.10$ should be expected. The lack, nowadays, of such a relevant
population of virtually ``zero-metal'' stars is actually the central issue that 
led theorists in the late 80's to invoke a composite formation for the Galaxy disk, 
through a ``buffering'' dynamical mechanism that led first to the formation of
an outer ``thin'' disk, followed by an inner``thick'' structure 
\citep{gilmore86,ferrini92}.

Translated into a linear metallicity scale, the undersized population of
stars with $[Fe/H] \ll -1$ can be matched by a roughly uniform star density along 
the $Z$ range, so that $dN_*/dZ \sim const$ (see, in this regard, \citealp{pagel89}, and 
especially the work of \citealp{beers85,beers92,sommerlarsen91,ryan96} and \citealp{cayrel96}).

The stellar metallicity distribution can easily be related with
the other distinctive properties of galaxy evolution in our framework, as
\begin{equation}
{dN_*(Z)\over dZ} = {dN_*(t)\over dt} \Bigg/ {dZ \over dt}.
\end{equation}
The first term in the r.h. side of the equation is, of course, proportional to
the SFR ($\propto t^{-\beta}$), while the second term is simply the time derivative
of the observed AMR ($\propto t^{\zeta-1}$). 
By replacing the relevant relations, we have
\begin{equation}
{dN_*(Z)\over dZ} \propto t^{1-\zeta-\beta} \propto Z^{(1-\zeta-\beta)/\zeta},
\end{equation}
or
\begin{equation}
{dN_*\over d[Fe/H]} \propto 10^{[Fe/H](1-\beta)/\zeta},
\label{eq:gg}
\end{equation}
if one better likes to put the relation in terms of [Fe/H].
For a roughly constant $dN_*/dZ$ distribution, it must be 
$(1-\zeta-\beta) \sim 0$, or $\zeta \simeq b$. According to Table~\ref{t4}), for
a current age of 13~Gyr, this implies that G-dwarf distribution may be reproduced
by a Galaxy birthrate of 
\begin{equation}
b \sim 0.6\pm 0.1.
\end{equation}

A graphical sketch of our conclusions is proposed in Fig.~\ref{gdwarfs}, by comparing 
the G-dwarf stellar sample of the Galaxy disk by \citet{rochapinto96} and \citet{nordstrom04}
with the expected metallicity distribution as obtained by applying eq.~(\ref{eq:gg}) 
for a range of SFR representative of the spiral-galaxy sequence in the \citet{buzzoni05} 
template galaxy models. Note that Milky Way observations closely match the standard case 
of the Sb/Sc morphological types (see Table~\ref{t2}). 

To a better analysis, however, we have to remark the somewhat marginal value of $b$ 
when compared to our previous conclusions, as
the tight relationship between Galaxy AMR and $\cal Z$ trend rather called for 
a lower figure (i.e.\ $b \lesssim 0.4$). This apparent dichotomy actually
summarizes the full essence of the G-dwarf problem emphasizing, from one hand,
the central role of the early stellar component in constraining the Galaxy AMR,
but requiring, on the other hand, for these stars to be born ``elsewhere''
(i.e.\ in the halo or in the ``thin'' disk).

\subsubsection{The ``$\alpha$''-element versus Iron-peak enrichment}

A further implication of the simple energetic arguments outlined in Sec.~2 is that metal 
enrichment should have proceeded selectively in the very early stages of Galaxy evolution, 
according to the composite interplay between ``quiescent'' luminosity evolution of stars and 
violent action of SNe. In fact, bright stars are to be regarded as the elective suppliers of lighter 
pre-Carbon elements, while SNe{\sc ii} further add to the synthesis of the $\alpha$-element 
chain, especially providing Oxygen and heavier metals \citep{woosley95}. For their different 
physical conditions, SNe{\sc i}a are meant, on the contrary, to be the prevailing donors
of Fe and other Iron-peak elements \citep{nomoto84} to the galaxy ISM.

Given the delayed appearence of SN{\sc i}a, one has to expect $\alpha$ elements
(and Oxygen, in particular) to cumulate earlier than Iron-peak elements into ISM of
the Milky Way. This feature directly calls, therefore, for some O-Fe decoupling among the
low-metallicity stellar component of our galaxy. A positive $[O/Fe]$ relative abundance is 
actually a widely recognized property when comparing halo and disk stars, and the effect is a 
supplementary piece of evidence dealing with the G-dwarf problem previously discussed. 
The study of the $[O/Fe]$ vs.\ $[Fe/H]$ relationship in the local framework prove therefore 
to set important constraints to the different feeding channels that eventually led to $Z$ enrichment
in the Galaxy.

\begin{figure}
\centerline{
\psfig{file=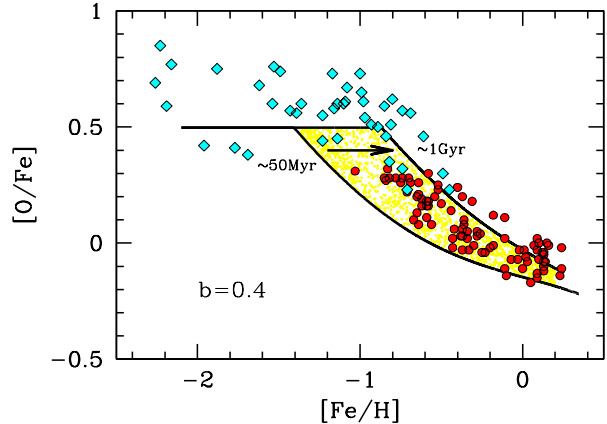,width=\hsize}
}
\caption{The $[O/Fe]$ distribution of stars in the Milky Way is displayed acording to
the \citet{jonsell05} (diamonds) and \citet{edvardsson93} (dots) data samples, respectively
for the halo and disk stellar populations. Our model output, according to eq.~(\ref{eq:ofe})
is superposed (solid lines), for a representative birthrate $b = 0.4$, and for a range of
SN{\sc i}a delay time $\Delta$ from $\sim 50$~Myr to $\sim 1$~Gyr, as discussed in Sec.~2.2.
}
\label{feo}
\end{figure}

We can further explore this issue by relying on our model to compare the expected $[O/Fe]$ 
evolution with the observed trend for the Galaxy stars. 
The work of \citet{pagel95} and \citet{scannapieco05} can be taken as useful references 
to set the relevant quantities for $Fe$ and $O$ yields in Type Ia and II SNe. Accordingly, we assume a 
typical abundance of $[O/Fe]_{II} = +0.50$~dex for the released mass by SNe{\sc ii} 
and $[O/Fe]_{I} = -1.51$~dex for the SN{\sc i}a case, being  $\log (O/Fe)_\odot = +1.33$~dex the 
reference figure (in number abundance) for the Sun \citep{grevesse98}. In addition, with 
\citet{scannapieco05}, we take a ratio $r = (m_{Fe}^{II} : m_{Fe}^{Ia}) = (0.062: 0.74)$ for the 
released Iron yields according to the SN type.\footnote{As Iron is the main output of any SNIa 
explosion, its yield must obey an inherent argument of energetic self-consistency being 
$m_{Fe}^{Ia} \approx (E_{51}^{SNI}\, 10^{51})/(W_{\rm SN}\, M_\odot)$. According to the relevant 
figures of Sec.~2.2.2, we actually have that $m_{Fe}^{Ia} \simeq 0.75 \pm 0.25$~M$_\odot$, 
in close agreement with the standard model predictions \citep[see, e.g.][their model W7]{nomoto84}.}

Equations (\ref{eq:sn20}), (\ref{eq:prompt}) and (\ref{eq:ext}) can still accomodate 
for our calculations providing to replace the $(E_{51}\,10^{51})/(W_{\rm SN}\, M_\odot)$ energy-release 
term with the relevant chemical yields. Recalling our SFR parameterization, and the notation of 
Sec.~2.2, the total mass released by Type {\sc ii} and {\sc i}a SNe for the element $X$ can be 
expressed as
\begin{equation}
m(X)  = m(X)_{II} +m(X)_{I}\,B(t,\Delta,b),
\end{equation}
where $B(t,\Delta,b)$ is a function of age ($t$) and stellar birthrate ($b$), and accounts for the 
SN{\sc i}a delay time ($\Delta$).
Like for eq.~(\ref{eq:a}), previously discussed, an analytical form can be devised also 
for this function, providing to express $\Delta$ in years, such as
\begin{equation}
B(t,\Delta,b) = \left\{
\begin{array}{l}
0 \hfill {\rm for~} \tau < 1 \\
~~~ \\
0.35\,(1-\tau^{-b}) + {\Delta \over {1.6\,10^{11}}} {{\tau\, [1-\tau^{-(1+b)}]}\over {1+b}},\\
\hfill {\rm for~} \tau \ge 1\\
\end{array}
\right.
\end{equation}
being, again, $\tau = t/\Delta$.

For the $[O/Fe]$ relative abundance we have therefore
\begin{equation}
\left[{O\over Fe}\right] = \log { {r\,10^{[O/Fe]II} + B\,10^{[O/Fe]I}} \over {r + B}}
\label{eq:ofe}
\end{equation}
Note, from eq.~(\ref{eq:ofe}), that $B = 0$ for $t\lesssim \Delta$, which implies that ISM $[O/Fe]$ abundance 
simply coincides with the SN{\sc ii} relative yields at the very early evolutionary stages of the
Galaxy. Iron enrichement is obviously favored by increasing $B$ in the equation. This naturally 
happens when getting the Galaxy older (that is by increasing $\tau$) and/or by decreasing 
the SN{\sc i}a delay time $\Delta$.

Our results are sketched in Fig.~\ref{feo}, comparing with the observation of disk and halo stars,
respectively from \citet{edvardsson93} and \citet{jonsell05}.
In the figure, a safe lower and upper envelope to the observed $[O/Fe]$ stellar distribution
can be placed for $\Delta \sim 50$~Myr and 1~Gyr, respectively. This range also consistently 
comprises the claimed ``glitch'' in the $[O/Fe]$ vs.\ $[Fe/H]$ trend, ideally located about
$[Fe/H] \approx -1$ \citep{clegg81}. Our output therefore confirms that most of the SN{\sc i}a 
impact on the Galaxy metal enrichment should have occurred very early within the first Gyr of life,
although just a glance to the data of Fig.~\ref{feo} indicates that some $10^8$~yr may have required 
for the SN{\sc i}a process to fully deploy within the Galaxy stellar environment. 
This conclusion is supported, indeed, by the empirical arguments of \citet{maoz10} 
and the theoretical ones by \citet{greggio05} and \citet{kobayashi98}.

\section{Chemo-photometric properties along the Hubble morphological sequence}

Our quite special lookout inside the Galaxy makes the study of our system certainly
favored but, at the same time, also subtly biased, compared to the analysis of external 
galaxies. From one hand, in fact, we can directly adress the problem of ``Galaxy metallicity'' 
in terms of the aggregated information from individual stars in the solar neighborhood; 
on the other hand, the criterion to define the star sample (either by fixed space volume, 
or apparent magnitude limit, or even selected spectral type etc.) may play a 
crucial role in driving our conclusions, as the ``representative'' metallicity value 
would eventually depend on the way we account for the contribution of individual stars.
This is a subtantial difference with respect to the study of distant galaxies, 
for which we usually cannot resolve individual stars at all. In the latter case,
one should forcedly rely on the integrated spectrophotometric properties of the 
system, as a whole, to derive somewhat ``effective'' distinctive parameters to
constrain galaxy properties.

\subsection{Matching the Arimoto \& Jablonka (1991) theoretical framework}

\begin{figure}
\centerline{
\psfig{file=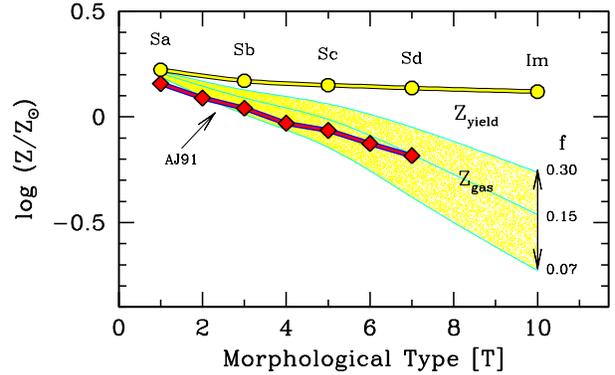,width=\hsize}
}
\caption{The gas metallicity of 15~Gyr disk stellar populations according to 
the \citet{arimoto91} (AJ91) models (solid line with diamonds) is displayed 
together with the expected predictions from eq.~(\ref{eq:zgas3}) (shaded area) 
for a range of returned mass fraction $f$ between 0.07 and 0.3, as labelled on 
the plot. For tighter self-consistency in our
comparison, the adopted gas fraction ${\cal G}_{\rm tot}$ for our model
sequence matches the corresponding figures of AJ91 models, while
yield metallicity derives from eq.~(\ref{eq:calz3}) for the relevant
birthrate values of the \citet{buzzoni05} template models 
along the Sa$\to$Im morphological sequence (solid line with dots).}
\label{ari}
\end{figure}

From the theoretical side, our general view of disk chemical properties along 
the Hubble morphological sequence may usefully complement the analysis of 
AJ91, certainly one of the few explicit attempts in the recent literature to 
self-consistently tackle the problem of chemo-photometric evolution of late-type 
galaxies 
\citep[see, in addition, also][to complete the theoretical picture]{arimoto86,arimoto87,koeppen90}. 
A first interesting comparison, in
this regard, is proposed in Fig.~\ref{ari} where the gas metallicity, 
$\log (Z_{\rm gas}/Z_\odot)$, for the 15 Gyr AJ91 disk models along the 
Sa~$\to$~Sd sequence is displayed together with the relevant values of our 
yield metallicity, $\log ({\cal Z}/Z_\odot)$. 

For its different IMF limits (stars are produced between 0.05 and $60~M_\odot$) 
the AJ91 stellar populations are slightly ``darker'' (that is with a larger $M/L$
ratio) for their larger fraction of dwarf stars, and with a less prominent 
contribution ($\sim -20\%$) of SNe to galaxy metal enrichment.
These differences with respect to our theoretical scheme become more important
when moving along the Sa~$\to$~Sd sequence, as stars of increasingly higher 
mass prevail as contributors to galaxy luminosity.

In spite of the different input physics, one has to remark however a notable
agreement in the [Fe/H] predictions, as shown in Fig.~\ref{ari}.
The case of Sa models is illustrative in this sense, as the vanishing residual 
gas (only 4\% of the total disk mass in the AJ91 model) makes $Z_{\rm gas}$ to 
closely approach ${\cal Z}$, as expected. Once accounting for the increasing fraction
of residual gas along the morphological galaxy sequence, the prediction of
eq.~(\ref{eq:zgas3}) tightly matches the AJ91 output. Facing the relatively
flat trend of $\cal Z$ vs.\ galaxy type, it is evident from the figure that
{\it the increasingly poorer gas metallicity along the Sa~$\to$~Im Hubble sequence 
is in fact mainly the result of the diluting process, that shares enriched 
stellar mass with a larger fraction of residual gas.}

Interestingly enough, in the quite delicate balancing mechanism that governs 
the [Fe/H] trend vs.\ Hubble type, the exact details of stellar mass return processes do
play a somewhat marginal role. This is evident, for instance, by comparing in 
Fig.~\ref{ari} the effect of changing the return mass fraction $f$ from its allowed 
maximum ($f = 0.3$) down to a factor of four lower value ($f = 0.07$). 
Rather, the amount of fresh residual gas, that survived past star formation, seems 
a much more central figure to modulate the enrichment process, and this directly 
calls for the distinctive birthrate $b$ as the ultimate parameter to constrain 
disk metallicity. We will further return in more detail on this important point 
in a moment (see Sec.~4.3).

\subsection{``Chemical'' vs. ``photometric'' metallicity estimates}

Due to the coexisting contribution of gas and stars inside a spiral galaxy,
any fair estimate of ``galaxy metallicity'' is a far more entangled problem.
Quoting \citet{roberts94}:
``intercomparison between abundances based on emission lines from H{\sc ii} regions
and on absorption lines and colors from a stellar population is still uncertain...
Atomic processes guide emission-line analysis, while the absorption lines must be
evaluated in terms of both the composite stellar population and the metallicity''.

From the observational point of view, the inherent difficulty in 
disentangling the genuine metallicity effects on the color properties 
of late-type galaxies has greatly favored the alternative approach relying on the 
study of H{\sc ii} regions as effective proxies of disk metallicity 
\citep[see, e.g.][]{peimbert78,zaritsky94,peimbert01,garnett02}. Broad-band colors 
and narrow-band spectrophotometric indices have been imposing, on the contrary, 
as election tools to probe chemical abundance in the galaxy bulges 
\citep{jablonka96,henry99,jablonka07,gorgas07}, while they mainly constrain galaxy SFR 
in the disks \citep[among others, see in particular][]{kennicutt94,gavazzi93,boselli01}.
In this framework, one remarkable effort toward a photometric approach 
to the metallicity problem is the work of \citet{perez09}, providing
Lick-index measurements of stellar populations in barred spirals.

\begin{figure}
\centerline{
\psfig{file=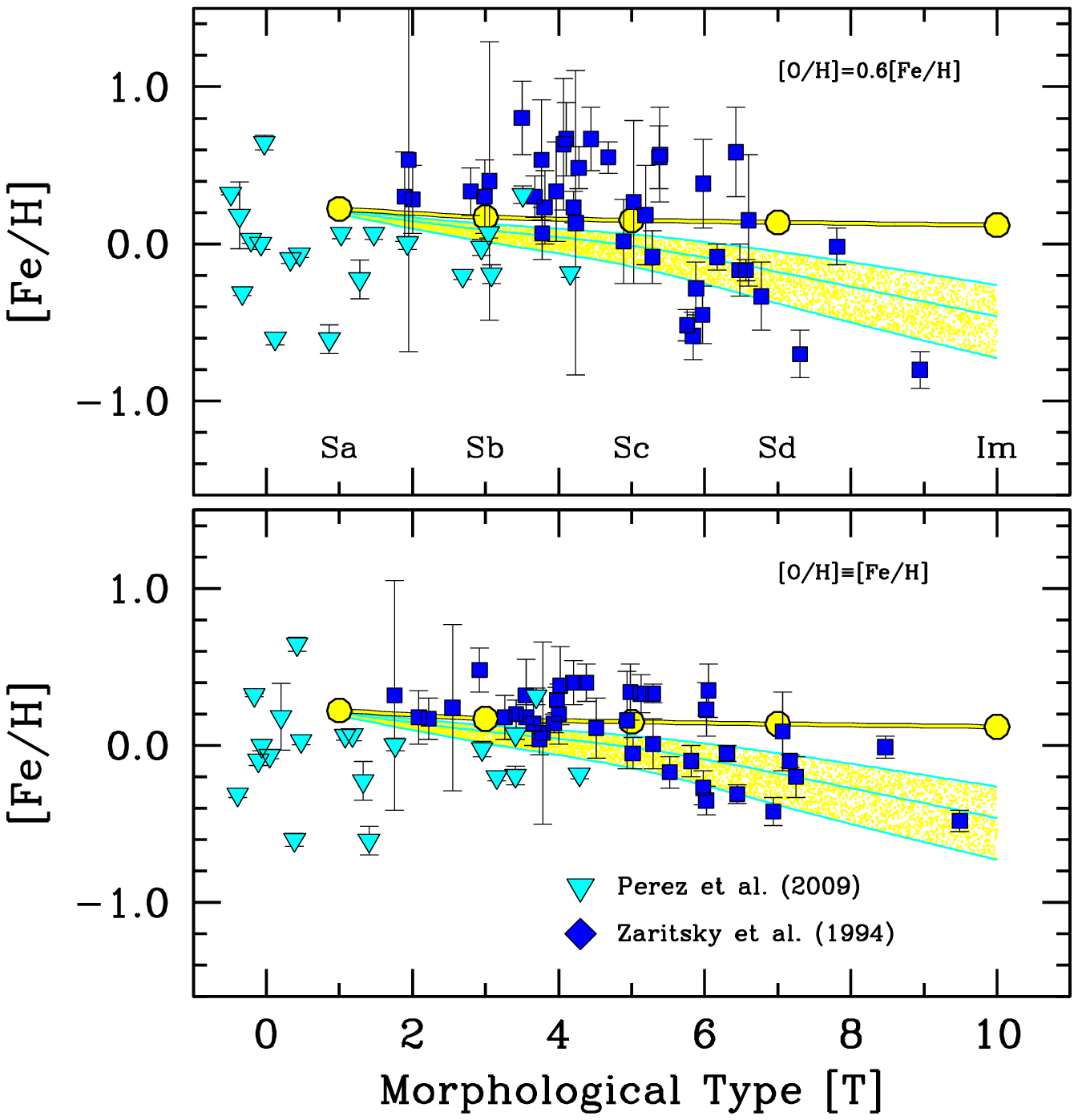,width=\hsize}
}
\caption{$[Fe/H]$ metallicity for the two galaxy samples of \citet{perez09}
(triangles) and \citet{zaritsky94} (squares), compared with the yield and gas
metallicity figures (solid line with dots and shaded area) as from Fig.~\ref{ari}.
While the \citet{perez09} estimates derive from Lick indices tracing the
galaxy stellar population, in case of \citet{zaritsky94} the $[Fe/H]$
derives from the Oxygen abundance of galaxy H{\sc ii} regions, by adopting
$\log (O/H)_\odot = 8.83$ for the Sun \citep{grevesse98}.
Two cases are envisaged, in this regard, assuming that Oxygen straightly 
traces Iron (and the full metallicity), as in lower panel or, more
realistically, that a scale conversion does exist such as $[O/H] = 0.6\,[Fe/H]$,
as suggested by the observation of disk stars in the Milky Way
\citep{clegg81,bessell91,edvardsson93}. A little random scatter has been
added to the morphological $T$ class of the points in each panel for better reading.
}
\label{zary}
\end{figure}

An instructive summary of observational galaxy diagnostics is proposed in Fig.~\ref{zary},
where the performance of both ``chemical'' (H{\sc ii} regions) and ``photometric''
(Lick indices from the aggregated stellar population) methods is compared for a set 
of spirals along the full Hubble
sequence. To the theoretical pattern of Fig.~\ref{ari}, repoted as a guideline,
we superposed the metallicity estimates for the galaxy set of \citet{zaritsky94}
(square markers), and for the \citet{perez09} S0/a sample (triangles).
As usual, the original gas metallicity in the \citet{zaritsky94} galaxies is given
in terms of $[O/H]$ abundance, so that two variants of the figure can be devised, 
depending whether we plainly assume Oxygen to trace global metallicity 
(so that $[O/H] = [Fe/H]$, as in lower panel of the figure) or, perhaps more 
realistically, whether decoupled $\alpha$-element enrichment should be considered 
to rescale chemical abundance (upper panel), as we have been discussing in previous
section. In the latter case, from different sources \citep{clegg81,bessell91,edvardsson93}, 
we set $[O/H] = 0.6[Fe/H]$. 

Compared with the theoretical predictions, it is evident from Fig.~\ref{zary}
that any empirical effort to constrain metallicity in spiral galaxies still 
suffers from very high uncertainty. This is especially evident for the H{\sc ii}
measurements of \citet{zaritsky94}, where point scatter is actually comparable to the
claimed internal error of individual measurements. Apparently, this does not
seem to be the case for the \citet{perez09} Lick data, although the important 
point-to-point spread casts evident doubts on the nominal accuracy claimed
for the individual [Fe/H] estimates. Definitely, we believe that a 
$\sigma[Fe/H] \simeq \pm 0.5$~dex might be regarded as a realistic figure for
current empirical (either ``chemical'' or ``photometric'') estimates of metal
abundance in late-type galaxy disks.

Once considering in more detail the overlapping range of morphological types 
between the two galaxy samples in  Fig.~\ref{zary}, it is interesting to note 
that ``photometric'' metallicity estimates appear to be systematically lower 
than ``chemical'' estimates. According to our previous arguments, this can be seen
as an eloquent example of ``biased'' sampling criteria; in fact, by probing
gas metallicity we retrieve the current status of galaxy chemical enrichment,
while through spectrophotometric Lick indices we are probing the 
galaxy stellar population ``averaging'' along the entire range of age of luminous
stars.  

\subsection{Mass, morphology \& birthrate (which drives galaxy metallicity?)}

Although morphological classification is the most immediate tool for 
a ``quick-look'' characterization of galaxy properties, it has been long 
questioned whether morphology univocally seats on intrinsic physical properties of
the galaxy system as a whole. This is especially true for late-type galaxies, where
the prominent spiral pattern actually involves a relatively small fraction of 
the galaxian total mass. As extensively discussed in the literature along
the latest decades \citep{whitford62,roberts94,kennicutt98}, the combined 
analysis of the galaxy leading properties led to a number of established 
correlations between ``primary'' physical parameters, the most popular one
being by far the \citet[][TF]{tully77} relation.

On the other hand, the obvious reference role of galaxy {\it luminosity}, as in the
TF relation, can easily be replaced by galaxy {\it stellar mass}, especially once
focussing on galaxy photometric properties in the mid infrared
range, as convincingly demonstrated by \citet{gavazzi93}. 
All these different pieces of evidence led eventually to identify galaxy (stellar)
mass as (one of) the leading parameters that may intimately mark the life and 
fate of the spiral systems as a whole \citep{tully82}. 

In this framework, the recent ultraviolet study  by \citet{marino10} may add 
further interesting arguments to this composite scenario. These authors 
collected {\sc Galex} observations of three spiral-rich aggregates, 
from the Lyon Groups of Galaxies \citep[LGG;][]{garcia93}. For their galaxy
population, these groups can be regarded as close analogs of the Local Group, 
displaying a quite ``clean'' sample of relatively unperturbed objects seen at 
comparable distances and covering the whole Hubble morphological sequence. 
While {\sc Galex} data allow us to explicitely assess current SFR in these galaxies, 
the match with long-wavelength magnitudes provide, on the contrary, a more
direct hint of the amount of stellar mass, through the appropriate M/L conversion ratio.

As a first step of our analysis, galaxy internal extinction has to be sized 
up relying on the {\sc Galex} $(FUV-NUV)$ color. The latter directly relates 
to the $\theta$ spectral slope of galaxy UV emission assuming, 
with \citet{calzetti99}, that $F(\lambda) \propto \lambda^\theta$. 
The intervening action of dust works in this sense of absorbing stellar 
luminosity at shorter wavelength making the spectral slope to flatten,
so that $\theta \to 0$. On the other hand, population synthesis models 
\citep[see e.g.][]{leitherer99,buzzoni02} firmly agree in predicting
a characteristic spectral slope $\theta_o \simeq -2.1$ in case of any dust-free
star-forming activity driven by the presence of high-mass ($M \ga 5$~M$_\odot$) stars.
Therefore, the amount of the $\theta$ flattening (i.e. $\Delta \theta = \theta -\theta_o$) 
leads to a straightforward measure of the monochromatic 
attenuation at the different photometric bands.\footnote{By adopting the 
\citet{calzetti99} attenuation curve, for the magnitude dimming
one derives \citep[cf. e.g.][]{buzzoni02} 
$A(B) = 1.31\,\Delta \theta$, $A(NUV) = 2.0\,\Delta \theta$ and 
$A(FUV) = 2.47\,\Delta \theta$, assuming FUV and NUV wavebands to be centered 
at 1520 \AA\ and 2300~\AA, respectively \citep{neff08}.\label{fred}}

\begin{table}
\scriptsize
\caption{Inferred physical properties for the \citet{marino10} LGG galaxy sample}
\label{t5}       
\begin{tabular}{lllrrrrr}
\hline
Galaxy ID & Morph & T & \multicolumn{1}{c}{$\theta$} & \multicolumn{1}{c}{$M_B$} & $\log M^*_{\rm gal}$ & \multicolumn{1}{c}{$\log$ SFR$_o$} & \multicolumn{1}{c}{$b$} \\
    &  \multicolumn{1}{c}{Type}   &    &    &	   &  &  & \\
\multicolumn{8}{c}{\hrulefill~~LGG 93~~\hrulefill} \\
NGC 1249   &   SBc  &  6.0  &  --1.02 &  --19.57 & 10.53 &    0.453 &  1.07   \\ 
NGC 1311   &   SBm  &  8.8  &  --1.07 &  --18.30 &  9.69 &    0.009 &  2.67   \\ 
IC 1933    &   Sc   &  6.1  &  --1.16 &  --18.66 & 10.16 &    0.226 &  1.52   \\ 
IC 1954    &   SBb  &  3.2  &  --0.96 &  --19.72 & 10.93 &    0.423 &  0.40   \\ 
IC 1959    &   SBd  &  8.4  &  --1.29 &  --18.21 &  9.70 &   -0.065 &  2.19   \\ 
\multicolumn{8}{c}{\hrulefill~~LGG 127~~\hrulefill} \\
NGC 1744   &   SBcd &  6.7  &  --1.18 &  --20.24 & 10.72 &    0.626 &  1.04  \\ 
NGC 1792   &   SBbc &  4.0  &  --0.24 &  --22.42 & 11.91 &    1.376 &  0.38  \\ 
NGC 1800   &   Sd   &  8.2  &  --1.22 &  --18.78 &  9.95 &    0.090 &  1.77  \\ 
NGC 1808   &   SABa &  1.2  &  --0.36 &  --22.21 & 12.16 &    1.035 &  0.10  \\ 
NGC 1827   &   SABc &  5.9  &  --0.22 &  --19.96 & 10.70 &    0.886 &  1.97  \\ 
ESO305-009 &   SBd  &  8.0  &  --1.31 &  --18.39 &  9.82 &   -0.022 &  1.85   \\ 
ESO305-017 &   IB   &  9.9  &  --1.29 &  --17.14 &  9.10 &   -0.504 &  3.25   \\ 
ESO362-011 &   Sbc  &  4.2  &  --2.85 &  --17.74 & 10.01 &   -1.000 &  0.13   \\ 
ESO362-019 &   SBm  &  8.9  &  --1.31 &  --17.72 &  9.45 &   -0.314 &  2.24   \\ 
\multicolumn{8}{c}{\hrulefill~~LGG 225~~\hrulefill} \\
NGC 3370    &	Sc   &  5.1  & --1.24 &  --19.76 & 10.72 &    0.207 &  0.40  \\ 
NGC 3443    &	Scd  &  6.6  & --1.18 &  --17.26 &  9.54 &   -0.208 &  2.32  \\ 
NGC 3447    &	Sm   &  8.8  & --1.40 &  --17.31 &  9.29 &   -0.014 &  6.33  \\ 
NGC 3447A   &	IB   &  9.9  & --1.76 &  --15.15 &  8.30 &   -0.768 &  11.0 \\  
NGC 3454    &	SBc  &  5.5  & --0.62 &  --19.08 & 10.39 &   -0.209 &  0.32  \\ 
NGC 3455    &	SABb &  3.1  & --0.73 &  --18.31 & 10.37 &    0.434 &  1.48  \\ 
NGC 3501    &	Sc   &  5.9  & --0.78 &  --18.99 & 10.31 &   -0.142 &  0.45  \\ 
NGC 3507    &	SBb  &  3.1  & --1.07 &  --20.10 & 11.09 &    0.422 &  0.28  \\ 
UGC 6022    &   I    &  9.9  & --1.31 &  --14.87 &  8.19 &   -1.127 &  6.27  \\       
UGC 6035    &	IB   &  9.9  & --1.27 &  --16.48 &  8.83 &   -0.695 &  3.82  \\ 
UGC 6083    &	Sbc  &  4.1  & --0.82 &  --17.39 &  9.88 &   -0.722 &  0.32  \\ 
UGC 6112    &	Scd  &  7.4  & --1.04 &  --17.49 &  9.54 &   -0.248 &  2.12  \\ 
UGC 6171    &	IB   &  9.9  & --1.60 &  --16.36 &  8.78 &   -0.847 &  3.02  \\ 
\hline
\noalign{Notes: Morphological-type classification from the \citet{marino10}
compilation; spectral slope $\theta$ from  $(FUV-NUV)$ color as
$\theta = -[2+0.4\,(FUV-NUV)/\log(1520/2300)]$; absolute $B$ magnitude $M_B$
from $B_T$ from RC3 \citep{rc3}, except UGC~6022,
which is from UGC \citep{nilson73}), after correction
for reddening (see footnote~\ref{fred}) and distance, the latter derived
from the mean redshift of the parent group, assuming a pure Hubble flow with 
$H_o=75$~km s$^{-1}$ Mpc$^{-1}$;
galaxy stellar mass in M$_\odot$ from $M/L_B$, according to \citet{buzzoni05}; SFR 
in  M$_\odot$\, yr$^{-1}$ from the dereddened NUV flux, according to 
footnote~\ref{fnuv}; birthrate $b$ assumes a current galaxy age of 13 Gyr.}
\end{tabular}
\end{table}

Internal reddening appears to be quite high in the UV bands, with the 
galaxies loosing typically 2-3 mag at 1500~\AA, a figure which is consistent with 
the more exhaustive studies of \citet{rifatto95} and \citet{gordon97}. 
Once accounted for reddening and distance, UV absolute luminosities directly translate
into actual SFR$_o$,\footnote{According to \citet{buzzoni02}, the inferred SFR 
(in M$_\odot$\, yr$^{-1}$) from {\sc Galex} NUV luminosities is
$\log\, {\rm SFR} = \log\, F_{\rm Galex} + 27.70$, providing to express flux 
density, in erg\,s$^{-1}$\,Hz$^{-1}$. This calibration is for a Salpeter IMF with 
stars between 0.1 and 120~M$_\odot$.\label{fnuv}}
while absolute B magnitude can eventually relate to galaxy stellar mass ($M^*_{\rm gal}$)
through the appropriate $M/L$ ratio from the \citet{buzzoni05} template galaxy models, 
as in Table~\ref{t2}. The galaxy characteristic birthrate can also be computed from 
our data, as $b = {\rm SFR}_o t_o/ M^*_{\rm gal}$, by assuming $t_o = 13$~Gyr.

\begin{figure}
\centerline{
\psfig{file=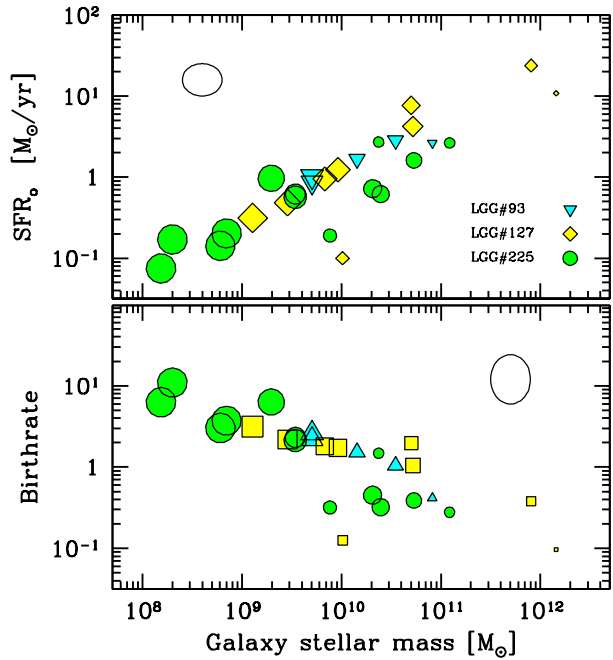,width=\hsize}
}
\caption{The inferred star-formation properties of the spiral-galaxy sample
by \citet{marino10}, referring to three Local-Group Analogs from the LGG
catalog \citet{garcia93}, as marked and labelled on the plot.
Marker size indicates galaxy morphology along the RC3 $T$-class sequence
(the biggest markers are for Sd/Im galaxies with $T \sim 10$, while the 
smallest are for Sa systems with $T \sim 1$). Current SFR derives
from {\sc Galex} NUV fluxes, dereddened as discussed in the text, and calibrated
according to \citet{buzzoni02}.
On the contrary, Galaxy stellar mass is computed from the relevant $M/L_B$ 
ratio, as from the \citet{buzzoni05} template galaxy models.
The birthrate assumes for all galaxies a current age of 13~Gyr.
The small ellipses in each panel report the typical uncertainty figures,
as discussed in the text. All the relevant data for this sample are summarized
in Table~\ref{t5}. Note the inverse relationship between $b$ and 
$M^*_{\rm gal}$, as in the down-sizing scenario for galaxy formation.
}
\label{lga}
\end{figure}

The results of our analysis are collected in Table~\ref{t5}, and summarized in Fig.~\ref{lga}.
Just as a reference, in each panel of the figure we also reported 
the typical error figures of the data (see the ellipses in each plot); in particular, 
we estimate a $\pm 0.2$~dex uncertainty on the inferred value of $\log {\rm SFR}_o$ 
(mainly as a consequence of the inherent uncertainty on the value of $\theta$), and 
a $\pm 0.2$~dex uncertainty for the derived galaxy mass (accounting for the reddening 
uncertainty and M/L calibration).This eventually leads to a $\Delta \log b = \pm 0.3$~dex 
for our birthrate estimates.

From both plots, a nice correlation ($\rho = 0.86$) is in place between star-formation 
properties and galaxy (stellar) mass. Indeed, the agreement might even be better 
if one could compare consistently with the {\it disk} stellar mass 
alone.\footnote{While the UV emission mainly traces star formation activity in 
the galaxy {\it disk}, our stellar mass derives, on the contrary, from the 
integrated $B$ magnitudes, thus including the bulge photometric contribution. 
For this reason, especially for Sa/b spirals, the value of $\log M^*_{\rm gal}$ may be 
overestimating the disk mass up to a factor of $\sim 4$ 
\citep[cf. Table 3 of][]{buzzoni05}, thus leading also to a correspondingly 
lower value of $b$.}

A nominal fit to the data in the upper panel of Fig.~\ref{lga} provides
\begin{equation}
\log {\rm SFR}_o = 
\begin{array}{l}
~~~\\
0.56\,\log M^*_{\rm gal} - 5.6 \qquad [M_\odot\,{\rm y}^{-1}]\\
\phantom{}\pm 7\phantom{....0000000} \pm 7
\end{array}
\label{eq:sfrm}
\end{equation}
with $\sigma(\log {\rm SFR}_o) = \pm 0.31$~dex.
As, by definition, SFR$_o \propto b\,M^*_{\rm gal}$, the deviation from the
one-to-one slope implies a dependence of the birthrate on the galaxy mass such as
\begin{equation}
b \propto M_{\rm gal}^{-0.44},
\end{equation}
as confirmed, indeed, by the point distribution in the lower panel.
In its essence, the inverse correlation of birthrate with galaxy mass
closely deals with the nowadays recognized effect of ``down-sizing'' 
\citep{cowie96,gavazzi96}, the imposing paradigm in theory of galaxy formation.

\begin{figure}
\centerline{
\psfig{file=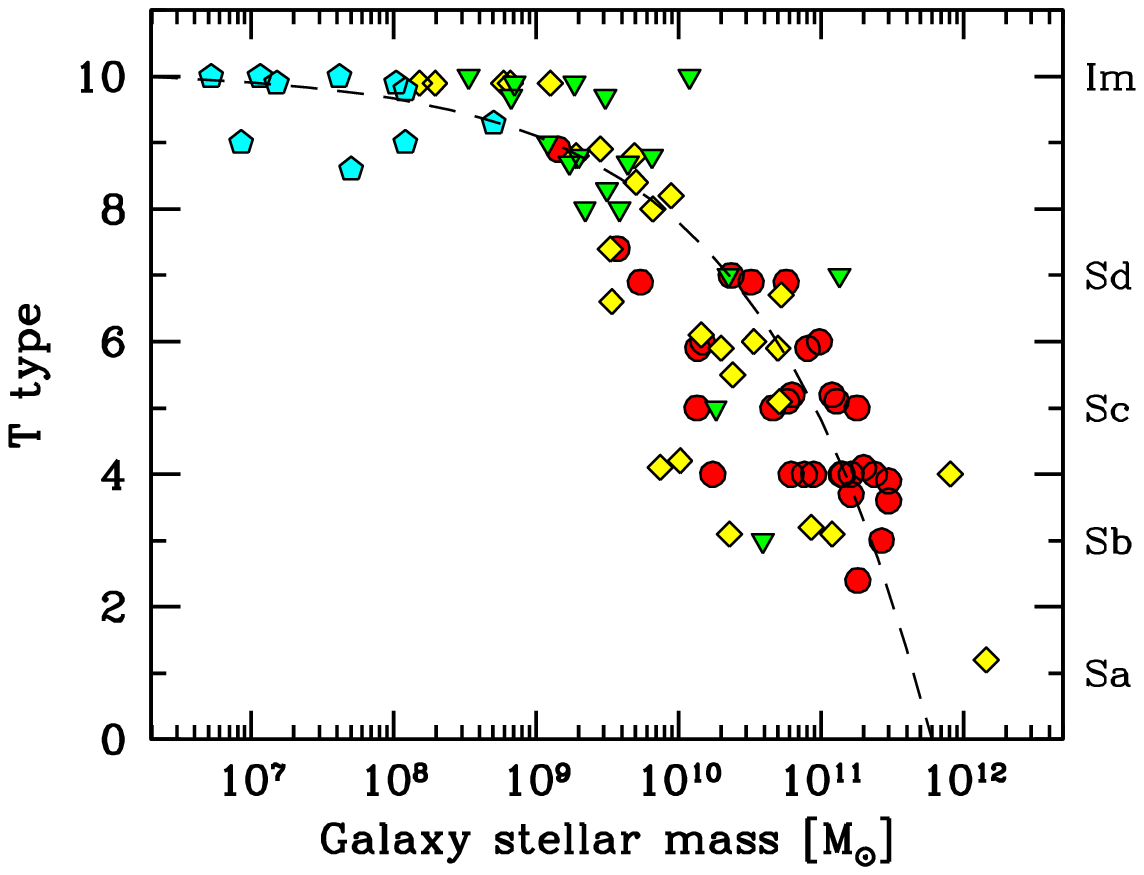,width=\hsize}
}
\caption{The observed relationship between morphological RC3 class $T$
and galaxy stellar mass according to four different galaxy samples
from the work of \citet{garnett02} (dots), \citet{marino10}
(diamonds), \citet{kuzio04} (triangles) and \citet{saviane08} (pentagons) 
(see Table~\ref{t5} and \ref{t6} for details).
The value of $M^*_{\rm gal}$ derives from $B$ photometry for
the \citet{garnett02}, \citet{kuzio04} and \citet{marino10} galaxies, and from $H$ absolute
magnitudes for the \citet{saviane08} sample, throught the appropriate
$M/L$ ratio, according to \citet{buzzoni05}.
Equation~(\ref{eq:tm}) provides a fair representation of the data, as
displayed by the dashed curve. As already pointed out also by \citet{roberts94}
one has to remark, however, the notable dispersion in mass among Sbc galaxies.
}
\label{mtype}
\end{figure}

The claimed link with galaxy morphology is another issue to be considered, of course.
In this case, however, a poorer correlation ($\rho = 0.59$) readily appears for 
the same data as far as the SFR$_o$ distribution versus the RC3 classification 
type ``T'' is dealt with. This is partly in consequence of 
a less tuned relationship between galaxy mass and morphological type,
as shown in Fig.~\ref{mtype}. 
In addition to the \citet{marino10} galaxies, we also joined to the figure
the sample of standard spirals studied by \citet{garnett02}, complemented by
the low-surface brightness galaxy sample of \citet{kuzio04}, and the dI
collection by \citet{saviane08}. For the different sets of data we 
derived $M^*_{\rm gal}$ with a similar procedure as for the \citet{marino10} data,
namely based on the integrated $B$ (for \citealp{garnett02} and \citealp{kuzio04}) 
and $H$ luminosity (for \citealp{saviane08}), respectively.
Again, a global fit to the data can be adjusted, tentatively in the form:
\begin{equation}
T = 10.1 - \left({{M^*_{\rm gal}}/10^9}\right)^{0.36}.
\label{eq:tm}
\end{equation}

While the general trend is evident from the data ($\rho = 0.83$),
one could easily verify that any attempt to use eq.(\ref{eq:tm}) for predicting 
galaxy morphology based on the mass, alone, is a nearly hopeless task.\footnote{The 
rms of the eq.(\ref{eq:tm}) fit can be written as $\sigma(T) \sim (8-0.8\,T)$.}
In particular, to a closer analysis, one has to report that
grand-design spirals (type Sbc or class $T \sim 4$) seem 
to set along a wider spread of mass (nearly two dex in $\log M^*_{\rm gal}$),
a feature that partially blurs the otherwise cleaner T-mass relationship,
as already pointed out by \citet{roberts94}.
Moreover, this picture cannot easily accomodate even dwarf ellipticals, 
that however seem to markedly characterize the high-density cosmic environment
\citep{cellone05,gavazzi10}. As well known, both these features directly call for 
a long-standing quest in the extragalactic debate about the role of primordial 
genesis and the environment ``nurture'' to modulate galaxy morphology
\citep[e.g.][]{dressler80,thuan92,blanton05,cooper07,tasca09}

\begin{figure}
\centerline{
\psfig{file=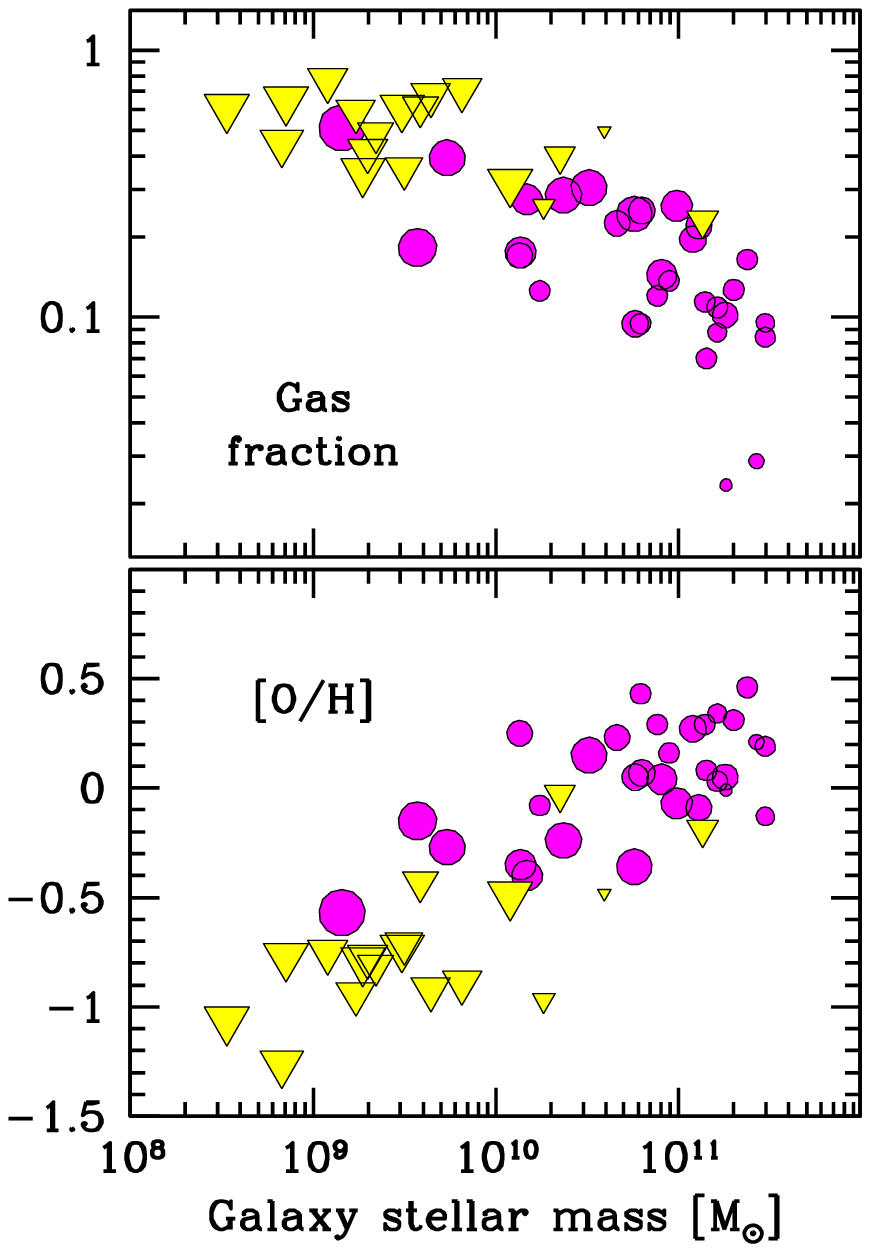,width=0.97\hsize}
}
\caption{Mass fraction of gas and Oxygen abundance of H{\sc ii} regions for the
\citet{garnett02} (dots) and \citet{kuzio04} (triangles) galaxy samples of Table~\ref{t6}. 
Marker size is propotional to
the galaxy morphological class $T$ (the biggest markers are for Sd/Im galaxies 
with $T \sim 10$, while the smallest are for Sa systems with $T \sim 1$).
Gas fraction accounts for both atomic and molecular phase. Note the inverse relationships 
of the data versus galaxy stellar mass, the latter as derived from the relevant $M/L$ ratio
(see, again, Table~\ref{t6} for details).
}
\label{gar}
\end{figure}

\begin{figure}
\centerline{
\psfig{file=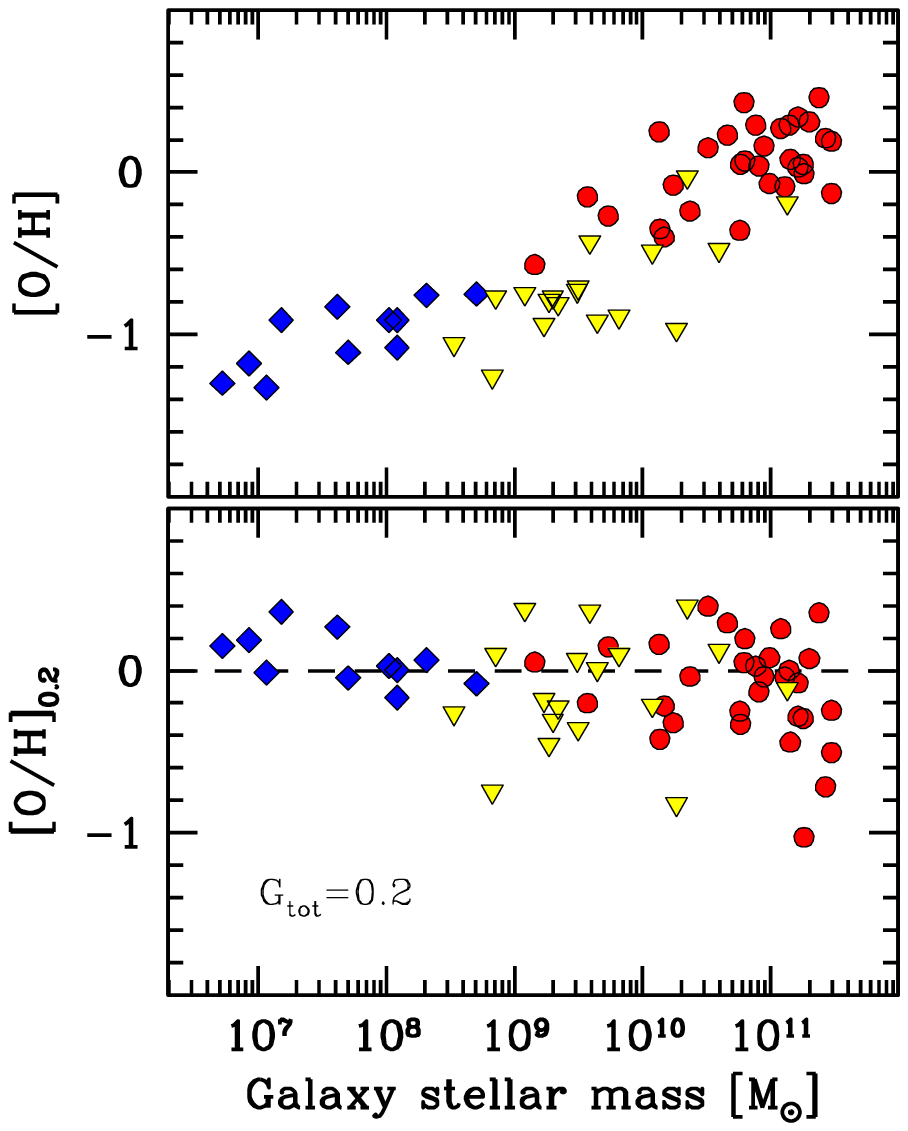,width=0.97\hsize}
}
\caption{{\it Upper panel:} Oxygen abundance from H{\sc ii} regions for the galaxy
samples of \citet{garnett02} (dots), \citet{kuzio04} (triangles) and \citet{saviane08} 
(diamonds).
A nice correlation is in place with galaxy stellar mass, where low-mass systems 
appear richer in gas (see Fig.~\ref{gar}) and poorer in metals.
Once accounting for the different gas fraction among the galaxies, and rescale
Oxygen abundance to the same gas fraction ($G_{\rm tot} = 0.2$, as in the indicative 
example displayed in the {\it lower panel}), note that most of the 
$[O/H]$ vs.\ $M^*_{\rm gal}$ trend is recovered, leaving a nearly 
flat $[O/H]$ distribution along the entire galaxy mass range, as predicted by
the nearly constant yield-metallicity of the systems. 
}
\label{sav}
\end{figure}

As far as galaxy metallicity is concerned, an interesting picture can be
devised, relying again on the \citet{garnett02} and \citet{kuzio04} contributions.
In Fig.~\ref{gar} we further elaborated their original data displaying,
in the two panels, the mass fraction of fresh gas (including
atomic and molecular Hydrogen), and the representative [O/H] abundance of
H{\sc ii} regions. For the reader's better convenience, a summary of these data
is also reported in Table~\ref{t6}.
The gas fraction ${\cal G}_{\rm tot} = M_{\rm gas}/(M_{\rm gas}+M^*_{\rm gal})$ can 
easily be derived from the data in force of the tight $M_{\rm gas}/M^*_{\rm gal}$ 
relationship:
\begin{equation}
\log M_{\rm gas} =  
\begin{array}{l}
~~~\\
0.60\,\log M^*_{\rm gal} +3.51,\\
\phantom{}\pm 8\phantom{..00000000} \pm 87
\end{array}
\label{eq:g}
\end{equation}
with $\sigma(\log M_{\rm gas}) = \pm 0.25$~dex, and $\rho = 0.81$.

The upper panel of the figure shows the somewhat unescapable consequence 
of the enhanced birthrate among low-mass systems. A higher value of $b$
implies that a larger fraction of fresh gas must still be available 
at present time.
Actually, one may even speculate that such a delayed mass processing 
is in fact the natural explanation also of the low-metallicity figures
that characterize, all the way, the dwarf-galaxy population in the Universe 
\citep{arimoto90,lee03}.
This scenario seems to find direct support once exploring,
as in the upper panel of Fig.~\ref{sav}, H{\sc ii} metallicity over the whole 
range of galaxy mass, for instance by extending again the \citet{garnett02} 
sequence for ``standard'' spirals with the low-mass samples of \citet{kuzio04}
and \citet{saviane08} (see also Table~\ref{t6}).

\begin{table}
\scriptsize
\caption{Inferred physical properties for the \citet{garnett02},
\citet{kuzio04}, and \citet{saviane08} galaxy samples}
\label{t6}       
\begin{tabular}{lccccc}
\hline
Galaxy ID & Morph. & $\log M^*_{\rm gal}$ & ${\cal G}_{\rm tot}^{(b)}$ & [O/H]$^{(c)}$ & [O/H]$_{0.2}^{(d)}$ \\
    & Type$^{(a)}$ & [$M_\odot$]   &  &  & \\
\hline
\multicolumn{6}{c}{\hrulefill~~\citet{garnett02}~~\hrulefill} \\
NGC~224   & 3.0 & 11.44 &  0.03 &	\phantom{+}0.21 &  --0.72 \\
NGC~253   & 5.1 & 10.80 &  0.09 &	\phantom{+}0.05 &  --0.33 \\
NGC~300   & 6.9 & \phantom{1}9.95 &  0.40 &  --0.27 &   \phantom{+}0.15 \\
NGC~598   & 5.9 & 10.22 &  0.17 &  --0.35  &       	   --0.42 \\  
NGC~628   & 5.2 & 10.92 &  0.25 &	\phantom{+}0.07 &   \phantom{+}0.20 \\
NGC~925   & 7.0 & 10.52 &  0.29 &  --0.24 &	           --0.04 \\  
NGC~1232  & 5.0 & 11.30 &  0.10 &	\phantom{+}0.05 &  --0.29 \\
NGC~1637  & 5.0 & 10.21 &  0.17 &	\phantom{+}0.25 &   \phantom{+}0.16 \\
NGC~2403  & 6.0 & 10.31 &  0.28 &  --0.40 &	           --0.22 \\  
NGC~2442  & 3.7 & 11.25 &  0.09 &	\phantom{+}0.34 &  --0.08 \\
NGC~2805  & 6.9 & 10.88 &  0.24 &  --0.36 &	           --0.25 \\  
NGC~2903  & 4.0 & 10.83 &  0.09 &	\phantom{+}0.43 &   \phantom{+}0.05 \\
NGC~3031  & 2.4 & 11.27 &  0.02 &  --0.01 &	           --1.03 \\ 
NGC~3344  & 4.0 & 10.30 &  0.13 &  --0.08 &	           --0.32 \\  
NGC~3521  & 4.0 & 10.94 &  0.12 &	\phantom{+}0.29 &   \phantom{+}0.03 \\
NGC~3621  & 6.9 & 10.67 &  0.31 &	\phantom{+}0.15 &   \phantom{+}0.40 \\
NGC~4254  & 5.2 & 11.17 &  0.20 &	\phantom{+}0.27 &   \phantom{+}0.26 \\
NGC~4258  & 4.0 & 11.18 &  0.07 &	\phantom{+}0.08 &  --0.44 \\
NGC~4303  & 4.0 & 11.26 &  0.11 &	\phantom{+}0.03 &  --0.28 \\
NGC~4321  & 4.1 & 11.36 &  0.13 &	\phantom{+}0.31 &   \phantom{+}0.07 \\
NGC~4395  & 8.9 & \phantom{1}9.47 &  0.51 &  --0.57 &   \phantom{+}0.05 \\
NGC~5033  & 5.1 & 11.22 &  0.22 &  --0.09 &	           --0.04 \\  
NGC~5055  & 4.0 & 11.01 &  0.14 &	\phantom{+}0.16 &  --0.04 \\
NGC~5194  & 4.0 & 11.20 &  0.11 &	\phantom{+}0.29 &   \phantom{+}0.00 \\
NGC~5236  & 5.0 & 10.77 &  0.22 &	\phantom{+}0.23 &   \phantom{+}0.29 \\
NGC~5457  & 6.0 & 11.12 &  0.26 &  --0.07 &	            \phantom{+}0.08 \\  
NGC~6384  & 3.6 & 11.52 &  0.10 &  --0.13 &	           --0.51 \\  
NGC~6744  & 4.0 & 11.45 &  0.16 &	\phantom{+}0.46 &   \phantom{+}0.36 \\
NGC~6946  & 5.9 & 10.97 &  0.14 &	\phantom{+}0.04 &  --0.13 \\
NGC~7331  & 3.9 & 11.51 &  0.08 &	\phantom{+}0.19 &  --0.25 \\
NGC~7793  & 7.4 & \phantom{1}9.66 &  0.18 &  --0.15 &  --0.20 \\
\multicolumn{6}{c}{\hrulefill~~\citet{kuzio04}~~\hrulefill} \\
F563-1	  & 8.0	 &  \phantom{1}9.34 &  \phantom{+}0.49 & --0.81 & --0.23 \\
F571-5	  & 9.0	 &  \phantom{1}9.08 &  \phantom{+}0.77 & --0.75 &  \phantom{+}0.38 \\
UGC~1230  & 8.7	 &  \phantom{1}9.64 &  \phantom{+}0.68 & --0.92 &  \phantom{+}0.01 \\
UGC~5005  & 9.9	 &  \phantom{1}9.27 &  \phantom{+}0.35 & --0.79 & --0.46 \\
UGC~9024  & 3.0	 & 10.59 &  \phantom{+}0.50 & --0.48 & 	    \phantom{+}0.12 \\  
UGC~12695 & 8.8	 &  \phantom{1}9.81 &  \phantom{+}0.71 & --0.89 &  \phantom{+}0.10 \\
F415-3	  & 9.9	 &  \phantom{1}8.85 &  \phantom{+}0.65 & --0.77 &  \phantom{+}0.10 \\
F469-2	  & 8.7	 &  \phantom{1}9.23 &  \phantom{+}0.59 & --0.94 & --0.18 \\
F530-3	  & 5.0	 & 10.26 &  \phantom{+}0.26 & --0.97 & 	   --0.82 \\  
F561-1	  & 8.3	 &  \phantom{1}9.50 &  \phantom{+}0.36 & --0.71 & --0.36 \\
F563-V1	  & 9.7	 &  \phantom{1}8.83 &  \phantom{+}0.45 & --1.26 & --0.75 \\
F563-V2	  & 9.7	 &  \phantom{1}9.49 &  \phantom{+}0.61 & --0.73 &  \phantom{+}0.07 \\
F568-6	  & 7.0	 & 11.13 &  \phantom{+}0.23 & -0.19 &  	   --0.11 \\  
F577-V1	  & 8.0	 &  \phantom{1}9.59 &  \phantom{+}0.61 & --0.43 &  \phantom{+}0.37 \\
F611-1	  & 10.0 &  \phantom{1}8.53 &  \phantom{+}0.61 & --1.06 & --0.26 \\
F746-1	  & 10.0 & 10.08 &  \phantom{+}0.32 & --0.49 & 	   --0.22 \\  
UGC~5709  & 7.0	 & 10.35 &  \phantom{+}0.40 & --0.03 & 	    \phantom{+}0.40 \\  
UGC~6151  & 8.8	 &  \phantom{1}9.30 &  \phantom{+}0.42 & --0.77 & --0.31 \\
\multicolumn{6}{c}{\hrulefill~~\citet{saviane08}~~\hrulefill} \\
ESO 347-G017 &  9.0 & \phantom{1}8.08 & {\it 0.67} &  --0.91 & \phantom{+}0.00 \\ 
UGC A442     &  8.6 & \phantom{1}7.70 & {\it 0.74} &  --1.11 & --0.05 \\ 
ESO348-G009  & 10.0 & \phantom{1}7.62 & {\it 0.76} &  --0.83 & \phantom{+}0.27 \\  
NGC 59 	     &--3.0 & \phantom{1}8.31 & {\it 0.62} &  --0.76 & \phantom{+}0.06 \\  
ESO473-G024  & 10.0 & \phantom{1}7.06 & {\it 0.84} &  --1.33 & --0.01 \\  
AM0106-382   &  9.0 & \phantom{1}6.93 & {\it 0.85} &  --1.18 & \phantom{+}0.19 \\ 
NGC 625      &  9.3 & \phantom{1}8.70 & {\it 0.54} &  --0.75 & --0.08 \\ 
ESO245-G005  &  9.9 & \phantom{1}8.02 & {\it 0.68} &  --0.91 & \phantom{+}0.03 \\ 
DDO42        &  9.8 & \phantom{1}8.08 & {\it 0.67} &  --1.08 & --0.17 \\ 
DDO53 	     &  9.9 & \phantom{1}7.18 & {\it 0.82} &  --0.91 & \phantom{+}0.36 \\ 
UGC4483      & 10.0 & \phantom{1}6.72 & {\it 0.88} &  --1.30 & \phantom{+}0.15 \\  
\hline
\noalign{$(a)$: morphological class T, as reported by {\sc Hyperleda} \citep{paturel03};} 
\noalign{$(b)$: gas fraction ${\cal G}_{\rm tot} = M_{\rm gas}/(M_{\rm gas}+ 
M^*_{\rm gal})$; for the \citet{saviane08} sample ${\cal G}$ is extrapolated from
eq.(\ref{eq:g}).}
\noalign{$(c)$: for the Sun we assume $\log (O/H)_\odot = 8.83$ \citep{grevesse98}.}
\noalign{$(d)$: $[O/H]$ abundance rescaled to ${\cal G}_{\rm tot} = 0.2$.}
\end{tabular}
\end{table}

On the other hand, a far more instructive experiment can be carried out
with the help of eq.~(\ref{eq:zgas3}). Facing a change of the reference
value for the gas fraction, ${\cal G}_{\rm tot}$, the equation allows us to compute
the induced change in the value of $Z_{\rm gas}$, by accounting for the 
different ISM dilution factor.
Accordingly, the observed gas metallicity could easily be rescaled
to any other reference value of ${\cal G}_{\rm ref}$ starting from the observed
value of ${\cal G}_{\rm obs}$.
As far as the H{\sc ii} metallicity is considered, the expected offset to 
the observed $[O/H]$ abundance can therefore be written as
\begin{equation}
\Delta [O/H]  = \log \left({{1-{\cal G}_{\rm ref}}\over{{\cal G}_{\rm ref}}}\right)-
\log \left({{1-{\cal G}_{\rm tot}}\over{{\cal G}_{\rm tot}}}\right).
\label{eq:ggg}
\end{equation}
By entering eq.~(\ref{eq:ggg}) with the values of ${\cal G}_{\rm tot}$
from Table~\ref{t6} we can eventually rescale galaxy metallicity to the same
reference gas fraction, say for example ${\cal G}_{\rm ref} = 0.2$, such 
as $[O/H]_{0.2} = [O/H]_{\rm obs} + \Delta [O/H]$.
The result of our excercise is shown in the lower panel of 
Fig.~\ref{sav} for the joint sample of galaxies. 

Definitely, once accounting for the increasing fraction of fresh
gas with decreasing galaxy mass, one sees that {\it most of the observed
trend of $[O/H]$ with $M^*_{\rm gal}$ may simply be recovered by the 
dilution effect of processed mass on the ISM.}

\subsection{The Roberts time \& the fate of spiral galaxies}

As discussed in Sec.~2.1, the rate at which the mass is chemically enriched
inside a galaxy follows a sort of universal law; calculations show in fact 
that the burning efficiency factor, $\cal F$ (see Table~\ref{t1}), is nearly
insensitive to the galaxy star formation history.
At present, about $11\%\pm 2\%$ of the processed mass in galaxies has been 
converted into heavy elements and, according to eq.~(\ref{eq:stars}), the 
process may go ahead for a supplementary factor of $(0.11)^{-1/0.23} \sim 10^4$  
of the Hubble time.\footnote{Curiously enough, 
the matter annihilation that accompanies nuclear burning 
processes may also result, in the long term, in a not quite negligible
effect for the total mass budget of a galaxy. Previous figures show, for instance, that
the Milky Way has already lost in luminosity about one thousandth (i.e.\ a factor 
of $0.11 \times 0.0078$, see footnote~\ref{foot2}) of its total mass. This is roughly the equivalent mass 
of the whole globular-cluster system currently surrounding our galaxy.}
At that time, Hydrogen will definitely vanish even inside stars, and a maximum 
metallicity $Z_{\rm max}$ has to be reached by the system for 
$(1+R)Z_{\rm max} = 1-Y_p$. In terms of the solar figures, galaxy chemical 
evolution will eventually end up with 
\begin{equation}
(Y_{\rm max}, Z_{\rm max})\simeq (3\,Y_\odot, 10\,Z_\odot).
\end{equation}

On the other hand, much earlier than the full process completion,
other outstanding events could mark the chemical history
of a galaxy. In particular, if one recalls that ${\cal S} \propto t^b$, and takes
the Hubble time ($t_H$) as a reference, then the
expected evolution of the fresh-gas content in the disk scales as
\begin{equation}
{{(1-{\cal G})}\over{{(1-{\cal G}_H)}}} = \left({t\over t_H}\right)^b.
\label{eq:trg}
\end{equation}
If we set ${\cal G} = 0$ and recall eq.~(\ref{eq:gtot}), then the timescale for the 
galaxy to consume its primordial-gas reservoir is
\begin{equation}
t_R = {t_H\over{{(1-{\cal G}_H)}^{1/b}}} = t_H\,\left({{1-f}\over{1-{\cal G}_{\rm tot}},}\right)^{1/b}
\label{eq:rob}
\end{equation}
being ${\cal G}_{\rm tot}$ and $f$ evaluated at the (current) Hubble time.
After a $t_R$ time, the so-called ``Roberts time'' \citep{roberts63},
star formation can only proceed further in 
a galaxy at cost of exploiting its own processed mass.
Since long, the problem of a realistic estimate of the Roberts timescale
has been investigated \citep{sandage86,kennicutt94}, leading to 
the quite puzzling conclusion that $t_R$ may not substantially depart from
one Hubble time. As a consequence, present-day galaxies might be all 
on the verge of exhausting their own gas resources.
To overcome this somewhat embarassing conclusion, it has been emphasized
that, after all, even at time $t_R$ a galaxy is not completely gas depleted,
still counting on a residual buffer $M_{\rm gas} \simeq f M^*_{\rm gal}$ of 
processed mass (see eq.~\ref{eq:zgas3}) still rich in Hydrogen and able, in 
principle, to sustain galaxy SFR for the Gyrs to come \citep{kennicutt98}.

\begin{figure}
\centerline{
\psfig{file=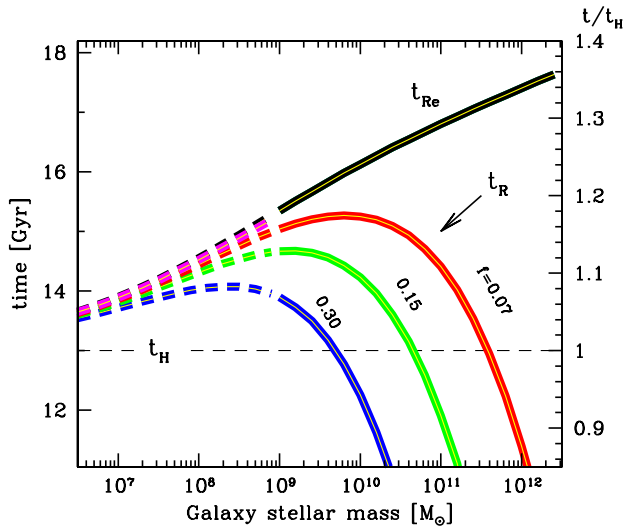,width=\hsize}
}
\caption{The lifetime of spiral galaxies as active star-forming systems,
better recognized as the \citet{roberts63} time, is computed along the full
mass range of the systems.
The definite consumption of the primordial gas reservoir, as in the
original definition of the timescale, $t_R$, derives from eq.~(\ref{eq:rob}),
by parameterizing with the current fraction of returned stellar mass, $f$.
Present-day galaxies are assumed to be one Hubble time old ($t_H \sim 13$~Gyr, see 
the left scale on the vertical axis). If returned stellar mass is also
included in the fuel budget, to further extend star formation, the the
``extended'' Roberts time, $t_{Re}$ derives from eq.~(\ref{eq:tre}).
In both equations, the distinctive birthrate relies on the observed down-sizing
relationship of eq.~(\ref{eq:sfrm}), while the gas fraction available to 
present-day galaxies is assumed to obey eq.~(\ref{eq:g}), eventually
extrapolated for very-low mass systems (i.e.\ $M^*_{\rm gal} \lesssim 10^9~M_\odot$)
(dashed curves in the plot). Note, for the latter, that they seem on the verge 
of definitely ceasing their star formation activity, while only
high-mass galaxies ($M^*_{\rm gal} \gtrsim 10^{11}~M_\odot$) may still have chance
to extend their active life for a supplementary 30-40\% lapse of $t_H$.
See text for a full discussion.
}
\label{trob}
\end{figure}

We can take advantage of our original approach to galaxy chemo-photometric
evolution to attempt a more confident assessment of the Roberts time, also
in its ``extended'' definition ($t_{Re}$), such as to account for the ``extra-time'' 
provided by processed-mass recycling.
First of all, according to its standard definition, the value of $t_R$ 
can easily be computed with our data along the different galaxy mass by means 
of eq.~(\ref{eq:g}). Once getting
the value of $M_{\rm gas}$ from the fitting equation, it is immediate to derive
${\cal G}_{\rm tot} = M_{\rm gas}/(M_{\rm gas}+M^*_{\rm gal})$.
Galaxy birthrate follows, as well, by relying on
eq.~(\ref{eq:sfrm}), and recalling that $b \simeq {\rm SFR}_H\,t_H/M^*_{\rm gal}$.
The results of eq.~(\ref{eq:rob}) are displayed in Fig.~\ref{trob},
maintaining the net value of the returned-mass fraction ($f$) as a free parameter
in our calculations. Evidently, current evolutionary scenario can typically be 
sustained by spiral galaxies only for a further 20\% of the Hubble time. 
For the most massive systems this is 2-3~Gyr ahead from now.

When returned stellar mass is included in our budget, too, then fresh stars 
can still form for a supplementary time lapse such as to lead to 
${\cal G}_{\rm tot} = 0$. Recalling again, eq.~(\ref{eq:gtot}), then
the ``extended'' Roberts time ($t_{Re}$) follows from eq.~(\ref{eq:trg}) as
\begin{equation}
t_{Re} = {t_H\over{{(1-{\cal G}_{tot})}^{1/b}}} = {t_R \over {(1-f)^{1/b}}},
\label{eq:tre}
\end{equation}
again with ${\cal G}_{\rm tot}$ and $f$ referring to the boundary conditions of
present-day galaxies (see Fig.~\ref{trob}).
Two important conclusions can be drawn from the analysis of these results.

{\it i)} For any realistic assumption about the returned mass fraction, 
{\it most of the galaxies more massive than $\sim 10^{11}~M_\odot$ have 
$t_R \lesssim t_H$. As a consequence, they may already have exceeded their 
Roberts time, being their current SFR mainly driven by processed mass rather than by 
primordial gas.} According to the morphology vs.\ mass relationship (see again
Fig.~\ref{mtype}) this seems to be the case for most of Sa/Sb spirals in 
the present-day Universe.

{\it ii)} When explicitely included in the gas budget, the processed-mass 
contribution certainly adds supplementary
chances to high-mass spirals  for continuing their star formation activity 
for a few Gyrs. However, the relative life extension is critically constrained 
by the balance of the two antagonic
processes of mass return ($f$) and mass consumption ($b$).
In absolute terms, eq.~(\ref{eq:tre}) shows that the ``extended'' Roberts time,
$t_{Re}$, is eventually driven by the ``down-sizing'' mechanism, which constrains 
the $b$ vs.\ mass relationship.
{\it The conclusion is that only a marked reduction of the present brithrate might allow 
spirals to sustain star formation far beyond one Hubble time.}
In any case, only a negligible ``bonus'' is awarded to low-mass systems (namely 
to Sd/Im galaxies less massive than $\sim 10^{11}~M_\odot$). For them, in fact, 
we definitely have $t_{Re} \simeq t_R \simeq t_H$, and just a couple of 
Gyrs, at most, still seem to be left ahead.

\section{Summary \& conclusions}

In this paper we tried an alternative approach to the study of chemo-photometric
properties of late-type galaxies. Following previous theoretical efforts in
this sense \citep[e.g.][]{pagel97}, our analysis relies on a 
basic criterion of energetic self consistency between chemical enhancement 
of galaxy mass, through nuclear processing, and the corresponding luminosity 
evolution of the system.

This plain physical constraint provides us with a brand new interpretative tool, 
as nearly all the leading theoretical codes along the last decades rather approached 
the problem of galaxy chemical evolution in terms of the sum of individual 
elemental contributions \citep[see, e.g.][]{tinsley80}. This required, in other words, to explicitely 
account for the chemical yields supplied by stars along the different range of mass. 
Clearly, this ``analytical'' approach is the unescapable way when we want to 
trace the specific enrichment history of single chemical elements.
On the other hand, as far as the global metallicity of a stellar aggregate
is concerned, the analytical process risks to carry along and magnify the 
combined uncertainty from stellar evolution theory making the emerging picture to
strongly depend on the model assumptions, often through a plethora of free tunable 
parameters.

On the contrary, in any ``synthetic'' approach, galaxy chemical evolution 
derives from other primary physical constraints. The works of \citet{schmidt63} 
and \citet{pagel75} are two outstanding examples in this sense, 
assessing the chemical history of a stellar system as an {\it implied consequence}
of its star-formation law. As a further variant on this line, 
our reasoning takes the move from the fact that a fixed amount of energy 
must always be released when matter is chemically enhanced through nuclear processing
inside stars.
Theoretical nucleosynthesis predicts, in fact, that $6.98\,10^{18}$~ergs are  
produced (see Sec.~2) for each gram of Hydrogen to be converted to heavier 
elements through non-explosive burning processes. A further amount of 
$1.03\,10^{18}$~ergs per gram should then be added if we want to account for 
the explosive nucleosynthesis dealing with the SN events. Overall, it is clear 
that a tight relationship has therefore to be expected between chemical 
enhancement of a galaxy and its past photometric history.

The theoretical fundamentals which provide the reference framework for
our analysis have been discussed in Sec.~2. In particular, we rely on the
\citet{buzzoni89,buzzoni95} original code for stellar population synthesis, 
further elaborated to build up three-zone galaxy templates along the Hubble 
morphological squence \citep{buzzoni02,buzzoni05}.
For its special relevance within the global properties of late-type systems,
our study is mainly focussed on the disk evolution, which broadly marks the macroscopic 
look of spiral and irregular galaxies.
The standard scenario of the present calculations considers a Salpeter IMF, 
with stars formed within a mass range between 0.1 and 120~M$_\odot$. In addition,
metals and Helium proceed from processed stellar mass assuming a fixed 
enrichment ratio $R = 3$, as discussed in Sec.~2.3. 

Following \citet{buzzoni05}, an important assumption of our models
is that stellar birthrate is a distinctive property of 
galaxy morphology and mass (the latter, in force of the observed 
$T$-$M^*_{\rm gal}$ relationship, as shown in Fig.~\ref{mtype}). 
Consistently with the down-sizing scenario for galaxy formation \citep{cowie96},
$b$ is therefore seen to decrease, in our models, with increasing 
$M^*_{\rm gal}$ \citep{gavazzi93}.
As a consequence, SFR scales with time according to a power law, and it
is not directly coupled to the mean gas density of the disk 
as imposed, on the contrary, by other classical schemes \citep{schmidt59}.

Chemical enhancement inside a galaxy is effectively assessed
by the burning efficiency factor, as displayed in 
eq.~(\ref{eq:stars}) of Sec.~2.1. Table~\ref{t1} reports a notable property of 
$\cal F$, being it nearly insensitive to the galaxy star formation history.
Along a full range of possible evolutionary scenarios, in fact, 
we always end up at present time with ${\cal F} = 0.11 \pm 0.02$.
This means, in other words that, within just small ($\pm 20\%$) individual differences, 
galaxies performed very similarly everywhere in the Universe as efficient engines
for energetic exploitation of the barionic matter through nuclear synthesis.
Accordingly, at any epoch we can univocally define a reference
figure, that we called ``yield metallicity'' ($\cal Z$, see eq.~\ref{eq:yield}
in Sec.~2.3), which traces the mean composition of the stellar mass.

The role of SNe in the more general context of disk chemical evolution 
has been assessed in some detail in Sec.~2.2. A comparison of the energetic 
budget involved in quiescent and explosive nucleosynthesis indicates 
that SNe are extremely powerful mass processors; for fixed amount of released 
energy, in fact, they burn a factor of seven more mass than ``normal'' stars.
Following yet classical arguments \citep[e.g.][]{matteucci86} we know that
the component of high-mass stars ($M \gtrsim 5~M_\odot$) enables a very quick 
chemical enrichment during early evolution of galaxies.
This ``prompt'' chemical enhancement is the result 
of the explosion of both core-collapsed objects, in the form of type {\sc ii} SNe, 
and accreted WDs as a result of binary interactions, like in type {\sc I}a SNe. 

Yield metallicity is therefore abruptly raised to $\log {\cal Z} \sim -3$ 
on a timescale of a few $10^7$~yrs, a figure that further increases by $\sim 0.15$~dex within the 
first Gyr of galaxy life (see Fig.~\ref{snrall}).
As, by definition, the cumulative action of ``prompt'' SNe{\sc ii} and {\sc i}a 
must be proportional to the galaxy stellar mass $M^*_{\rm gal}$, the net effect
of their contribution is to set a {\it steady offset} (that we quantified 
in Sec.~3.2 in $[Fe/H] \simeq -1.10$~dex) to galaxy yield metallicity.

In this framework, for a Salpeter IMF, the implied yield metallicity evolves
as ${\cal Z} \propto t^{0.23}$ (see eq.~\ref{eq:calz}). For the previous arguments, 
this can be seen as a nearly universal law, only marginally dependent on the 
galaxy birthrate (see, for instance, Fig.~\ref{zevol}). 
An immediate consequence of this important property is that IMF, rather than SFR,
is the key player to set metallicity at primeval epochs, when $t \to 0$ and 
SNe were governing chemical evolution \citep[e.g.][]{scannapieco03}.
In particular, non-Salpeter stellar populations, strongly biased toward high-mass
stars as in a flatter IMF, have to be invoked to fit with the peculiar presence 
of bright super-solar galaxies at high redshift, as sometimes observed \citep[e.g.][]{maiolino06}
and also envisaged by some updated theoretical schemes (see, for instance,
the interesting explorations of \citealp{kroupa03}, and \citealp{koeppen07}).
Figure~\ref{f1} is an illustrative example of this mechanism when tracing the contribution 
of SNe{\sc ii} to metal enhancement with varying the IMF power-law index.

A major advantage of taking yield metallicity as a reference marker in our
analysis is that 
$\cal Z$ can easily be related to other key physical processes that 
constrain galaxy chemical evolution in its different facets. 
The ISM metal enrichment is an evident example in this sense. 
Chemical composition of the gaseous phase inside a galaxy is the resulting 
balance of two basic mechanisms. From one hand, one should carefully consider  
the bulk of processed mass returned by stars to the ISM through SN events 
and quiescent stellar wind. On the other hand, this input has to be
properly assessed in the physical context of dynamical evolution of the disk,
such as to trace the way processed mass is eventually ``diluted'' within 
the fresh primordial gas. 

In tackling the problem (Sec.~3.1), we tried to maintain our formal 
treatment as much as possible free from 
any {\it ``ab initio''} assumptions that might bias our conclusions. 
The returned mass flow is therefore dealt with in eq.~(\ref{eq:zgas})
in terms of the parameterized fraction $f(t)$ of net stellar mass returned 
to the ISM within the time $t$ (see footnote~\ref{foot8}). In our notation,
this component is always traced separately from the
primordial gas component ${\cal G}(t)$, so that the global gas amount in the disk comes throughout 
as a sum of the two nominal contributions, as in eq.~(\ref{eq:gtot}).
In its formal elaboration (see eq.~\ref{eq:zgas2} and \ref{eq:zgas3}), 
the ISM metallicity ($Z_{\rm gas}$) can eventually be set in terms 
of yield metallicity, and derives from just two leading parameters, namely 
$\cal G$ and $f$ (see Fig.~\ref{gf}). For the latter, a safe upper limit
has been placed, such as $f \lesssim 0.3$ at any age for a Salpeter IMF,
according to a full discussion of the SSP case, as in Sec.~3.1.3.

The case of the Milky Way must be the obvious and preeminent testbed
for our theoretical framework. In particular, the expected evolution of the yield
metallicity is compared, in Fig.~\ref{tutti}, with the observed AMR, as traced
by different stellar samples in the solar neighborhood.
By definition, at every epoch $\cal Z$ caps the value of the ISM
metallicity, and therefrom edges the composition of newly formed stars.
As discussed in Sec.~3.1, the study of the 
$\Delta [Fe/H] \simeq \log (Z_{\rm gas}/{\cal Z})$ difference is 
richful of important information to constrain the Galaxy star formation history.
We have been able, for instance, to set a firm upper limit to the Galaxy birthrate, such
as $b \lesssim 0.5$ (see eq.~\ref{eq:birthb}), while the current contribution
of the primordial gas cannot exceed roughly one-third of the local mass density
of the disk (see eq.~\ref{eq:go}). A further boundary condition can also be
posed to the chemical enrichment ratio, that must be $R \lesssim 5$.

A similar comparison, performed between ``analytical'' model 
predictions from several reference codes in the literature 
and observed AMR, is summarized in Fig.~\ref{slopes}. The broad
range of envisaged scenarios predicted by theory leads to a spread of up 
to 0.5~dex in the value of stellar $[Fe/H]$ at a given age. We want to look 
at this figure as a measure of the intrinsic limit of theory 
to master the combined interplay among the many leading mechanisms 
that modulate chemical evolution. To a closer analysis, however, it has to 
be recognized a clear tendency of models to predict a sharper chemical 
evolution of Galaxy disk, such as to exceed, in most cases, the observed stellar 
metallicity at the present epoch. Evidently, the shallower
enrichment rate indicated by the observations points to a still unsettled 
problem with the primeval steps of disk evolution, and more specifically 
with the pre-enrichment processes related to the well known G-dwarf problem.
It is interesting to see, in this regard, that when the observed stellar 
$[Fe/H]$ distribution is explicitely dealt with in our analysis, as in 
Sec.~3.2.1, the implied birthrate ($b = 0.6\pm 0.1$) tends to oversize the 
corresponding figure derived from the AMR. Such a more ``delayed'' star formation
is in fact a sign for primeval stars to have formed ``elsewhere'' (alias
in the ``thin'' disk, as claimed by the standard established scenario).

An outstanding dyscrasia of any theoretical scheme for Galaxy chemical
evolution is that no spread in $[Fe/H]$ can be admitted among coeval stars
populating the same region of the system,
being the AMR univocally settled by the boundary physical conditions of
the model. This is evidently at odds with what we observe, at least in the 
solar neighborhood, where an intrinsic $\sigma [Fe/H]$ up to 0.2~dex 
seems an unquestionable feature for the local stellar population (see 
Fig.~\ref{allstars2}). As an important consequence of the metal dispersion,
one may actually conclude that, disregarding age, about four out of five 
stars in the solar vicinity approach the expected yield metallicity 
within a factor of two. When interpreted in terms of mixing properties 
between enriched and primordial gas, such a tuned distribution clearly
demonstrates that star formation in the Galaxy only proceeded, all the time,
in a highly contaminated environment. This claim can be better quantified
by means of Fig.~\ref{gf}, which confirms that the processed stellar mass
returned to the ISM must in fact be the prevailing component to
gas density in star-forming regions if we want $[Fe/H]$ of fresh stars 
at any epoch to closely match (i.e.\ within $\Delta [Fe/H] \simeq 0.3$~dex)
the yield metallicity, as observed.

This evidence severely tackles the classical scheme
{\it \`a la Schmidt}, where star-formation strength is fully driven 
by just the mean gas density. On the contrary, what it might be
the real case, is that star formation process in spiral 
galaxies behaves like a wild ``outbreak'', that stem from previously 
contaminated regions to progressively spread across the rest of galaxy body
\citep{nepveu88,nepveu89,barnes04}.

The possible implication of the Milky Way scenario to the more
general case of late-type galaxy evolution along the Hubble morphological
sequence has been the focus of Sec.~4.
As far as closeby external galaxies in the local Universe are
concerned, observations still hamper any firm assessment of galaxy metal
abundance. As a matter of fact
(see Fig.~\ref{zary}), the galaxy-to-galaxy scatter of $[Fe/H]$ distribution 
seems to largely overcome the nominal accuracy of any individual estimate,
a feature that makes evident the intrinsic difficulty in defining
a fair ``representative'' metallicity when such a composite stellar 
environment is dealt with, as in the late-type galaxy disks \citep{arimoto91}.

On the other hand, the combined study of the other leading parameters
can add important pieces of information, that greatly help constraining  
the problem. In this regard, our discussion along Sec.~4.3 basically grounds 
on three relevant relationships, that emerge from the observational
properties of different galaxy samples. In particular,

{\it i)} the down-sizing mechanism appears to govern star formation
in the local Universe, as well, with a clear relationship between current 
SFR and galaxy stellar mass (see Fig.~\ref{lga});

{\it ii)} the ``delayed'' star formation among low-mass galaxies, as 
implied by the inverse $b$-$M^*_{\rm gal}$ dependendence, naturally leads 
to a more copious gas fraction when moving from giant to dwarf galaxies, 
as observed indeed (see Fig.~\ref{gar});

{\it iii)} the physical relationship of star formation mechanisms
with galaxy stellar mass can also aproximately fit into a straight 
morphological scheme (Fig.~\ref{mtype}), where galaxies with decreasing mass
more likely take the look of later-type spirals.
It has to remain clear, however, that mass {\it not} morphology is the
primary parameter that govern late-type galaxy properties \citep{boselli01}.

Once accounting for galaxy chemical properties, by taking for instance
Oxygen abundance of H{\sc ii} regions as a tracer, the blurring trend 
with morphology leaves space to a much cleaner relationship with
galaxy stellar mass, as in the upper panel of Fig.~\ref{sav}.
Facing the fact that low-mass systems also appear
richer in gas and poorer in metals, one might even speculate on a 
simple scenario where the observed trend is in fact merely the result 
of the softening mechanism, that leads enriched stellar mass to mix
into fresh unprocessed gas in the ISM.
Actually, when galaxy data are properly rescaled to a fixed 
reference value of ${\cal G}_{\rm tot}$, through eq.~(\ref{eq:ggg}), 
one sees (lower panel of Fig.~\ref{sav}) that most of the 
$[O/H]$ vs.\ $M^*_{\rm gal}$ trend is recovered, leaving a nearly 
flat $[O/H]$ distribution along the entire galaxy mass range. This result 
is consistent with our theoretical predictions for a nearly constant 
yield metallicity along the entire late-type galaxy sequence.

A striking evidence, when sampling the local Universe, is that spiral 
galaxies appear all dangerously close to shut down their venture as stars-forming 
systems, being on the verge of exhausting their own gas resources. A so 
special case, and our somewhat suspicious temporal 
location as privileged observers, did not escape the analysis of many 
authors in the literature, leading \citet{roberts63} first to wonder
about the real ``residual life'' left to spiral systems according to their 
current gas capabilities. 
Relying on our notation, the so-called Roberts time ($t_R$) easily derives 
in its original definition, namely as the timescale 
for the primordial gas to vanish (see eq.~\ref{eq:rob}). On the other 
hand, and more interestingly, one would like to assess the even more general 
case, when the global gas consumption (i.e.\ including returned stellar mass) 
eventually sets (or overcome) the ultimate limit to galaxy star formation 
\citep{kennicutt94}.

Equation~(\ref{eq:tre}) shows that, as far a the down-sizing prescriptions set 
the reference relationship between birthrate and galaxy mass, 
one has to conclude that spiral galaxies cannot escape an age
where no stars will form at all. In fact, if returned mass is left as
the only gas supplier to the ISM, then the maximum birthrate that a galaxy
can sustain is of the order of $b_{\rm max} \simeq \dot{f}\,t_H/(1-f) \ll f/(1-f)$,
which must evidently be much less than $\sim 0.45$, for a Salpeter IMF.
As a result, only massive ($M_{\rm gal} \gtrsim 10^{11}~M_\odot$) Sa/Sb 
spirals may have some chance to extend their active life for a while
(namely for $\sim 0.3\,t_H$, as indicated by Fig.~\ref{trob}). On the 
contrary, no way out seems to be envisaged for dwarf systems, that will soon
cease their star formation activity unless to drastically 
reduce their apparent birthrate below the $b_{\rm max}$ figure.

\section*{Acknowledgments}
I'd like to thank Luis Carrasco and Peppo Gavazzi for so many inspiring 
discussions along all these years, when gazing together at stars and galaxies atop 
the Mexican mountains. Thanks are also due to Robert Kennicutt, for his wise 
comments to the draft of this paper, and to Giovanni Carraro and Ivo Saviane, 
for their warm hospitality during my visit at the ESO premises in Santiago de 
Chile, where part of this work has been conceived.
Finally, the anonymous referee is especially acknowledged for a number of very 
constructive suggestions, which greatly helped better focus the key issues of the paper.

This work has made extensive use of different on-line extragalactic databases, 
namely the NASA/IPAC Extragalactic Database (NED), operated by JPL/CIT under 
contract with NASA, the Hyper-Linked Extragalactic Databases and Archives
(HyperLeda) based at the Lyon University, and the VizieR catalog
service of the Centre de Donn\'ees astronomiques de Strasbourg. Part of the
data retrieval has been eased by the DEXTER graphic interface, hosted
at the German Astrophysical Virtual Observatory (GAVO) in Heidelberg.

\label{lastpage}
\bsp

\begin{thebibliography}{}
\bibitem[\protect\citeauthoryear{Alib{\'e}s, Labay, \& Canal}{2001}]{alibes01} Alib{\'e}s A., Labay J., Canal R., 2001, A\&A, 370, 1103 
\bibitem[\protect\citeauthoryear{Arimoto \& Jablonka}{1991}]{arimoto91} Arimoto N., Jablonka P., 1991, A\&A, 249, 374 (AJ91)
\bibitem[\protect\citeauthoryear{Arimoto \& Tarrab}{1990}]{arimoto90} Arimoto N., Tarrab I., 1990, A\&A, 228, 6 
\bibitem[\protect\citeauthoryear{Arimoto \& Yoshii}{1986}]{arimoto86} Arimoto N., Yoshii Y., 1986, A\&A, 164, 260 
\bibitem[\protect\citeauthoryear{Arimoto \& Yoshii}{1987}]{arimoto87} Arimoto N., Yoshii Y., 1987, A\&A, 173, 23 
\bibitem[\protect\citeauthoryear{Baade}{1944}]{baade44} Baade W., 1944, ApJ, 100, 137 
\bibitem[\protect\citeauthoryear{Barnes}{2004}]{barnes04} Barnes J.~E., 2004, MNRAS, 350, 798 
\bibitem[\protect\citeauthoryear{Beers, Preston, \& Shectman}{1985}]{beers85} Beers T.~C., Preston G.~W., Shectman S.~A., 1985, AJ, 90, 2089 
\bibitem[\protect\citeauthoryear{Beers, Preston, \& Shectman}{1992}]{beers92} Beers T.~C., Preston G.~W., Shectman S.~A., 1992, AJ, 103, 1987 
\bibitem[\protect\citeauthoryear{Bessell, Sutherland, \& Ruan}{1991}]{bessell91} Bessell M.~S., Sutherland R.~S., Ruan K., 1991, ApJ, 383, L71 
\bibitem[\protect\citeauthoryear{Blanton et al.}{2005}]{blanton05} Blanton M.~R., Eisenstein D., Hogg D.~W., Schlegel D.~J., Brinkmann J., 2005, ApJ, 629, 143 
\bibitem[\protect\citeauthoryear{Boissier \& Prantzos}{1999}]{boissier99} Boissier S., Prantzos N., 1999, MNRAS, 307, 857 
\bibitem[\protect\citeauthoryear{Boissier et al.}{2003}]{boissier03} Boissier S., Prantzos N., Boselli A., Gavazzi G., 2003, MNRAS, 346, 1215 
\bibitem[\protect\citeauthoryear{Boselli et al.}{2001}]{boselli01} Boselli A., Gavazzi G., Donas J., Scodeggio M., 2001, AJ, 121, 753 
\bibitem[\protect\citeauthoryear{Brandt et al.}{2010}]{brandt10} Brandt T.~D., Tojeiro R., Aubourg {\'E}., Heavens A., Jimenez R., Strauss M.~A., 2010, AJ, 140, 804 
\bibitem[\protect\citeauthoryear{Buzzoni}{1989}]{buzzoni89} Buzzoni A., 1989, ApJS, 71, 817 
\bibitem[\protect\citeauthoryear{Buzzoni}{1995}]{buzzoni95} Buzzoni A., 1995, ApJS, 98, 69 
\bibitem[\protect\citeauthoryear{Buzzoni}{2002}]{buzzoni02} Buzzoni A., 2002, AJ, 123, 1188 
\bibitem[\protect\citeauthoryear{Buzzoni}{2005}]{buzzoni05} Buzzoni A., 2005, MNRAS, 361, 725 
\bibitem[\protect\citeauthoryear{Calzetti}{1999}]{calzetti99} Calzetti D., 1999, MmSAI, 70, 715 
\bibitem[\protect\citeauthoryear{Carigi}{1994}]{carigi94} Carigi L., 1994, ApJ, 424, 181 
\bibitem[\protect\citeauthoryear{Carlberg et al.}{1985}]{carlberg85} Carlberg R.~G., Dawson P.~C., Hsu T., Vandenberg D.~A., 1985, ApJ, 294, 674 
\bibitem[\protect\citeauthoryear{Carraro, Ng, \& Portinari}{1998}]{carraro98} Carraro G., Ng Y.~K., Portinari L., 1998, MNRAS, 296, 1045 
\bibitem[\protect\citeauthoryear{Casagrande et al.}{2007}]{casagrande07} Casagrande L., Flynn C., Portinari L., Girardi L., Jimenez R., 2007, MNRAS, 382, 1516 
\bibitem[\protect\citeauthoryear{Cayrel}{1996}]{cayrel96} Cayrel R., 1996, A\&ARv, 7, 217 
\bibitem[\protect\citeauthoryear{Cellone \& Buzzoni}{2005}]{cellone05} Cellone S.~A., Buzzoni A., 2005, MNRAS, 356, 41 
\bibitem[\protect\citeauthoryear{Chevalier}{1976}]{chevalier76} Chevalier R.~A., 1976, Nature, 260, 689 
\bibitem[\protect\citeauthoryear{Chiappini, Matteucci, \& Gratton}{1997}]{chiappini97} Chiappini C., Matteucci F., Gratton R., 1997, ApJ, 477, 765 
\bibitem[\protect\citeauthoryear{Clayton}{1983}]{clayton83} Clayton, D.D. 1983, Principles of stellar evolution and nuclear nucleosynthesis, (Univ. of Chicago Press: Chicago)
\bibitem[\protect\citeauthoryear{Clegg, Tomkin, \& Lambert}{1981}]{clegg81} Clegg R.~E.~S., Tomkin J., Lambert D.~L., 1981, ApJ, 250, 262 
\bibitem[\protect\citeauthoryear{Cooper et al.}{2007}]{cooper07} Cooper M.~C., et al., 2007, MNRAS, 376, 1445 
\bibitem[\protect\citeauthoryear{Cowie et al.}{1996}]{cowie96} Cowie L.~L., Songaila A., Hu E.~M., Cohen J.~G., 1996, AJ, 112, 839 
\bibitem[\protect\citeauthoryear{Dahlen et al.}{2004}]{dahlen04} Dahlen T., et al., 2004, ApJ, 613, 189 
\bibitem[\protect\citeauthoryear{Dallaporta}{1973}]{dallaporta73} Dallaporta N., 1973, A\&A, 29, 393 
\bibitem[\protect\citeauthoryear{de Vaucouleurs et al.}{1991}]{rc3} de Vaucouleurs G., de Vaucouleurs A., Corwin H.~G., Jr., Buta R.~J., Paturel G., Fouque P., 1991, Third Reference Catalog of Bright Galaxies. Springer, Heidelberg
\bibitem[\protect\citeauthoryear{Dopita \& Ryder}{1994}]{dopita94} Dopita M.~A., Ryder S.~D., 1994, ApJ, 430, 163 
\bibitem[\protect\citeauthoryear{Dressler}{1980}]{dressler80} Dressler A., 1980, ApJ, 236, 351 
\bibitem[\protect\citeauthoryear{Edvardsson et al.}{1993}]{edvardsson93} Edvardsson B., Andersen J., Gustafsson B., Lambert D.~L., Nissen P.~E., Tomkin J., 1993, A\&A, 275, 101 
\bibitem[\protect\citeauthoryear{Ferrini et al.}{1992}]{ferrini92} Ferrini F., Matteucci F., Pardi C., Penco U., 1992, ApJ, 387, 138 
\bibitem[\protect\citeauthoryear{Firmani \& Tutukov}{1992}]{firmani92} Firmani C., Tutukov A., 1992, A\&A, 264, 37 
\bibitem[\protect\citeauthoryear{Flynn et al.}{2006}]{flynn06} Flynn C., Holmberg J., Portinari L., Fuchs B., Jahrei{\ss} H., 2006, MNRAS, 372, 1149 
\bibitem[\protect\citeauthoryear{Friel}{1995}]{friel95} Friel E.~D., 1995, ARA\&A, 33, 381 
\bibitem[\protect\citeauthoryear{Fukugita \& Kawasaki}{2006}]{fukugita06} Fukugita M., Kawasaki M., 2006, ApJ, 646, 691 
\bibitem[\protect\citeauthoryear{Garcia}{1993}]{garcia93} Garcia A.~M., 1993, A\&AS, 100, 47 
\bibitem[\protect\citeauthoryear{Garnett}{2002}]{garnett02} Garnett D.~R., 2002, ApJ, 581, 1019 
\bibitem[\protect\citeauthoryear{Gavazzi}{1993}]{gavazzi93} Gavazzi G., 1993, ApJ, 419, 469 
\bibitem[\protect\citeauthoryear{Gavazzi, Pierini, \& Boselli}{1996}]{gavazzi96} Gavazzi G., Pierini D., Boselli A., 1996, A\&A, 312, 397 
\bibitem[\protect\citeauthoryear{Gavazzi et al.}{2010}]{gavazzi10} Gavazzi G., Fumagalli M., Cucciati O., Boselli A., 2010, A\&A, 517, A73 
\bibitem[\protect\citeauthoryear{Giavalisco et al.}{2004}]{giavalisco04} Giavalisco M., et al., 2004, ApJ, 600, L103 
\bibitem[\protect\citeauthoryear{Gilmore \& Wyse}{1986}]{gilmore86} Gilmore G., Wyse R.~F.~G., 1986, Nature, 322, 806
\bibitem[\protect\citeauthoryear{Giovagnoli \& Tosi}{1995}]{giovagnoli95} Giovagnoli A., Tosi M., 1995, MNRAS, 273, 499 
\bibitem[\protect\citeauthoryear{Gordon, Calzetti, \& Witt}{1997}]{gordon97} Gordon K.~D., Calzetti D., Witt A.~N., 1997, ApJ, 487, 625 
\bibitem[\protect\citeauthoryear{Gorgas, Jablonka, \& Goudfrooij}{2007}]{gorgas07} Gorgas J., Jablonka P., Goudfrooij P., 2007, A\&A, 474, 1081 
\bibitem[\protect\citeauthoryear{Greggio}{1997}]{greggio97} Greggio L., 1997, MNRAS, 285, 151 
\bibitem[\protect\citeauthoryear{Greggio}{2005}]{greggio05} Greggio L., 2005, A\&A, 441, 1055 
\bibitem[\protect\citeauthoryear{Greggio \& Renzini}{1983}]{greggio83} Greggio L., Renzini A., 1983, A\&A, 118, 217 
\bibitem[\protect\citeauthoryear{Grevesse \& Sauval}{1998}]{grevesse98} Grevesse N., Sauval A.~J., 1998, SSRv, 85, 161 
\bibitem[\protect\citeauthoryear{Hashimoto}{1995}]{hashimoto95} Hashimoto M., 1995, PThPh, 94, 663 
\bibitem[\protect\citeauthoryear{Henry \& Worthey}{1999}]{henry99} Henry R.~B.~C., Worthey G., 1999, PASP, 111, 919 
\bibitem[\protect\citeauthoryear{Hillebrandt et al.}{2003}]{hillebrandt03} Hillebrandt W., Niemeyer J.~C., Reinecke M., Travaglio C., 2003, MmSAI, 74, 942 
\bibitem[\protect\citeauthoryear{Holmberg, Nordstr{\"o}m, \& Andersen}{2007}]{holmberg07} Holmberg J., Nordstr{\"o}m B., Andersen J., 2007, A\&A, 475, 519 
\bibitem[\protect\citeauthoryear{Jablonka, Martin, \& Arimoto}{1996}]{jablonka96} Jablonka P., Martin P., Arimoto N., 1996, AJ, 112, 1415 
\bibitem[\protect\citeauthoryear{Jablonka, Gorgas, \& Goudfrooij}{2007}]{jablonka07} Jablonka P., Gorgas J., Goudfrooij P., 2007, A\&A, 474, 763 
\bibitem[\protect\citeauthoryear{Jonsell et al.}{2005}]{jonsell05} Jonsell K., Edvardsson B., Gustafsson B., Magain P., Nissen P.~E., Asplund M., 2005, A\&A, 440, 321 
\bibitem[\protect\citeauthoryear{Kasen \& Plewa}{2007}]{kasen07} Kasen D., Plewa T., 2007, ApJ, 662, 459 
\bibitem[\protect\citeauthoryear{Kennicutt}{1983}]{kennicutt83} Kennicutt R.~C., Jr., 1983, ApJ, 272, 54 
\bibitem[\protect\citeauthoryear{Kennicutt}{1998}]{kennicutt98} Kennicutt R.~C., Jr., 1998, ARA\&A, 36, 189 
\bibitem[\protect\citeauthoryear{Kennicutt, Tamblyn, \& Congdon}{1994}]{kennicutt94} Kennicutt R.~C., Jr., Tamblyn P., Congdon C.~E., 1994, ApJ, 435, 22 
\bibitem[\protect\citeauthoryear{Khokhlov, Mueller, \& Hoeflich}{1993}]{khokhlov93} Khokhlov A., Mueller E., Hoeflich P., 1993, A\&A, 270, 223 
\bibitem[\protect\citeauthoryear{Kobayashi et al.}{1998}]{kobayashi98} Kobayashi C., Tsujimoto T., Nomoto K., Hachisu I., Kato M., 1998, ApJ, 503, L155 
\bibitem[\protect\citeauthoryear{Koeppen \& Arimoto}{1990}]{koeppen90} Koeppen J., Arimoto N., 1990, A\&A, 240, 22 
\bibitem[\protect\citeauthoryear{K{\"o}ppen, Weidner, \& Kroupa}{2007}]{koeppen07} K{\"o}ppen J., Weidner C., Kroupa P., 2007, MNRAS, 375, 673 
\bibitem[\protect\citeauthoryear{Kroupa, Tout, \& Gilmore}{1993}]{kroupa93} Kroupa P., Tout C.~A., Gilmore G., 1993, MNRAS, 262, 545 
\bibitem[\protect\citeauthoryear{Kroupa \& Weidner}{2003}]{kroupa03} Kroupa P., Weidner C., 2003, ApJ, 598, 1076 
\bibitem[\protect\citeauthoryear{Kuzio de Naray, McGaugh, \& de Blok}{2004}]{kuzio04} Kuzio de Naray R., McGaugh S.~S., de Blok W.~J.~G., 2004, MNRAS, 355, 887 
\bibitem[\protect\citeauthoryear{Larson}{1976}]{larson76} Larson R.~B., 1976, MNRAS, 176, 31 
\bibitem[\protect\citeauthoryear{Larson \& Tinsley}{1978}]{larson78} Larson R.~B., Tinsley B.~M., 1978, ApJ, 219, 46 
\bibitem[\protect\citeauthoryear{Larson, Tinsley, \& Caldwell}{1980}]{larson80} Larson R.~B., Tinsley B.~M., Caldwell C.~N., 1980, ApJ, 237, 692 
\bibitem[\protect\citeauthoryear{Lee, Grebel, \& Hodge}{2003}]{lee03} Lee H., Grebel E.~K., Hodge P.~W., 2003, A\&A, 401, 141 
\bibitem[\protect\citeauthoryear{Leitherer et al.}{1999}]{leitherer99} Leitherer C., et al., 1999, ApJS, 123, 3 
\bibitem[\protect\citeauthoryear{Lequeux et al.}{1979}]{lequeux79} Lequeux J., Peimbert M., Rayo J.~F., Serrano A., Torres-Peimbert S., 1979, A\&A, 80, 155 
\bibitem[\protect\citeauthoryear{Maciel}{2001}]{maciel01} Maciel W.~J., 2001, Ap\&SS, 277, 545 
\bibitem[\protect\citeauthoryear{Maeder}{1992}]{maeder92} Maeder A., 1992, A\&A, 264, 105 
\bibitem[\protect\citeauthoryear{Mallik \& Mallik}{1985}]{mallik85} Mallik D.~C.~V., Mallik S.~V., 1985, JApA, 6, 113 
\bibitem[\protect\citeauthoryear{Maiolino et al.}{2006}]{maiolino06} Maiolino R., et al., 2006, MmSAI, 77, 643 
\bibitem[\protect\citeauthoryear{Mannucci et al.}{2005}]{mannucci05} Mannucci F., Della Valle M., Panagia N., Cappellaro E., Cresci G., Maiolino R., Petrosian A., Turatto M., 2005, A\&A, 433, 807 
\bibitem[\protect\citeauthoryear{Maoz, Sharon,\& Gal-Yam}{2010}]{maoz10} Maoz D., Sharon K., Gal-Yam A., 2010, ApJ, 722, 1879
\bibitem[\protect\citeauthoryear{Marigo, Chiosi, \& Kudritzki}{2003}]{marigo03} Marigo P., Chiosi C., Kudritzki R.-P., 2003, A\&A, 399, 617 
\bibitem[\protect\citeauthoryear{Marino et al.}{2010}]{marino10} Marino A., Bianchi L., Rampazzo R., Buson L.~M., Bettoni D., 2010, A\&A, 511, A29 
\bibitem[\protect\citeauthoryear{Matteucci}{2003}]{matteucci03} Matteucci F., 2003, The Chemical Evolution of the Galaxy, Astrophysics \& Space Science Library (Dordrecht: Kluwer Academic Publ.) 
\bibitem[\protect\citeauthoryear{Matteucci}{2004}]{matteucci04} Matteucci F., 2004, in Origin and Evolution of the Elements, from the Carnegie Obs. Centennial Symp., ed. A. McWilliam \& M. Rauch (Cambridge Univ. Press), p. 85 
\bibitem[\protect\citeauthoryear{Matteucci \& Fran\c{c}ois}{1989}]{matteucci89} Matteucci F., Fran\c{c}ois P., 1989, MNRAS, 239, 885 
\bibitem[\protect\citeauthoryear{Matteucci \& Greggio}{1986}]{matteucci86} Matteucci F., Greggio L., 1986, A\&A, 154, 279 
\bibitem[\protect\citeauthoryear{Matteucci \& Recchi}{2001}]{matteucci01} Matteucci F., Recchi S., 2001, ApJ, 558, 351
\bibitem[\protect\citeauthoryear{Melchior, Combes, \& Gould}{2007}]{melchior07} Melchior A.-L., Combes F., Gould A., 2007, A\&A, 462, 965 
\bibitem[\protect\citeauthoryear{Meusinger, Stecklum, \& Reimann}{1991}]{meusinger91} Meusinger H., Stecklum B., Reimann H.-G., 1991, A\&A, 245, 57 
\bibitem[\protect\citeauthoryear{Mihara \& Takahara}{1996}]{mihara96} Mihara K., Takahara F., 1996, PASJ, 48, 467 
\bibitem[\protect\citeauthoryear{Miller \& Scalo}{1979}]{miller79} Miller G.~E., Scalo J.~M., 1979, ApJS, 41, 513 
\bibitem[\protect\citeauthoryear{Milone \& Milone}{1988}]{milone88} Milone L.~A., Milone A.~A.~E., 1988, Ap\&SS, 150, 299 
\bibitem[\protect\citeauthoryear{Neff et al.}{2008}]{neff08} Neff, S.G., Hollis, J.E. \& Offenberg, J.D.\ 2008 ``Galex Observer's Guide'', Web edition available on line at {\sf http://galexgi.gsfc.nasa.gov/docs/galex/Documents}
\bibitem[\protect\citeauthoryear{Nepveu}{1989}]{nepveu89} Nepveu M., 1989, A\&A, 224, 86 
\bibitem[\protect\citeauthoryear{Nepveu}{1988}]{nepveu88} Nepveu M., 1988, A\&A, 193, 173 
\bibitem[\protect\citeauthoryear{Nilson}{1973}]{nilson73} Nilson P., 1973, Uppsala General Catalogue of Galaxies, Acta Universitatis Upsalienis, Nova Regiae Societatis Upsaliensis, Series v
\bibitem[\protect\citeauthoryear{Nittler}{2005}]{nittler05} Nittler L.~R., 2005, ApJ, 618, 281 
\bibitem[\protect\citeauthoryear{Nomoto}{1980}]{nomoto80} Nomoto K., 1980, in Type I supernovae, proc. of the Texas Workshop (Austin, Univ. of Texas) p. 164 
\bibitem[\protect\citeauthoryear{Nomoto, Thielemann, \& Yokoi}{1984}]{nomoto84} Nomoto K., Thielemann F.-K., Yokoi K., 1984, ApJ, 286, 644 
\bibitem[\protect\citeauthoryear{Nordstr{\"o}m et al.}{2004}]{nordstrom04} Nordstr{\"o}m B., et al., 2004, A\&A, 418, 989 
\bibitem[\protect\citeauthoryear{Oey}{2000}]{oey00} Oey M.~S., 2000, ApJ, 542, L25 
\bibitem[\protect\citeauthoryear{Pagel}{1989}]{pagel89} Pagel B.~E.~J., 1989, in Evolutionary phenomena in galaxies, eds. J.E. Beckman \& B.E.J. Pagel (Cambridge University Press) p. 201 
\bibitem[\protect\citeauthoryear{Pagel}{1997}]{pagel97} Pagel B.E.J., 1997, Nucleosynthesis and chemical evolution of galaxies, (Cambridge Univ. Press: Cambridge)
\bibitem[\protect\citeauthoryear{Pagel \& Patchett}{1975}]{pagel75} Pagel B.~E.~J., Patchett B.~E., 1975, MNRAS, 172, 13 
\bibitem[\protect\citeauthoryear{Pagel \& Tautvaisiene}{1995}]{pagel95} Pagel B.~E.~J., Tautvaisiene G., 1995, MNRAS, 276, 505 
\bibitem[\protect\citeauthoryear{Pagel et al.}{1992}]{pagel92} Pagel B.~E.~J., Simonson E.~A., Terlevich R.~J., Edmunds M.~G., 1992, MNRAS, 255, 325 
\bibitem[\protect\citeauthoryear{Panagia, Della Valle, \& Mannucci}{2007}]{panagia07} Panagia N., Della Valle M., Mannucci F., 2007, in The multicolored landscape of compact objects and their explosive origins, eds. L.A. Antonelli et al., AIP Conf. Proc., Vol 924 (American Inst. Phys., New York) p. 373 
\bibitem[\protect\citeauthoryear{Pardi \& Ferrini}{1994}]{pardi94} Pardi M.~C., Ferrini F., 1994, ApJ, 421, 491 
\bibitem[\protect\citeauthoryear{Paturel et al.}{2003}]{paturel03} Paturel G., Petit C., Prugniel P., Theureau G., Rousseau J., Brouty M., Dubois P., Cambr{\'e}sy L., 2003, A\&A, 412, 45 
\bibitem[\protect\citeauthoryear{Peimbert, Carigi, \& Peimbert}{2001}]{peimbert01} Peimbert M., Carigi L., Peimbert A., 2001, ApSSS, 277, 147 
\bibitem[\protect\citeauthoryear{Peimbert, Peimbert, \& Luridiana}{2002}]{peimbert02} Peimbert A., Peimbert M., Luridiana V., 2002, ApJ, 565, 668 
\bibitem[\protect\citeauthoryear{Peimbert, Rayo, \& Torres-Peimbert}{1978}]{peimbert78} Peimbert M., Rayo J.~F., Torres-Peimbert S., 1978, ApJ, 220, 516 
\bibitem[\protect\citeauthoryear{P{\'e}rez, S{\'a}nchez-Bl{\'a}zquez, \& Zurita}{2009}]{perez09} P{\'e}rez I., S{\'a}nchez-Bl{\'a}zquez P., Zurita A., 2009, A\&A, 495, 775 
\bibitem[\protect\citeauthoryear{Pilyugin \& Edmunds}{1996}]{pilyugin96} Pilyugin L.~S., Edmunds M.~G., 1996, A\&A, 313, 783 
\bibitem[\protect\citeauthoryear{Portinari, Chiosi, \& Bressan}{1998}]{portinari98} Portinari L., Chiosi C., Bressan A., 1998, A\&A, 334, 505 
\bibitem[\protect\citeauthoryear{Prantzos}{2008}]{prantzos08} Prantzos N., 2008, in Stellar nucleosynthesis: 50 years after B$^2$FH, eds. C. Charbonnel \& J.-P. Zahn, EAS Pub. Ser., Vol. 32 p. 311 
\bibitem[\protect\citeauthoryear{Prantzos \& Aubert}{1995}]{prantzos95} Prantzos N., Aubert O., 1995, A\&A, 302, 69 
\bibitem[\protect\citeauthoryear{Reinecke, Hillebrandt, \& Niemeyer}{2002}]{reinecke02} Reinecke M., Hillebrandt W., Niemeyer J.~C., 2002, A\&A, 386, 936 
\bibitem[\protect\citeauthoryear{Renzini \& Buzzoni}{1983}]{rb83} Renzini A., Buzzoni A., 1983, MemSAIt, 54, 739 
\bibitem[\protect\citeauthoryear{Renzini \& Buzzoni}{1986}]{rb86} Renzini A., Buzzoni A., 1986, ASSL, 122, 195 
\bibitem[\protect\citeauthoryear{Rifatto, Longo, \& Capaccioli}{1995}]{rifatto95} Rifatto A., Longo G., Capaccioli M., 1995, A\&AS, 114, 527 
\bibitem[\protect\citeauthoryear{Roberts}{1963}]{roberts63} Roberts M.~S., 1963, ARA\&A, 1, 149 
\bibitem[\protect\citeauthoryear{Roberts \& Haynes}{1994}]{roberts94} Roberts M.~S., Haynes M.~P., 1994, ARA\&A, 32, 115 
\bibitem[\protect\citeauthoryear{Rocha-Pinto \& Maciel}{1996}]{rochapinto96} Rocha-Pinto H.~J., Maciel W.~J., 1996, MNRAS, 279, 447 
\bibitem[\protect\citeauthoryear{Rocha-Pinto et al.}{2000}]{rochapinto00} Rocha-Pinto H.~J., Maciel W.~J., Scalo J., Flynn C., 2000, A\&A, 358, 850 
\bibitem[\protect\citeauthoryear{Romano et al.}{2005}]{romano05} Romano D., Chiappini C., Matteucci F., Tosi M., 2005, A\&A, 430, 491 
\bibitem[\protect\citeauthoryear{Ryan, Norris, \& Beers}{1996}]{ryan96} Ryan S.~G., Norris J.~E., Beers T.~C., 1996, ApJ, 471, 254 
\bibitem[\protect\citeauthoryear{Salpeter}{1955}]{salpeter55} Salpeter E.~E., 1955, ApJ, 121, 161 
\bibitem[\protect\citeauthoryear{Sandage}{1986}]{sandage86} Sandage A., 1986, A\&A, 161, 89 
\bibitem[\protect\citeauthoryear{Sandage \& Fouts}{1987}]{sandage87} Sandage A., Fouts G., 1987, AJ, 93, 74 
\bibitem[\protect\citeauthoryear{Saviane et al.}{2008}]{saviane08} Saviane I., Ivanov V.~D., Held E.~V., Alloin D., Rich R.~M., Bresolin F., Rizzi L., 2008, A\&A, 487, 901 
\bibitem[\protect\citeauthoryear{Scannapieco \& Bildsten}{2005}]{scannapieco05} Scannapieco E., Bildsten L., 2005, ApJ, 629, L85 
\bibitem[\protect\citeauthoryear{Scannapieco, Schneider, \& Ferrara}{2003}]{scannapieco03} Scannapieco E., Schneider R., Ferrara A., 2003, ApJ, 589, 35 
\bibitem[\protect\citeauthoryear{Serrano \& Peimbert}{1981}]{serrano81} Serrano A., Peimbert M., 1981, RMxAA, 5, 109 
\bibitem[\protect\citeauthoryear{Scalo \& Elmegreen}{2004}]{scalo04} Scalo J., Elmegreen B.~G., 2004, ARA\&A, 42, 275 
\bibitem[\protect\citeauthoryear{Schmidt}{1959}]{schmidt59} Schmidt M., 1959, ApJ, 129, 243 
\bibitem[\protect\citeauthoryear{Schmidt}{1963}]{schmidt63} Schmidt M., 1963, ApJ, 137, 758 
\bibitem[\protect\citeauthoryear{Smartt et al.}{2009}]{smartt09} Smartt S.~J., Eldridge J.~J., Crockett R.~M., Maund J.~R., 2009, MNRAS, 395, 1409 
\bibitem[\protect\citeauthoryear{Sommer-Larsen}{1991}]{sommerlarsen91} Sommer-Larsen J., 1991, MNRAS, 249, 368 
\bibitem[\protect\citeauthoryear{Sullivan et al.}{2006}]{sullivan06} Sullivan M., et al., 2006, ApJ, 648, 868 
\bibitem[\protect\citeauthoryear{Tasca et al.}{2009}]{tasca09} Tasca L.~A.~M., et al., 2009, A\&A, 503, 379 
\bibitem[\protect\citeauthoryear{Thielemann, Nomoto, \& Hashimoto}{1996}]{thielemann96} Thielemann F.-K., Nomoto K., Hashimoto M.-A., 1996, ApJ, 460, 408 
\bibitem[\protect\citeauthoryear{Thuan, Balkowski, \& Tran Thanh Van}{1992}]{thuan92} Thuan T.~X., Balkowski C., Tran Thanh Van J., 1992, Physics of Nearby Galaxies: Nature or Nurture? Proc. of the 27th Rencontre de Moriond, Ed. T. Xuan Thuan, Ch. Balkowski, and J. Tran Thanh Van (Gif-sur-Yvette: Ed. Frontieres), p.225
\bibitem[\protect\citeauthoryear{Timmes, Woosley, \& Weaver}{1995}]{timmes95} Timmes F.~X., Woosley S.~E., Weaver T.~A., 1995, ApJS, 98, 617 
\bibitem[\protect\citeauthoryear{Tinsley}{1980}]{tinsley80} Tinsley B.M., 1980, Fund. Cosm. Physics, 5, 287
\bibitem[\protect\citeauthoryear{Tornamb{\'e}\ \& Matteucci}{1986}]{tornambe86} Tornamb{\'e}\ A., Matteucci F., 1986, MNRAS, 223, 69 
\bibitem[\protect\citeauthoryear{Tosi}{1996}]{tosi96} Tosi M., 1996, in From stars to galaxies: the impact of stellar physics on galaxy evolution, eds. C. Leitherer, U. Fritze-von-Alvensleben, \& J. Huchra ASP Conf. Ser., Vol. 98,  (Astron. Soc. Pac., San Francisco) p. 299 
\bibitem[\protect\citeauthoryear{Trimble}{1991}]{trimble91} Trimble V., 1991, A\&ARv, 3, 1 
\bibitem[\protect\citeauthoryear{Tully \& Fisher}{1977}]{tully77} Tully R.~B., Fisher J.~R., 1977, A\&A, 54, 661 
\bibitem[\protect\citeauthoryear{Tully, Mould, \& Aaronson}{1982}]{tully82} Tully R.~B., Mould J.~R., Aaronson M., 1982, ApJ, 257, 527 
\bibitem[\protect\citeauthoryear{Twarog}{1980}]{twarog80} Twarog B.~A., 1980, ApJ, 242, 242 
\bibitem[\protect\citeauthoryear{van den Bergh}{1962}]{vandenbergh62} van den Bergh S., 1962, AJ, 67, 486 
\bibitem[\protect\citeauthoryear{van der Kruit}{1986}]{vanderkruit86} van der Kruit P.~C., 1986, A\&A, 157, 230 
\bibitem[\protect\citeauthoryear{van Zee \& Haynes}{2006}]{vanzee06} van Zee L., Haynes M.~P., 2006, ApJ, 636, 214 
\bibitem[\protect\citeauthoryear{Weidemann}{2000}]{weidemann00} Weidemann V., 2000, A\&A, 363, 647 
\bibitem[\protect\citeauthoryear{Whitford}{1962}]{whitford62} Whitford A.~E., 1962, in Problems of extra-galactic research, proc. of IAU Symp. no. 15, ed. G. Cunliffe McVittie (Macmillan Press, New York), p. 27 
\bibitem[\protect\citeauthoryear{Wirth}{1981}]{wirth81} Wirth A., 1981, AJ, 86, 981 
\bibitem[\protect\citeauthoryear{Woosley \& Weaver}{1995}]{woosley95} Woosley S.~E., Weaver T.~A., 1995, ApJS, 101, 181 
\bibitem[\protect\citeauthoryear{Woosley \& Janka}{2005}]{woosley05} Woosley S., Janka T., 2005, NatPh, 1, 147 
\bibitem[\protect\citeauthoryear{Wyse}{2006}]{wyse06} Wyse R.~F.~G., 2006, MemSAIt, 77, 1036 
\bibitem[\protect\citeauthoryear{Wyse \& Silk}{1989}]{wyse89} Wyse R.~F.~G., Silk J., 1989, ApJ, 339, 700 
\bibitem[\protect\citeauthoryear{Zaritsky, Kennicutt, \& Huchra}{1994}]{zaritsky94} Zaritsky D., Kennicutt R.~C., Jr., Huchra J.~P., 1994, ApJ, 420, 87 
\end{thebibliography}
\end{document}